\documentclass[conference]{IEEEtran}

\ifCLASSOPTIONcompsoc
  \usepackage[nocompress]{cite}
\else
  \usepackage{cite}
\fi
\ifCLASSINFOpdf
  \usepackage[pdftex]{graphicx}
  \graphicspath{ {./images/} }
\else
\fi
\usepackage{amsmath}
\usepackage{amssymb}
\usepackage{amsthm}
\usepackage{amsfonts}
\usepackage{algorithmicx}
\usepackage{algpseudocode}
\usepackage{fixltx2e}
\usepackage{stfloats}
\usepackage{url}
\usepackage{booktabs}
\usepackage{textcomp}
\usepackage{xcolor}
\usepackage{subcaption}
\usepackage{tcolorbox}
\usepackage{makecell}
\usepackage{paralist}
\usepackage{multirow}
\usepackage{multicol}
\usepackage{enumitem}
\usepackage{float}
\usepackage{listings}
\usepackage{pgfplots}

\usepackage[absolute,overlay]{textpos}
\pgfplotsset{compat=1.18}
\definecolor{myblue}{HTML}{1f77b4}
\definecolor{myorange}{HTML}{ff7f0e}

\newtheorem{theorem}{Theorem}
\newtheorem{definition}{Definition}

\newtheorem{remark}{Remark}

\newtheorem{proposition}{Proposition}

\newcommand{\Icorr}{\mathbb{I}_{corr}}
\newcommand{\Abrute}{\mathcal{A}_{brute}^{c,\kappa}}
\newcommand{\AbruteOpt}[1]{\mathcal{A}_{brute}^{c,\kappa,#1}}
\newcommand{\Afa}{\mathcal{A}_{FA}}
\newcommand{\adv}{\mathcal{A}}
\newcommand{\GameCRI}[2]{\mathcal{G}ame^{#1, #2}_{CR-I}}
\newcommand{\GameCRP}[2]{\mathcal{G}ame^{#1, #2}_{CR-P}}
\newcommand{\Pideal}{Ideal}
\newcommand{\G}{\mathcal{G}ame}
\newcommand{\DR}{\mathcal{D}_R}
\newcommand{\DT}{\mathcal{D}_T}
\newcommand{\DO}{\mathcal{D}_O}
\newcommand{\PBB}{\mathbb{PBB}}

\algdef{SE}{UponReceiving}{EndUponReceiving}[1]{\textbf{upon receiving} \mbox{#1}}{}%

\algdef{SE}{WhileReceiving}{EndWhileReceiving}[1]{\textbf{while receiving} \mbox{#1} \textbf{do}}{\textbf{end while}}%

\newcommand{\JQ}[1]{#1}
\newcommand{\JQC}[1]{\textcolor{purple}{#1}}
\newcommand{\JQdone}[1]{}

\newcommand{\TZdone}[1]{}

\newcommand{\MAdone}[1]{}

\begin{document}

\begin{textblock*}{\textwidth}(1in,0.5in)
    \centering
    \footnotesize\itshape
    This is the extended version of a paper accepted at IEEE Computer Security Foundations Symposium 2026 (CSF'26).
\end{textblock*}

%
\title{On the Necessity of Pre-agreed Secrets for Thwarting Last-minute Coercion: Vulnerabilities and Lessons From the Loki E-voting Protocol}


\author{
\IEEEauthorblockN{Jingxin Qiao}
\IEEEauthorblockA{School of Informatics \\
The University of Edinburgh\\
Edinburgh, UK \\
J.Qiao-3@sms.ed.ac.uk}
\and
\IEEEauthorblockN{Myrto Arapinis}
\IEEEauthorblockA{School of Informatics \\
The University of Edinburgh\\
Edinburgh, UK \\
marapini@inf.ed.ac.uk}
\and
\IEEEauthorblockN{Thomas Zacharias}
\IEEEauthorblockA{School of Computing Science \\
University of Glasgow\\
Glasgow, UK \\
Thomas.Zacharias@glasgow.ac.uk}
}


%

\maketitle

\begin{abstract}
Coercion-resistance (CR) is a crucial security property in e-voting systems. It ensures that an attacker cannot compel a voter to vote in a specific way by using threats or rewards. The Loki e-voting protocol, proposed by Giustolisi \emph{et al.} at IEEE S\&P (2024), introduces a novel design that mitigates last-minute coercion through a re-voting mechanism. It also aims to address the usability issues of the seminal JCJ e-voting protocol, specifically: i) the requirement that voters can store and hide pre-agreed credentials, and ii) the ability of voters to convincingly lie while being coerced. 

In this work, we identify two vulnerabilities in Loki. The first is a brute-force attack that compromises the integrity of the evasion strategy. Specifically, this attack allows an adversary to cast a ballot on behalf of their victim in a way that the evasion strategy cannot defend against, rendering it ineffective. The second vulnerability is a forced abstention attack, which allows an adversary to detect when their victim has complied with their instruction not to vote. We generalise the integrity attack to reveal a fundamental dilemma: without pre-agreed secret credentials, it is not possible to prevent last-minute coercion. Finally, we show how reverting to pre-agreed secret credentials fixes the aforementioned vulnerabilities and discuss the trade-off between tallying efficiency and stronger trust assumptions.
\end{abstract}


%

\section{Introduction}
\label{sec: introduction}

The promise of secure and efficient electronic voting (e-voting) is continually challenged by the persistent threat of coercion, a vulnerability that strikes at the core of democratic elections. Unlike traditional voting methods, e-voting's remote accessibility and digital nature introduce unique avenues for large-scale coercion and bribery, potentially undermining electoral integrity and eroding public trust. Therefore, achieving robust coercion-resistance is paramount for the successful deployment of e-voting systems in critical elections.

The evolution of coercion-resistant e-voting protocols has seen a shift from early fake-credential approaches, pioneered by Juels \emph{et al.}~\cite{juels2005coercion}, to more recent re-voting strategies. While fake-credential protocols, subsequently refined in works like JCJ*~\cite{cortier2022jcj}, Civitas~\cite{clarkson2008civitas}, \JQ{efficiency-focused improvements~\cite{araujo2016remote,grontas2019towards,spycher2011new},} and blockchain-based optimizations~\cite{yin2023scalable}, offer strong security guarantees, they often encounter usability and scalability issues. Re-voting protocols, such as DeVoS~\cite{muller2024devos} and VoteAgain~\cite{lueks2020voteagain}, aim to enhance practical viability by allowing voters to replace coerced ballots. However, these designs frequently introduce new complexities, including reliance on last-minute coercion-free windows and deniability challenges.

In this context, Giustolisi \emph{et al.} introduced Loki~\cite{giustolisi2023thwarting} in 2024, a re-voting protocol that aims to eliminate the need for a last-minute coercion-free period by leveraging a trusted Voting Server. Loki's core mechanism relies on the voter’s ``secret'' (re-)voting history, enabling the server to distinguish between freely cast ballots and those cast under coercion. The security of Loki is argued against an ad hoc definition of coercion-resistance, tailored specifically for its analysis.

In this work, we analyse the Loki protocol using the general definition proposed by Küsters \emph{et al.}~\cite{kusters2011verifiability, kusters2012game}, which captures coercion resistance through two conditions: (CR-Integrity) the high probability of successfully evading coercion, and (CR-Privacy) the $\delta$-bounded probability for the coercer of distinguishing between compliance and evasion. Our analysis reveals two vulnerabilities. Specifically, the Loki design, while innovative, is susceptible to attacks that exploit the predictability of voting patterns and the adversary's ability to infer voting histories, particularly in early election stages. These vulnerabilities arise from hard-coded assumptions in the original Loki paper's definition of coercion-resistance and the absence of a comprehensive model for voting pattern distributions, which adversaries can exploit in real-world scenarios.

Furthermore, we establish a no-go result, showing that last-minute coercion resistance is impossible for a large class of e-voting protocols—encompassing all previously proposed designs for coercion resistance—unless the voter and election authorities share pre-agreed secrets. Fixing Loki can thus only be achieved by reverting to fake credentials. For completeness, we show how re-introducing pre-agreed secrets into Loki's architectural design achieves coercion resistance. We name this resulting protocol CR-Loki, and conduct a formal security analysis of it. We discuss the trade-off between its enhanced efficiency and the stronger trust assumption it requires compared to the JCJ family of e-voting protocols.
\smallskip{}

\noindent\textbf{Summary of the contributions.}
\begin{enumerate}[leftmargin=*]
\item We employ Küsters \emph{et al.}'s definition of coercion-resistance to analyse the Loki protocol, formalizing its two requirements as a reachability and an indistinguishability property, \JQ{referred to as \emph{CR-Integrity} and \emph{CR-Privacy}}, respectively.
\item \JQ{We identify a brute-force attack that compromises CR-Integrity by exploiting weaknesses in Loki's signalling scheme. This attack allows a coercer to bypass its evasion strategy, rendering it ineffective. We support this analysis with real-world data drawn from elections in Estonia.}
\item \JQ{We examine Loki's noise distribution configuration, and uncover a forced-abstention attack breaking CR-Privacy.}
\item We establish a no-go result, demonstrating that a large class of protocols, including all well-known designs, cannot simultaneously satisfy the arbitrary coercion-free window assumption and avoid the need for pre-agreed secrets shared between voters and election authorities.
\end{enumerate}

\noindent\textbf{Related work.}
Given the critical importance of coercion-resistance in electronic voting, numerous protocols have been developed to incorporate this essential property. The seminal work by Juels \emph{et al.} in 2005~\cite{juels2005coercion} introduces an innovative fake-credential mechanism for achieving coercion-resistance. In this approach, voters use fake credentials when under coercion, and ballots cast with these fake credentials are subsequently removed during tallying through pairwise plaintext equivalence tests (PETs). This work has spurred extensive research. For example, Civitas~\cite{clarkson2008civitas} implements JCJ, while Yin \emph{et al.}~\cite{yin2023scalable} identify the computational complexity of $\mathcal{O}(n^2)$ as a scalability bottleneck and improve it to $\mathcal{O}(n)$ for blockchain-based decision-making. However, Cortier \emph{et al.}~\cite{cortier2022jcj} discover a statistical attack that compromises the JCJ protocol, particularly leaking information during revoting. Although they propose a fix, it proved to be computationally more expensive than the original JCJ. Park \emph{et al.}~\cite{park2024zkvoting} utilize a fake keys approach based on a novel nullifiable commitment scheme. All of these designs employ the fake-credential strategy. In contrast, Achenbach \emph{et al.}~\cite{achenbach2015improved} extend JCJ to enable deniable revoting, a different defense mechanism based on re-voting.

Re-voting allows voters to cast multiple ballots, providing an opportunity to replace a coerced ballot with a genuine one. This approach must ensure deniability, meaning that coerced voters should be able to plausibly deny having revoted. DeVoS~\cite{muller2024devos} achieves deniable revoting by dividing the voting phase into several rounds and introducing a trusted party to re-encrypt ballots for those who did not vote in a particular round, thereby preventing the coercer from distinguishing whether a ballot is new or a re-encryption of a previous one. However, DeVoS offers only deniable revoting, not full coercion-resistance, as voters could still sell their votes if bribed, violating the weaker notion of receipt-freeness. VoteAgain~\cite{lueks2020voteagain} is another protocol utilizing the re-voting strategy but was later found to have flawed assumptions, as noted by Haines \emph{et al.}~\cite{haines2023scalable}. Both DeVoS and VoteAgain require a coercion-free window before the end of the voting phase. Loki~\cite{giustolisi2023thwarting}, which also falls into this category, addresses a significant challenge by eliminating the need for such last-minute coercion-free periods, a common issue in many re-voting designs.

The definitions of coercion-resistance can be broadly categorized into two main classes. The first class focuses on providing generic definitions applicable to a wide range of problems, not just voting protocols but also multi-party computation. The earliest attempt was made by Canetti and Gennaro~\cite{canetti1996incoercible} in 1996, but their definition was limited to a weak form of coercion. Canetti \emph{et al.}~\cite{canetti2015adaptively} later modified the Universal Composability (UC) framework~\cite{canetti2001universally} to include an ideal incoercible message transmission functionality. However, this model requires the coerced party to be aware of the identities of other coerced or corrupted parties, a limitation also present in Unruh \emph{et al.}'s work~\cite{unruh2010universally}. Building on the UC framework, Alwen \emph{et al.}~\cite{alwen2015incoercible} proposed a model with four worlds: ideal deception, ideal coercion, real deception, and real coercion. The model requires that if the best environment's advantage in distinguishing between ideal deception and ideal coercion is $\Delta$, then the advantage in distinguishing between real deception and real coercion should also be $\Delta' = \Delta$. 
Küsters \emph{et al.}~\cite{kusters2012game, kusters2011verifiability} proposed a generic game-based definition for coercion resistance, parameterizing crucial factors that determine whether a voting protocol is coercion-resistant. This requires each protocol to instantiate its own definition for proving coercion resistance. JCJ and its subsequent revisions have been analysed against it.

The other approach in defining coercion-resistance is more ad hoc, with each protocol design typically accompanied by a tailored definition of coercion-resistance. For example, VoteAgain and Loki both present oracle game-based definitions inspired by the receipt-freeness definition in~\cite{chaidos2016beleniosrf}, where the coercer's capabilities are modeled through oracle access.
\section{Preliminaries}
\label{sec: preliminaries}

Throughout the text, we use $\lambda$ as the security parameter. \subsection{Coercion-Resistant E-Voting Scheme Syntax}
\label{subsec:coercion_resistant_syntax}

An e-voting scheme is defined with respect to a result function $\rho: (\mathbb{I}\times\mathbb{O}^*) \to \mathbb{W}$, where $\mathbb{I} = \{id_1, id_2, \dots, id_{n_v}\}$ represents the set of $n_v$ eligible voters, $\mathbb{O}$ denotes the set of voting options, and $\mathbb{W}$ is the result space of the election. The set of voting options $\mathbb{O}=\{o_1,\dots,o_{n_o}\}$ includes the abstention option, denoted by $\Phi\in\mathbb{O}$.

A typical e-voting scheme involves the following key parties: the voters $\mathbb{I}$, a trusted Voting Server $VS$, the talliers $\mathbb{T}$, the Ballot Box $\mathbb{BB}$, and the Public Bulletin Board $\PBB$. 
For ease of notation, we also use  $\mathbb{BB}$ and $\PBB$ to denote the set of messages maintained by the respective parties.
Some voters may be corrupted, and the remaining ones are honest but potentially coerced. We denote the set of $n_d$ corrupted voters by $\mathbb{I}_{corr} \subset \mathbb{I}$. The Voting Server is responsible for casting noise ballots (not counted in the final result) and is trusted solely for coercion-resistance. 
An e-voting scheme comprises distinct phases, typically \emph{registration}, \emph{casting}, and \emph{tallying}. The casting phase spans a discrete time interval of duration $t_{end} \in \mathbb{N}^*$. \MAdone{In the remainder of the paper we only talk about the options, so we do not need to introduce $\mathbb{C}$. The set of $n_c$ candidates is given by $\mathbb{C} = \{c_1, c_2, \dots, c_{n_c}\}$.} 

Furthermore, five discrete probability distributions specify an election. The voters' choice of election option follows distribution $\mathcal{D}_O$. The number of votes cast by each voter ($r^{id}_v$) and by the Voting Server ($r^{id}_{vs}$), follow distributions $\mathcal{D}^v_R$ and $\mathcal{D}^{vs}_R$, respectively, where $1 \leq r^{id}_v, r^{id}_{vs} \leq t_{end} + 1$. The times of casting votes for each voter ($\vec{t}^{id}_v$) and the Voting Server ($\vec{t}^{id}_{vs}$) follow distributions $\mathcal{D}^v_T$ and $\mathcal{D}^{vs}_T$, respectively, such that for all times $t$ in $\vec{t}^{id}_v$ and $\vec{t}^{id}_{vs}$, $t \in [0, t_{end}]$. 

\begin{definition}
\label{def:e-voting_scheme}
A coercion-resistant e-voting scheme $\Gamma^{\DR^{vs},\DT^{vs}}$ for a result function $\rho$ consists of a tuple of probabilistic polynomial-time (PPT) algorithms:
$(\mathsf{Setup}, \mathsf{Register}, \mathsf{Fake}, \mathsf{Vote}, \mathsf{Noise}, \mathsf{Include}, \mathsf{Publish}, \mathsf{Tally},$ $\mathsf{VerifyTally},$$\mathsf{VerifyVote})$
that operate as follows:
\begin{itemize}[leftmargin=*]
    \item $\mathsf{Setup}(1^{\lambda}, \mathbf{param})$: Given the security parameter $\lambda$ and public parameters $\mathbf{param} = (\mathbb{I}, \mathbb{O}, \DO, \DR^v, \DT^{v}, t_{end})$, this algorithm generates a tally key pair $(sk_T, pk_T)$\footnote{The secret tallying key is generated in a distributed way.} and a key pair for the Voting Server $(sk_{VS}, pk_{VS})$. 

    \item $\mathsf{Register}(id)$: Given a voter identity $id$, this algorithm outputs a public/private key pair $(usk_{id}, upk_{id})$ and a credential pair $(CR_{id}, cr_{id})$.

    \item $\mathsf{Fake}(id, \PBB, CR_{id}, cr_{id})$: Given a voter identity $id$, the public bulletin board, and their credentials, this algorithm outputs fake credential $\tilde{cr}_{id}$. 

    \item $\mathsf{Vote}(id, usk_{id}, pk_T, pk_{VS}, o, cr_{id})$: Given a voter identity $id$, private key $usk_{id}$, public keys of the tallier $pk_T$ and the Voting Server $pk_{VS}$, voting intention $o$, and secret credential $cr_{id}$, this algorithm outputs a ballot $\beta$.

    \item $\mathsf{Noise}(id, sk_{VS}, pk_T, \PBB)$: Given a voter identity $id$, private key of the Voting Server $sk_{VS}$, tallier's public key $pk_T$, and the public bulletin board $\PBB$, this algorithm outputs a noise ballot $\beta$. The number and timing of noise ballots follow the distributions $\DR^{vs}$ and $\DT^{vs}$ respectively.

    \item $\mathsf{Include}(id, sk_{VS}, \mathbb{BB}, \beta)$: Given a voter identity $id$, ballot box $\mathbb{BB}$, and ballot $\beta$, this algorithm updates the ballot box. Typically, it includes ballot validity checks, such as signal or format checks. 
    
    \item $\mathsf{Publish}(\mathbb{BB})$: Given the ballot box $\mathbb{BB}$, this algorithm publishes the public bulletin board $\PBB$.

    \item $\mathsf{Tally}(\PBB, sk_T)$: Given the public bulletin board $\PBB$ and tallying key $sk_T$, this algorithm computes and outputs the election result $W$ along with a proof $\Pi$ demonstrating the correctness of the tallied result.
    
    \item $\mathsf{VerifyTally}(\PBB, W, \Pi)$: Given $\PBB$ \JQ{(that contains public keys used for verification)}, result $W$, proof $\Pi$, this algorithm outputs $1$ if the election result is valid, and $0$ otherwise.

    \item \JQ{$\mathsf{VerifyVote}(id,usk_{id},upk_{id},\PBB,\beta)$: Given a voter identity $id$, public bulletin board $\PBB$ and a ballot $\beta$, this algorithm is intended for voters and outputs $1$ if their ballots will be included in the tally.} 

\end{itemize}
\end{definition}


\subsection{Coercion-Resistance Definition}
\label{subsec:coercion_resistance_definition}

In this work, we adopt the definition of coercion-resistance proposed by Küsters \emph{et al.} in~\cite{kusters2012game, kusters2011verifiability}. According to this definition, a coercer attempts to compel a coerced voter to execute a dummy strategy, denoted as $\mathtt{dum}$. This dummy strategy blindly follows the coercer's instructions, rather than executing the protocol's intended program. In this context, coercion-resistance requires the existence of a counter-strategy, denoted as $\alpha$. This counter-strategy must satisfy two conditions as per the following definition.

\begin{definition}[Informal]
\label{def:kusters_cr_definition}
Let $\Gamma^{\DR^{vs},\DT^{vs}}$ be an e-voting scheme, and $\mathbf{param} = (\mathbb{I}, \mathbb{O},$ $\DO, \DR^v, \DT^{v}, t_{end})$ be some voting parameters. Let $\delta \in [0,1]$.
The election system $\Gamma$ is $\delta$-coercion-resistant with respect to $\mathbf{param}$, 
if there exists a counter-strategy $\alpha$ such that for all adversaries:
\begin{enumerate}[leftmargin=*]
    \item Coerced voters achieve their goal of successfully evading coercion with overwhelming probability;
    \item The coercer distinguishes whether the coerced voter executed '$\mathsf{dum}$' (indicating compliance) or $\alpha$ (indicating evasion) with \JQ{$\delta$-bounded probability}.
\end{enumerate}
$\Gamma$ is said to be $\delta$-coercion-resistant if it is $\delta$-coercion-resistant with respect to all $\mathbb{I}$, $\mathbb{O}$, $\DO$, $\DR^v$, $\DT^{v}$, and $t_{end}$.
\end{definition}
In other words, to protect a voter under coercion, two conditions are essential. Firstly, the counter-strategy $\alpha$ must guarantee the voter's success with a very high probability, regardless of the coercer's chosen strategy. Secondly, the coercer must be effectively unable to distinguish whether the voter is executing $\alpha$ or $\mathtt{dum}$ with no more advantage than when just observing the tally result.\JQdone{no more than observing the voting result} 
\JQ{In this work, we adhere to Definition~\ref{def:kusters_cr_definition} and port it to a game-based execution model.}
We will refer to the first condition as \emph{CR-Integrity}, and will further formalize it as a reachability game, $\GameCRI{\Gamma}{\cdot}$, in Section~\ref{subsec: modeling of result integrity}. The second condition will be referred to as \emph{CR-Privacy}, and will further formalize it as an indistinguishability game, $\GameCRP{\Gamma}{\cdot}$, in Section~\ref{subsec: forced-abstention instantiate on loki}. 
\JQ{We emphasize that coercion resistance is stronger than \emph{receipt-freeness} (see classification by Alwen \emph{et al.}~\cite{alwen2015incoercible}). Receipt-freeness prevents voters from obtaining proof of their vote, thereby mitigating vote-buying or forced vote disclosure. Coercion resistance, however, goes further by enabling voters to defy a coercer's instructions—such as voting in a specific manner or abstaining—without the coercer being able to verify compliance. Fundamentally, receipt-freeness defends against passive attacks, while coercion resistance counters active adversaries who can issue instructions to coerced voters during the casting process.}

\subsection{Overview of Loki}
\label{subsec: overview of loki}

\begin{figure}[!t]
    \centering
    \includegraphics[width=\linewidth]{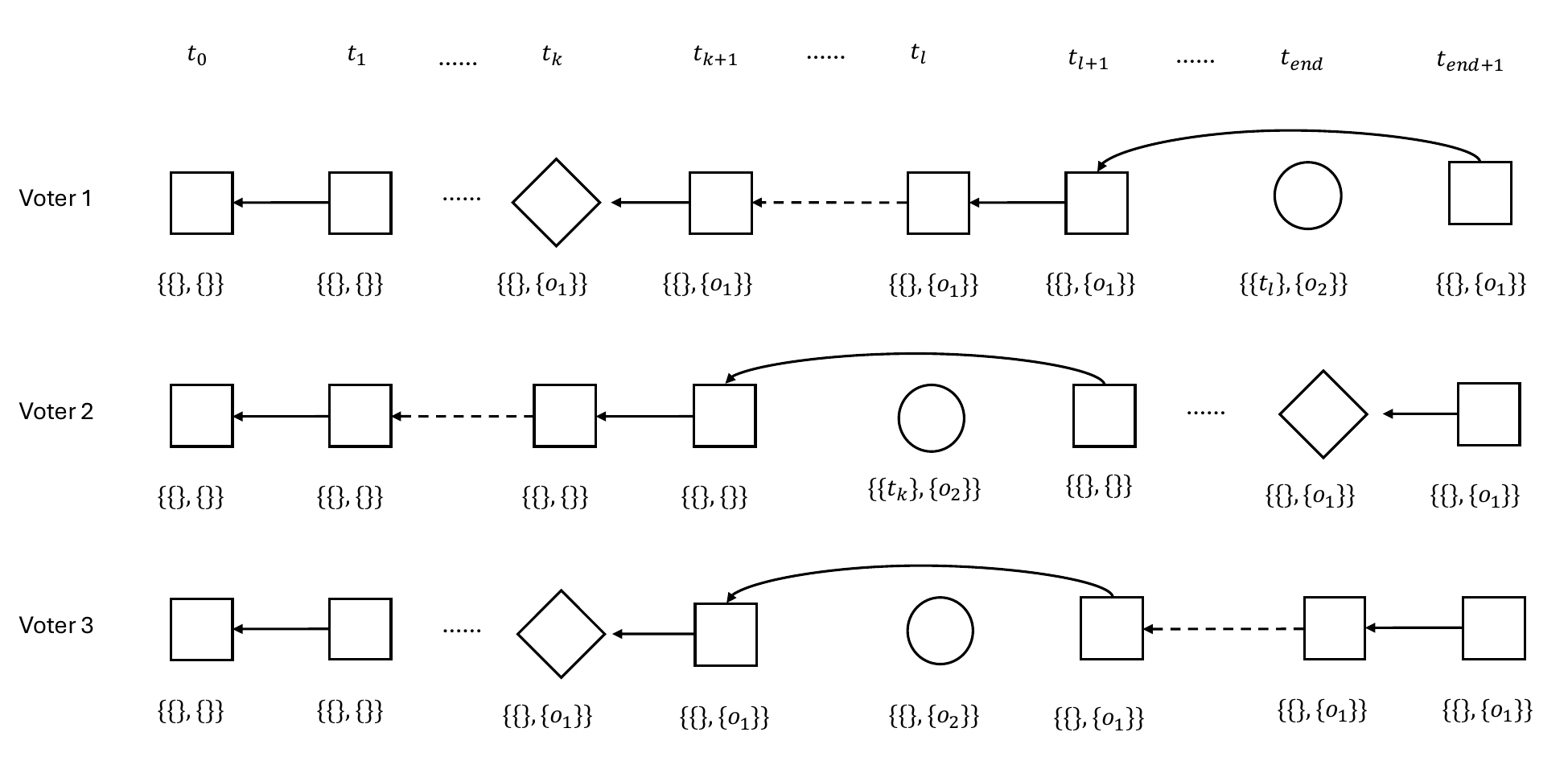}
    \caption{Loki framework. Circles denote ballots under coercion. Diamonds are genuine ballots. Squares are noise ballots from VS. The malicious option is $o_2$ and genuine one is $o_1$.}
    \label{fig: Loki structure}
\end{figure}

The Loki protocol aims to provide coercion-resistance in electronic voting, specifically against last-minute coercion. To achieve this, it introduces a signalling mechanism that allows voters to indicate to the Voting Server which ballots were generated and cast freely, and which were cast under coercion. The voter's ballots—whether cast freely, under coercion, or as noise ballots generated by the Voting Server—are chronologically linked in what the authors term the voter's \emph{cast ballot record}. Both the voter and the Voting Server maintain a record of indices that correspond to the freely cast ballots in the voter's cast ballot record. Every time a ballot is cast, an encrypted version of this index list is included. If a voter who has already voted is subject to coercion, they must present a falsified index list to their coercer, which will allow the Voting Server to detect the coercion taking place. The Voting Server generates noise ballots, but also ensures that the last ballot in a voter's cast ballot record is always a re-encryption of the voter's last freely cast ballot. Figure~\ref{fig: Loki structure} illustrates the concept of voter cast ballot records. These records are maintained on the public bulletin board, $\PBB$. We will refer to voter $id$'s casting ballot record, as published on $\PBB$, as $\PBB^{id}$. 

\noindent\textbf{Threat Model.}
\JQ{Loki assumes an over-the-shoulder coercer~\cite{clark2011selections}, capable of forcing voters to cast a specific ballot or to abstain. While voters are aware whenever the coercer submits a ballot on their behalf (due to inalienable authentication), they may not know which candidate it encodes—since the coercer can privately forge the ballot—or whether the ballot is even valid. Loki employs the Voting Server to prevent the adversary from distinguishing genuine ballots from noise or tracing re-encryptions. This server is trusted for coercion-resistance, which aligns with the e-voting scheme syntax defined in Section~\ref{subsec:coercion_resistant_syntax}.}

Below, we provide an overview of the algorithms defining the Loki protocol.

\begin{description}[style=unboxed,leftmargin=0.3cm]
    \item[$\mathsf{Setup}(1^\lambda, \mathbf{param})$ -]
    This function is to run the encryption key generation algorithms to generate the encryption key pairs $(pk_{VS}, sk_{VS})$ and $(pk_{T},sk_{T})$ for the Voting Server and the talliers respectively.

    \item[$\mathsf{Register}(id)$ -] 
    This function produces the signing key pair for every eligible voter by running the signing key generation algorithm. The private credential $l_{id}$ is an empty list by design, and the public credential is the encryption of zeros.

    \item[$\mathsf{Fake}(id,\PBB, L_{id}, l_{id})$ -] 
    This function generates a falsified list of indices (i.e., a list that differs from $l_{id}$) intended to be disclosed to the coercer. To ensure plausibility, the forged list must match the length of the original ballot record $L_{id}$, which is therefore required as input. \MAdone{For Loki we need the list of the CBR at the time of coercion for the algorithm to produce a plausible list of indices. Fix accordingly in the previous section.} 
    
    \item[$\mathsf{Vote}(id, usk_{id}, pk_T, pk_{VS}, o, l)$ -] This function generates $id$'s ballot $\beta$ for $o$ with format $ \beta\ =\ ct_o \mathbin\Vert ct_l  \mathbin \Vert ct_{l_{id}} \mathbin \Vert \pi $, where
    \begin{itemize}[leftmargin=0pt,labelsep=0.5em,itemindent=\dimexpr\labelwidth+\labelsep\relax]
        \item $ct_o = \mathsf{Enc}(pk_T, o; r_o)$ is the encryption of the chosen option $o$ under the tallying public key $pk_T$.
        \item $ct_l = \mathsf{Enc}(pk_{VS}, l; r_l)$ is the encryption of a list of indices $l$ under the Voting Server's public key $pk_{VS}$.
        \item $ct_{l_{id}}= \mathsf{Enc}(pk_{VS}, l_{id}; r_{l_{id}})$ is the encryption of $l_{id}$, the Voting Server's view of the voter's $id$'s history of freely cast ballots, under its public key $pk_{VS}$. This ciphertext is computed by re-encrypting the corresponding ciphertext from the previous ballot on the voter's ballot record.
        \item $\pi$ is a zero-knowledge proof demonstrating the correct construction of the three ciphertexts. It proves three types of relations: (i) the correct construction of the voter's ballot, denoted $R^{id}$; (ii) the correct re-encryption of the received ballot when the index $l$ matches the Voting Server's index $l_{id}$ of voter $id$, denoted $R^{\mathsf{pred}}$; (iii) the correct re-encryption that skips the received ballot when $l \neq l_{id}$, denoted $R^{\mathsf{pred2}}$.
    \end{itemize}
    
    The list of indices $l$ is used to signal the ballot's validity. In a freely cast ballot, the voter includes a list of valid indices $l$ (corresponding to all previously freely cast ballots) to signal to the Voting Server that the ballot is legitimate, as $l$ matches $l_{id}$. In a coerced ballot or a ballot generated by a coercer on behalf of a coerced voter, this index list $l$ is deliberately invalid. This signals to the Voting Server that the ballot should be disregarded. The coerced voter provides a false $l \not= l_{id}$ to the coercer. This is achieved using the $\mathsf{Fake}$ algorithm. Furthermore, when the $\mathsf{Vote}$ algorithm is executed freely by voter $id$ (\emph{i.e.}, not under coercion), the list $l_{id}$ is updated to reflect the current length of the voter's cast ballot record. This update ensures accurate tracking for future re-votes. Voters submit their ballots to the Voting Server for inclusion in the public bulletin board ($\mathbb{PBB}$). This ballot format allows the Voting Server to decrypt $ct_l$ and $ct_{l_{id}}$ to obtain the lists of indices $l$ and $l_{id}$, while at the same time preserving the content of $ct_o$, that is the actual choice of the voter.
    
    \item[$\mathsf{Noise}(id, sk_{VS}, pk_T, \PBB)$ -] This function allows the Voting Server to produce noise ballots on behalf of voter $id$. It retrieves the last ballot on $\PBB^{id}$ and re-encrypts it to produce a new ballot for the same election option. These noise ballots (the number and the time of casting of these ballots) are specified by the two distributions $\mathcal{D}^{vs}_R$ and $\mathcal{D}^{vs}_T$. The Loki paper dictates that to ensure these noise ballots are indistinguishable from genuine voter ballots, these distributions should be the same as those specifying the voter's voting behaviour, \emph{i.e.} $\mathcal{D}^{v}_R$ and $\mathcal{D}^{v}_T$.

    \item[$\mathsf{Include}(id, sk_{VS}, \mathbb{BB}, \beta)$ -] The Voting Server runs this algorithm to include in the ballot box $\mathbb{BB}$ a ballot $\beta$. This ballot $\beta$ might have been cast by voter $id$ (freely or under coercion), or be a noise ballot generated by the Voting Server itself on behalf of voter $id$. If $\beta$ was cast by voter $id$ under coercion it should encode an invalid indices list. If $\beta$ was freely generated by voter $id$ it should contain a valid list of indices.     
    Before appending $\beta$ to $\mathbb{BB}^{id}$, the Voting Server checks the proof $\pi$ in the ballot and if it verifies successfully. \JQ{Observe that the proof includes a proof of knowledge of the secret signing key (i.e., $usk_{id}$). Since the adversary does not have access to the secret signing keys of honest voters, any proofs accompanying adversarial ballots that attempt to impersonate honest voters will fail verification. And therefore, only ballots submitted on behalf of corrupted or coerced voters will successfully pass the proof check.} 
    It then checks the list of indices encoded in $\beta$. If the list matches its own record of voter $id$'s valid indices, it will mark it with 1 and update its record of voter $id$'s valid indices with the length of $\mathbb{BB}^{id}+1$; otherwise, it will mark it with 0 and leave its record of voter $id$'s valid indices unchanged. 
    The Voting Server then generates a second noise ballot, denoted $\beta'$. If the received ballot $\beta$ is identified as a genuine submission—originating either from the voter $\mathit{id}$ or the server itself—then $\beta'$ is constructed as a re-encryption of $\beta$. However, if the server detects that $\beta$ is a coerced ballot, it instead generates $\beta'$ as a noise ballot by re-encrypting the last valid ballot posted on $\mathsf{PBB}_{\mathit{id}}$.
    In this way, this second ballot $\beta'$ is always a re-encryption of the last ballot that voter $id$ cast freely. This ballot $\beta'$ is marked with 0. We call $\beta'$ the twin ballot of ballot $\beta$. 
    Finally, the Voting Server appends $\beta$ to $\mathbb{BB}^{id}$, and then it appends $\beta'$ to $\mathbb{BB}^{id}$. 
    If the proof $\pi$ in ballot $\beta$ does not verify successfully, the ballot is ignored and the ballot box $\mathbb{BB}$ is unchanged. The $\mathsf{Include}$ algorithm ensures that coerced ballots are ``by-passed'', \emph{i.e.} the last ballot in $\mathbb{BB}^{id}$ is always a re-encryption of the last ballot cast freely by $id$.
    
    \item[$\mathsf{Publish}(\mathbb{BB})$ -] This algorithm publishes all the ballots in $\mathbb{BB}$ while stripping them from the flags that mark each ballot as coerced or genuine. 

    \item[$\mathsf{Tally}(\PBB, sk_T)$ -] The tally algorithm is jointly run by the talliers. They first retrieve the last ballot on each voter's cast ballot record, i.e. the set $\{\beta_{id}\ |\ id\in\mathbb{I},\ \PBB^{id} = \_\mathbin\Vert[\beta_{id}]\}$, \MAdone{This needs fixing} and then compute the result of the election $W$, and a proof of correct tallying $\Pi$. Loki supports homomorphic tallying and other common tallying techniques used in cryptographic voting protocols. Thanks to the structure of Loki's public bulletin board, only one of each voter's ballot is decrypted, the complexity of tallying is thus linear in the number of voters, \emph{i.e.} $\mathcal{O}(n_v)$.

    \item[$\mathsf{VerifyTally}(\PBB, W, \Pi)$ -] This algorithm allows anyone to verify the correctness of $(W,\Pi)$ using the talliers' public key $pk_{T}$, and that the published tally result W is valid with respect to the result space $\mathbb{W}$.

    \item[$\mathsf{VerifyVote}(id,usk_{id},upk_{id},\PBB,\beta)$ -] \JQ{This algorithm allows voters to verify that their ballot will be counted in the tally.}

    \end{description}

\begin{remark} Unlike other re-vote schemes, which require a coercion-free window after active coercion, Loki relaxes this assumption and allows for the coercion-free window to occur at any point — either before or after active coercion. As shown in Figure \ref{fig: Loki structure}, voter $1$ is an example of last-minute coercion, voters $2$ and $3$ illustrate active coercion taking place before and after the moment-of-privacy respectively.
\end{remark}
\section{Attack on CR-Integrity}
\label{sec: cr-integrity attack}

\subsection{Formalising CR-integrity}
\label{subsec: modeling of result integrity}

Following Küsters \emph{et al.}'s approach (see Definition~\ref{def:kusters_cr_definition}), we define a security game that captures CR-Integrity. 

\subsubsection{Effective evasion} We define $\gamma_{ee}$ to capture the effectiveness of the evasion strategy\footnote{The definition of $\gamma_{ee}$ is similar to the definition of the quantitative goal for verifiability proposed in \cite{cortier2016sok}.}. In a protocol run, the evasion strategy is considered to have been effective if a coerced voter, by following it, was able to ensure their intended vote option was included in the final tally. Roughly, this requires the \JQ{difference} between the produced result and the ``ideal’’ result (obtained when the intended choices of honest voters are counted and one choice for each dishonest voter) to be zero. Essentially, this ensures that the result $W$ reflects all non-adversarial votes (including the coerced one) and could have been produced just by the adversary casting adversarial votes $o_1',\cdots o'_{n_d}$.
Formally, the goal $\gamma_{ee}(\mathbb{I}, \Icorr, \rho, \mathcal{O}, W)$, where $\mathcal{O} = [(id_1, o_1), \dots,$ $(id_{n_v}, o_{n_v})]$, is satisfied in a protocol run yielding the election result $W$ if there exist valid options $o_1',\cdots o'_{n_d}$ (possible choices of corrupted voters) such that:
\JQ{
\begin{align*}
        \rho((id_{i_1},o_{i_1}), \cdots,(id_{i_{n_v-n_d}}, o_{i_{n_v-n_d}}), \\(id_1, o_1'),\cdots, (id_{n_d} ,o'_{n_d})) = W
\end{align*}
}
where $\{id_{i_1}, \dots,  id_{i_{n_v-n_d}}\} = \mathbb{I}\setminus\Icorr$ are the honest voters.
\TZdone{Is the above consistent with the domain of $\rho$?} \MAdone{$\rho$ should be stable under permutation of the inputs}\JQdone{The permutation stable is only required for function $d$, and the way we define it is already stable.}

\subsubsection{The CR-Integrity game} CR-Integrity is specified as a reachability game $\GameCRI{\Gamma}{\adv}$ between a Challenger, who controls the honest parties, and an adversary $\mathcal{A}$, who controls the corrupted ones. $\GameCRI{\Gamma}{\adv}$ is formally defined in Figure~\ref{algo: game CR-I Gamma}. $\mathcal{A}$ corrupts voters (line~\ref{line: CR-I corruption}), while all non-corrupted voters are honest, except for voter $id_j$, who is coerced (line~\ref{line: CR-I coercion}). \JQ{By receiving the list of adversarial ballots $\mathbb{L}$ (line~\ref{line: receive adversarial ballots}), the challenger becomes aware of when the coercer casts a ballot. This is consistent with the over-the-shoulder coercer assumption.}

Considering the impact of distributions of voting time, we introduce the concept of a clock (initialized at line~\ref{line: CR-I init clock}). At each clock tick, only one ballot per voter is allowed, whether cast by the voter, the coercer on their behalf, or the Voting Server on their behalf. 

\MAdone{Can we discuss the following. Not sure we deal with this correctly in the game.} $\adv$ can only win over traces for which effective evasion is meaningful, that is when  
\begin{inparaenum}[(a)]
\item there exists a time $t$ within the sequence $\vec{t}^{id_j}_{v}$ such that the adversary $\mathcal{A}$ does not insert a ballot into $\PBB^{id_j}$, ensuring the existence of a moment of privacy ($used=\bot$ at line~\ref{line: CR-I used}), and
\item the coerced voter chooses to evade coercion during this moment of privacy, as indicated by the variable $flag$ ($flag:=\top$ at line~\ref{line: CR-I flag}).
\end{inparaenum}
$\mathcal{A}$ wins the CR-Integrity game if the tally result verifies successfully and the evasion strategy has not been effective, i.e., $\gamma_{ee}(\mathbb{I}, \Icorr, \rho, \mathcal{O},W)$ is not satisfied (line~\ref{line: CR-I not ee}). 
\MAdone{I don't understand the argument on VerifyTally, and I cannot recall our discussion.} \JQdone{In our syntax, there is no mention on validity check for encrypted candidate. A corrupted voter could encrypt for an option that is not within our option space. } Note that $\mathcal{A}$'s winning condition requires $\mathsf{VerifyTally}(\cdot)$ to hold true. This constrains choices from corrupted voters to be within the valid option space. A stronger integrity property might guarantee that invalid ballots are dropped; however, CR-Integrity is within the scope of coercion-resistance \JQ{in the setting where $\adv$ can win only with valid ballots.}

We say that a protocol satisfies CR-Integrity if the probability of any adversary $\mathcal{A}$ winning $\GameCRI{\Gamma}{\adv}$ is negligible.
\begin{definition}[CR-Integrity]
\label{def: result integrity} 
Let $\Gamma^{\DR^{vs}, \DT^{vs}}$ be an e-voting scheme, and let 
$\mathbf{param} = (\mathbb{I}, \mathbb{O}, \DO, \DR^v, \DT^v, t_{\mathrm{end}})$ 
denote the voting parameters. 
We say that $\Gamma^{\DR^{vs}, \DT^{vs}}$ satisfies \emph{CR-Integrity} with respect to voting parameters $\mathbf{param}$, if for all PPT adversaries $\mathcal{A}$, there exists a negligible function $\mu(\cdot)$ such that:
\begin{equation*}
    \begin{split}
        \Pr[\GameCRI{\Gamma^{\DR^{vs},\DT^{vs}}}{\adv}(1^{\lambda},\mathbf{param})=1]\leq \mu(\lambda).
    \end{split}
\end{equation*}
\end{definition}
\TZdone{Is it for all parameters? Or is it with respect to parameters? Could there exist extreme cases? E.g., is there a system that is secure for all distributions?}\MAdone{We would want the scheme to be secure (and thus to satisfy CR-Integrity)  irrespective of how voters vote though (i.e. for all $\mathcal{D}^v_R, \mathcal{D}^v_T, \DO$) and of the casting duration (i.e. for all $t_{end}$). No ?}
\JQdone{In the kuster definition, it is with respect to the intention mapping $\DO$ (I am considering switch the type of $\DO$ from mapping to distribution for the sake of consistency with Kuster definition.), number of voters, number of options and number of corrupted voters. However, our forced abstention attack requires that prob. of abstention is non-nelg. If we loosen it, our attack may not stand.}

\MAdone{At lines 28/29, 39, 42, and 54 of Figure~\ref{algo: RI game} are we not hard-coding in the security definition that the VS's VrfSgn algorithm is correct?}

\MAdone{In Loki, for the tally to be computed the VS needs to decrypt the outter layer of encryption of the last ballot of each CBR. Is this even compatible with our syntax ? See line 58 of Figure~\ref{algo: RI game}}

\begin{figure}[!t]
\hrule
\vspace*{3pt}
$\GameCRI{\Gamma^{\DR^{vs},\DT^{vs}}}{\adv}$
\vspace*{3pt}
\noindent\hrule
\begin{algorithmic}[1]
\scriptsize
\Require{$1^{\lambda},\mathbf{param}(\mathbb{I}, \mathbb{O}, \DO, \DR^v, \DT^v, t_{end})$}
\State\label{line: CR-I init clock} $clk:=0$ \algorithmiccomment{initialize the clock}
\State $\{(sk_T,pk_T),(sk_{VS},pk_{VS})\}\leftarrow \mathsf{Setup}(1^{\lambda},\mathbf{param})$
\State Initialize $\mathbb{BB}$ and $\PBB$ as $[\ ]$
\State Initialize the list $\mathcal{O}$ as $[\ ]$ 
\State $\{(upk_{id},usk_{id}),CR_{id},cr_{id}\}\leftarrow\mathsf{Register}(id)_{id\in\mathbb{I}}$
\State\label{line: CR-I corruption} Send $(pk_T,pk_{VS},\{upk_{id}, CR_{id}\}_{id\in\mathbb{I}}, \PBB, \mathbf{param})$ to $\mathcal{A}$
\State Receive $\Icorr\in\mathbb{I}$ from $\mathcal{A}$ 
\algorithmiccomment{corruption}
\For{$id\in\mathbb{I}\backslash\Icorr$}
\State $o_{id}\leftarrow \DO(id)$
\State $\mathcal{O} := \mathcal{O} \mathbin\Vert [(id, o_{id})$]
\If{$o_{id} \neq \Phi$} \algorithmiccomment{voter's intention is not to abstain}
\State $r^{id}_{v} \leftarrow\mathcal{D}^v_R(id)$, $\vec{t}^{id}_{v}\leftarrow\mathcal{D}^v_T(r_v^{id}, id)$
\Else
\State $r^{id}_v:= 0$, $\vec{t}^{id}_{v} := [\ ]$ \algorithmiccomment{distributions never map to 0}
\EndIf
\State $r^{id}_{vs} \leftarrow\mathcal{D}^{vs}_R(id)$, $\vec{t}^{id}_{vs}\leftarrow\mathcal{D}^{vs}_T(r_{vs}^{id}, id)$
\algorithmiccomment{third party voting distribution}
\EndFor
\State Send $(\{usk_{id},cr_{id}\}_{id\in\Icorr})$ to $\mathcal{A}$ 
\State\label{line: CR-I coercion} Receive $(id_j,o)$ from $\mathcal{A}$
\algorithmiccomment{coerce voter $id_j$ whose intention is $o$}
\If{$id_j\notin\mathbb{I}\backslash \Icorr \lor o \notin \mathbb{O} $}
\State \Return 0
\EndIf
\State Send $usk_{id_j}$ to $\mathcal{A}$.
\State $flag:=\bot$
\While{\JQ{$clk\leq t_{end}$}}
\State $used:=\bot$ \algorithmiccomment{mark of moment-of-privacy}
\For {$id\in\mathbb{I}\backslash \Icorr$}
\State Send $(\PBB,clk)$ to $\mathcal{A}$ and receive list of ballots $\mathbb{L}$
\label{line: receive adversarial ballots}
\For{tuples $(id',\beta_{\mathcal{A}}) \in \mathbb{L}$}
\If{\JQ{no ballot added for $id'$ to $\mathbb{BB}$ at $clk$}}
\State\label{line: CR-I accepted} $\PBB\leftarrow\mathsf{Publish}(\mathsf{Include}(id', sk_{VS}, \mathbb{BB}, \beta_{\mathcal{A}}))$
\EndIf
\If{$id'=id_j$}
\State $used := \top$ \algorithmiccomment{not moment-of-privacy}
\label{line: CR-I moment of privacy}
\EndIf
\EndFor
\If{$clk\in\vec{t}^{id}_{v}$} \algorithmiccomment{voter $id$ casts ballot}
\If{$id=id_j \land used = \bot\land o \neq \Phi$}\label{line: CR-I used}
\State\label{line: CR-I flag} $flag := \top$\algorithmiccomment{use moment-of-privacy to cast freely}
\State $\beta\leftarrow \mathsf{Vote}(id_j,usk_{id_j},pk_T, pk_{VS},o,cr_{id_j})$
\State $\PBB\leftarrow\mathsf{Publish}(\mathsf{Include}(id_j, sk_{VS}, \mathbb{BB}, \beta))$
\ElsIf{$id\neq id_j$}
\State $\beta\leftarrow\mathsf{Vote}(id,usk_{id},pk_T,pk_{VS},o_{id},cr_{id})$
\State $\PBB\leftarrow\mathsf{Publish}(\mathsf{Include}(id, sk_{VS}, \mathbb{BB},\beta))$
\EndIf
\EndIf
\EndFor
\If{$clk\in\vec{t}^{id}_{vs}\land clk\notin\vec{t}^{id}_{v} \land \adv$ did not cast for $id$} \State\algorithmiccomment{third party noise for voter $id$}
\State $\PBB\leftarrow\mathsf{Publish}(\mathsf{Include}(id, sk_{VS}, \mathbb{BB}, \mathsf{Noise}(id, sk_{VS},pk_T,\PBB)))$
\EndIf
\State $clk++$ \algorithmiccomment{clock advances}
\EndWhile
\State $(W,\Pi)\leftarrow\mathsf{Tally}(\PBB,sk_{T})$ 
\If{$\mathsf{VerifyTally}(\PBB, W, \Pi)\land flag \land \neg\gamma_{ee}(\mathbb{I},\Icorr,\rho,\mathcal{O},W)$}
\label{line: CR-I not ee}
\State \algorithmiccomment{$id_j$ used moment of privacy ($flag$), but true intention not counted in $W$ ($\neg\gamma_{ee}$)}
\State\Return 1 \algorithmiccomment{$\mathcal{A}$ wins}
\Else
\State\Return 0 \algorithmiccomment{$\mathcal{A}$ fails}
\EndIf
\end{algorithmic}
\vspace*{3pt}\hrule\vspace*{3pt}
\caption{CR-Integrity game for protocol $\Gamma^{\DR^{vs},\DT^{vs}}$ and adversary.}
\label{algo: game CR-I Gamma}
\end{figure}

\subsection{Brute-Force Attack on Loki's CR-Integrity}

Let $m$ be the length of coerced voter id's cast ballot record at the time where $\adv$ launches its active coercion against $id$, that is the length of $\PBB^{id}$ at that time. 
Our brute force adversary tries to guess the indices in $\PBB^{id}$ of the ballots freely cast by voter $id$, and casts as many ballots encoding as many possible guesses. If any of the cast ballots encodes the valid list of indices for voter $id$, then the Voting Server will accept it as a legitimate freely cast ballot from voter $id$. To optimize the attack, our adversary might leverage their knowledge of $\mathcal{D}^v_R$, the distribution of re-vote counts. Specifically, the adversary only tests lists of indices that correspond to a number of re-votes with non-zero probability in $\mathcal{D}^v_R$. We define the \emph{difficulty} of this brute-force attack as: \MAdone{Why do we use $F$ for the difficulty ? Is this from the Loki paper ? }\JQdone{What is $n$?}
\[
F_{\DR^v}(m, id, t_{end}) \triangleq \sum_{r \in \mathcal{R}} \binom{m}{r}
\]
where $\mathcal{R} \triangleq \{ r\ |\ \Pr[r\leftarrow\DR^v(id, t_{end})]>0
\text{ and } r < m\}\ \cup\ \{m^*\ |\ 
\exists r\ge m: \Pr[r\leftarrow\DR^v(id, t_{end})]>0 \text{ and } m^*=m\}$. \MAdone{Simplify the definition of $\mathcal{R}$}\JQdone{if m is time, does $r<m$ make sense?}\MAdone{$m$ is not time in the difficulty definition. If you see the proof, we use the time $\kappa(\lambda)$ to bound the length of the CBR at coercion time.} We note that for any distribution $\DR^v$ we have that:
\[
F_{\DR^v}(m, id, t_{end})\ \le\ F_{\mathcal{D}_{uni}}(m, id, t_{end})\ =\ 2^m.
\]
\JQ{where $\mathcal{D}_{\mathsf{uni}}$ is the uniform distribution over $[0,t_{end}]$.}


\MAdone{Informally describe what $\Abrute$ does to help readability of Figure~\ref{algo: A brute}.}



\begin{figure}[!t]
\hrule
\vspace*{3pt}
 $\AbruteOpt{\mathcal{D}}$ \JQdone{this is attacker against Loki, so probably we keep the notation as $l_{id}$ and $L_{id}$ and add some explaination to avoid confusion?}
\vspace*{3pt}
\noindent\hrule
\begin{algorithmic}[1]
\scriptsize
\Require{$\lambda$}
\UponReceiving{$(pk_T,pk_{VS},\{upk_{id}, L_{id}\}_{id\in\mathbb{I}}, \PBB, \mathbf{param})$}
\State return $[\ ]$ as $\Icorr$
\algorithmiccomment{no voters corrupted}
\EndUponReceiving

\UponReceiving{$(\{usk_{id},l_{id}\}_{id\in\Icorr})$}
\State $(id_j,o)\leftarrow_{\$}(\mathbb{I}, \mathbb{O})$
\algorithmiccomment{select coerced voter $id_j$ and their intention $o$}
\State send back $(id_j,o)$
\EndUponReceiving

\UponReceiving{$usk_{id_j}$}
\State $\tilde{o} \leftarrow_{\$}\mathbb{O}\backslash o$
\algorithmiccomment{choose a coercion option that differs from $o$}
\State $cnt:=0$
\State $atk:=\bot$
\EndUponReceiving

\WhileReceiving{ $(\PBB,clk)$}
\If{$clk=\kappa(\lambda)$}
\State $m:= |\PBB^{id_j}|$
\State $\mathrm{G}uess := \{0^{m}\}$
\algorithmiccomment{set of guesses}
\For{$i\in[0,F_{\mathcal{D}}(m, id_j, t_{end})]\ \land\ cnt < \lambda^c$}
\State $cnt:=cnt+1$
\State $t\leftarrow \mathcal{D}$ \JQdone{better way to express?}
\algorithmiccomment{learn possible cast number from distributions}
\State $l^i \leftarrow_{\$} \{l \in \{0,1\}^{m}\ |\ \mathsf{HammingWeight}(l) = t\}$
\MAdone{These are the guesses for $(\mathcal{D}_{uni}, \mathcal{D}_{uni})$}
\MAdone{$l^i \leftarrow_{\$} \{0,1\}^{\kappa(\lambda)}$}
\State $\mathrm{G}uess :=\mathrm{G}uess \cup [l^i]$
\EndFor
\State $atk:=\top$ 
\EndIf
\If{$atk=\top\ \land\ \mathrm{G}uess\not=\emptyset$}
\State $l_j\leftarrow_{\$}\mathrm{G}uess$
\State $\mathrm{G}uess:=\mathrm{G}uess\backslash\{l_j\}$
\algorithmiccomment{\JQ{update set to remove attempts}}
\State $\beta_{\mathcal{A}}\leftarrow\mathsf{Vote}(id_j,usk_{id_j},pk_T,pk_{VS},\tilde{o},l_j)$
\State Send $(id_j,\beta_{\mathcal{A}})$
\Else
\State Send [ ]
\EndIf
\EndWhileReceiving
\end{algorithmic}
\vspace*{3pt}\hrule\vspace*{3pt}
\caption{Attack algorithm of $\AbruteOpt{\mathcal{D}}$. Attacker $\AbruteOpt{\mathcal{D}}(1^\lambda)$ runs in $\lambda^c$ steps \MAdone{$\lambda^c$?}, and launches its attack at time $\kappa(\lambda)$. The Hamming weight of a string is the number of symbols that are different from the zero-symbol of the alphabet used.}
\label{algo: A brute}
\end{figure}

\subsubsection{$\DR^v$-agnostic brute-force adversaries}
\label{subsubsec: early stage vulnerability}

The following result establishes that a large class of \emph{naive brute-force adversaries} can succeed in the CR-Integrity game—effectively defeating the coerced voter's evasion strategy—by initiating their active coercion phase sufficiently early, specifically within a time frame that is \JQ{logarithmic} in the security parameter
(see Appendix~\ref{apx: proof of thm CR-I attack without DR} for proofs). 
We characterize as \emph{naive adversaries} those that do not exploit their knowledge of the voters' voting behaviour, \emph{i.e.} they do not exploit knowledge of $\DR^v$, $\DT^v$, or $\DO$. 

\MAdone{In the theorem below, $\kappa(\lambda)$ is the time of coercion rather than the length of the CBR at coercion time. I am not sure what the constraint on the length of the CBR could be.}
\begin{theorem}
\label{thm: CR-I attack without DR}
    Let $\mathbf{param} = (\mathbb{I}, \mathbb{O}, \DO, \DR^v, \DT^{v}, t_{end})$ be any voting parameters and $\DR^{vs}, \DT^{vs}$ be any distributions. Let $\kappa(\cdot) = \Theta(\log(\cdot))$, then there exists a non-negligible function $\alpha(\cdot)$, and $c\in\mathbb{N}$ such that for all $\lambda \in\mathbb{N}$ such that $\lambda > 1$, and for all $\delta\in\mathbb{N}$:
    \begin{equation*}
    \begin{split}
    &t_{end} = \kappa(\lambda) + \delta\Rightarrow\\
    &\Pr[\mathcal{G}ame^{Loki^{\DR^{vs},\DT^{vs}}, \mathcal{A}_{brute}^{c,\kappa, \mathcal{D}_{uni}}}_{CR-I}(1^\lambda, \mathbf{param})=1]= \\
    &\hspace{6.5cm} \min(\delta \cdot \alpha(\lambda), 1).
    \end{split}
    \end{equation*}
\end{theorem}

\begin{remark}
    We observe that in particular, if $\delta = F_{\mathcal{D}_{uni}}(\kappa(\lambda), id, t_{end})$, then the adversary's probability of winning the CR-Integrity game is 1. That is, the adversary launched his attack early enough before the end of the casting period, to allow him to cast ballots for all possible guesses of the list of valid indices.
\end{remark}

Conversely, we show that \emph{lazy} brute-force adversaries, who do not actively coerce voters until after their cast ballot record is sufficiently long, are not successful in their attack. This holds true only when considering unrealistic voter behaviours (see Appendix~\ref{apx: proof of clm super ploy m has bf security} for details).

\subsubsection{$\DR^v$-aware brute-force adversaries}
\label{subsec: inforamtion leaked in distribution}

Here, we examine revoting distributions that correspond real-world voters. Specifically, we focus on the class $\mathbb{D}_{p,q}$ of distributions describing voters who vote at most $q$ times with probability $p$. In Appendix~\ref{app:distribution affects difficulty}, we demonstrate how the voter's revoting behaviour significantly affects the difficulty of launching brute-force attack.

We then explore the advantage gained by our brute-force adversary by exploiting their knowledge of this distribution. Notably, we demonstrate that this slightly more sophisticated adversary can break CR-Integrity with non-negligible probability, \emph{regardless of the time $\kappa(\lambda)$} when they launch their active coercion 
(see Appendix~\ref{apx: proof of thm CR-I attack with DR} for proofs).

\begin{theorem}
\label{thm: CR-I attack with DR}
Let $\mathbf{param} = (\mathbb{I}, \mathbb{O}, \DO, \DR^v, \DT^{v}, t_{end})$ be voting parameters such that $\mathcal{D}^v_R \in \mathbb{D}_{p,q}$, where $p,q$ are constants. Let $\DR^{vs}, \DT^{vs}$ be any distributions. For every polynomial $\kappa(\cdot)$, there exists a non-negligible function $\alpha(\cdot)$, and $c\in\mathbb{N}$ such that for all $\lambda\in\mathbb{N}$ such that $\lambda > 1$, and for all $\delta\in\mathbb{N}$: 
    \begin{equation*}    
    \begin{split}
    &t_{end} = \kappa(\lambda) + \delta\Rightarrow \\ 
    &\Pr[\mathcal{G}ame^{Loki^{\DR^{vs},\DT^{vs}}, \mathcal{A}_{brute}^{c,\kappa, \mathcal{D}^*_{1, q}}}_{CR-I}(1^\lambda, \mathbf{param})=1]\\
    &\ge p\cdot\min(\delta\cdot \alpha(\lambda), 1),
    \end{split}
    \end{equation*}
    where $\mathcal{D}^*_{1,q}$ is uniform over $\{0,1,\ldots,q\}$.\JQdone{Actually, our attacker doesn't exploit the information of p, but only q. Is that the reason why we feed attacker with $\mathcal{D}^*_{1,q}$, rather than $\mathcal{D}^*_{p,q}$?}
\end{theorem}

This attack can, of course, be further optimized by exploiting the adversary's knowledge of $\DT^v$, $\DR^{vs}$, $\DR^{vs}$, and $\DO$. Specifically, the current bound is calculated without considering the Voting Server's distribution or the abstention rate. However, if we further constrain the Voting Server's distribution, $\DR^{vs}$, we can obtain a tighter bound. For example, Loki suggests that the Voting Server behaves similarly to the voters, meaning that $\DR^{vs} = \DR^{v}$ or, at least, that $\DR^{vs}$ belongs to a distribution family $\mathbb{D}_{p',q'}$ for some constants $p'$ and $q'$. We will see how this dramatically increases our adversary's advantage on our use case data in the next section. \MAdone{We can actually characterize this generally, but do we want? This is what we do in the the Estonia case study.}

This class of distributions is, in fact, more realistic than a uniform random distribution for two reasons. First, this phenomenon has been observed in numerous large-scale real-world elections. (Our later example of Estonia will also illustrate this.) Second, the inherent inertia of human behaviour theoretically imposes an upper bound on voters' voting times.

\subsection{Loki for Real-World Elections: the Case of Estonia}
\label{subsec: analysis with estonia}


To illustrate, let's examine Estonia which has deployed internet voting since 2005. Research suggests that there, re-voting was infrequent. Specifically, about 98\% of those who voted online did so only once~\cite{ehin2022internet}. Notably, the abstention rate was also substantial, reaching at least 35\%. The statistical data on Estonian voters' re-voting behaviour over the last five elections can be found in Table IV of Appendix~\ref{apx: Estonia data}.

To examine the scenario least \JQ{susceptible} to brute-force attacks, let's consider the 2021 Local Election (LE21), which had the highest number of re-votes. We can model LE21 voters using two distributions: $\DO^{LE21}$, where the abstention rate ($\DO^{LE21}[\Phi]$) is 45.21\%, and $\DR^{LE21}$, where 95.47\% of voters cast one vote ($\DR^{LE21}[1]$) and 4.53\% cast two or more ($\DR^{LE21}[2+]$). See Appendix~\ref{apx: calculation of DR LE21} for calculation details of $\DR^{LE21}$. Two distributions are assumed for the general public (\emph{i.e.} $\forall id, \mathcal{D}_{O}^{LE21}(id) = \mathcal{D}_{O}^{LE21}, \DR^{LE21}(id) = \DR^{LE21} $).

The distribution $\DR^{LE21}$ belongs to the family $\mathbb{D}_{0.9547,1}$. So, according to Theorem~\ref{thm: CR-I attack with DR}, depending on the time $t$ at which a brute-force adversary launches their attack, they can break CR-Integrity with probability at least: 
\[
0.9547\cdot\min(\frac{t_{end}-t}{t+1}, 1).
\]
According to the Loki paper~\cite{giustolisi2023thwarting}, the Voting Server should also use the $\DR^{LE21}$ distribution when generating noise ballots. However, with this configuration, the adversary's advantage does not decrease over time as expected because the Voting Server adds very few noise ballots. To show this, we calculate (see Appendix~\ref{apx: calculation of PBB id length} for computation details) and summarize in Table~\ref{tb: cbr length distribution} the probability distribution of cast ballot record lengths, assuming both voters and the server follow the $\DR^{LE21}$ distribution.
\begin{table}[!t]
\centering
\begin{tabular}{|c|c|c|c|}
\toprule
\multicolumn{1}{|c|}{} & \multicolumn{3}{c|}{$|\PBB^{id}|$} \\ \cline{2-4}
                          & 1 & 2 & $\geq 3$ \\ \midrule
 $\mathcal{D}_{|\PBB^{id}|}$     & 43.16\% & 41.20\% & 15.64\% \\ 
\bottomrule
\end{tabular}%
\caption{$\mathcal{D}_{|\PBB^{id}|}$: voters' cast ballot record length distribution for the 2021 Local Election in Estonia}
\label{tb: cbr length distribution}
\end{table}

\MAdone{The detailed calculation for $\mathcal{D}_{|\PBB^{id}|}$ do not need to be here.}

With this at hand, we can finally compute a tighter bound on the adversary's advantage of winning the CR-Integrity game. Indeed, given that with more than 84\% probability the cast ballot record of any voter is strictly smaller than 3, we can deduce that provided the adversary allows for their attack at least three clock ticks ($\delta\geq3$ to cast three coerced ballots) their success probability is 80.53\%.
More generally, at any time $t \leq t_{end}$ the adversary can decide the coerced voter's tallied vote with the probabilities reported in Table~\ref{tb: attacking estonian voters}.
See Appendix~\ref{apx: calculation of prob. A brute wins in Estonia} for calculation details.


\begin{table}[!t]
\centering
\begin{tabular}{|c|c|c|c|}
\toprule
\multicolumn{1}{|c|}{} & \multicolumn{3}{c|}{Active coercion time} \\ \cline{2-4}
                          & $\leq t_{end}-3$ & $t_{end}-2$ & $t_{end}-1$ \\ \midrule
      Success probability & 80.53\% & 53.69\% & 26.84\% \\ 
\bottomrule
\end{tabular}%
\caption{Lower bound on success probability of any adversary that can compute at least three ballots, depending on the time of active coercion.}
\label{tb: attacking estonian voters}
\end{table}

\begin{figure}[!t]
    \centering
    \def\tend{9}

\pgfplotstableread[col sep=space]{
t               dpq                  dran
1 1 1
2 0.8835849200000001 0.5
3 0.82249117 0.109375
4 0.8090505450000001 0.0234375
5 0.8061485918750001 0.0048828125
7 0.8054135576953125 0.00018310546875
8 0.5369280529492187 3.0517578125e-05
9 0.2684622366186524 3.814697265625e-06
}\mydata

\begin{tikzpicture}
    \begin{axis}[
        xlabel={Time of launching attack},
        xlabel style={font=\footnotesize},
        ylabel={Attack Success Rate},
        ylabel style={font=\footnotesize},
        legend pos=north east,
        legend cell align={left},
        legend style={font=\footnotesize},
        grid=major,
        width=0.96\linewidth,
        height=0.75\linewidth,
        xtick={1,2,3,4,5,7,8,9,10},
        xticklabels={1,2,3,4,5,{$\scriptstyle t_{end}-3$},{$\scriptstyle t_{end}-2$},{$\scriptstyle t_{end}-1$}},
        extra x ticks={6},
        extra x tick labels={$\boldsymbol{\dots}$},
        extra x tick style={
            tick label style={
                font=\footnotesize,
                yshift=-.5ex
            }
        },
        xticklabel style={
            font=\small, 
            /utils/exec={ 
                \ifdim\tick pt > 6pt 
                    \pgfkeysalso{rotate=45, anchor=east}
                \fi
            }
        },
        yticklabel style={font=\footnotesize},
        outer sep=0pt,
    ]

        \addplot[
            color=myblue,
            mark=*,
            line width=1pt,
        ] table [x=t, y=dpq] {\mydata};
        \addlegendentry{$\mathcal{D}^{LE21}$}

        \addplot[
            color=myorange,
            mark=*,
            line width=1pt,
        ] table [x=t, y=dran] {\mydata};
        \addlegendentry{$\mathcal{D}_{\text{uni}}$}

    \end{axis}
\end{tikzpicture}
    \caption{The success rate gap between ideal and real voter re-voting distributions $\mathcal{D}_{uni}$ and $\mathcal{D}^{LE1}_R$ respectively.}
    \label{fig: Gap between Dran and DLE21}
\end{figure}
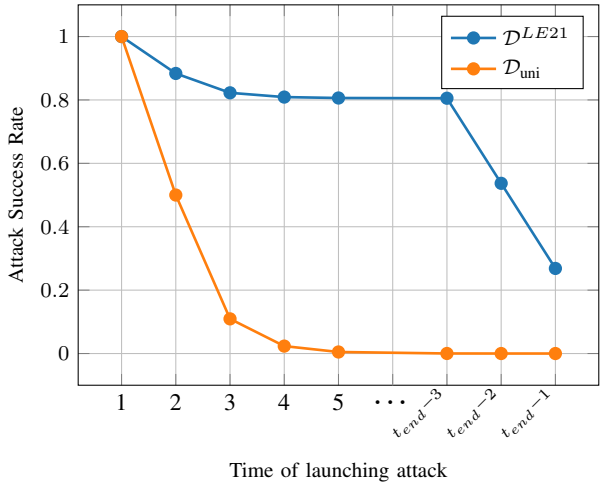

\JQ{We compare in Figure~\ref{fig: Gap between Dran and DLE21} the attack's success with respect to the empirical distribution \(\DR^{\mathrm{LE21}}\) and the ideal distribution \(\mathcal{D}_{\mathrm{uni}}\). For the empirical distribution \(\DR^{\mathrm{LE21}}\) the success rate remains above $80\%$ until $t=t_{end}-3$, in accordance with Table~\ref{tb: attacking estonian voters}. 
This makes it clear that if the adversary allows sufficient time for their attack (three clock ticks), their success probability remains very high. This is in stark contrast to the exponential decrease in the attacker's success probability if voters were, in fact, re-voting uniformly at random, as the gap between the plots of the two cases suggests.}

This section demonstrates that the distribution of the noise ballot significantly affects the complexity of the attack. However, as established by Theorem~\ref{thm: CR-I attack with DR}, the protocol remains insecure regardless of the Voting Server's distribution. Therefore, while the noise ballot's distribution impacts the attack's complexity, it does not prevent the attack itself.

\subsection{Impact Discussion}

The aforementioned brute-force attacks not only compromise CR-Integrity but, more importantly, forbid any recovery because the coercer can cast enough ballots to invalidate any subsequent ballots submitted by the voter.

Specifically, if an adversary, $\Abrute$, successfully guesses the voter's index list, the Voting Server accepts the malicious ballot, resulting in a tampered tally.
However, a more significant issue is the resulting state desynchronization. Once the server accepts a malicious ballot, its state diverges from that of the coerced voter, $id_j$. To cast any valid subsequent ballots, the voter would then need to guess the entire malicious voting history submitted by the coercer—effectively requiring them to perform their own brute-force attack. The adversary can make this recovery even harder by submitting repetitive ballots, increasing the difficulty of guessing the history.


This second consequence highlights a fundamental vulnerability in Loki: even if we re-introduced the last-minute coercion-free window assumption, the system would remain insecure due to this state desynchronization.

Furthermore, as demonstrated by Jamroga \emph{et al.}~\cite{jamroga2024you}, Loki is vulnerable to forced participation attacks whereby a coerced voter intending to abstain is forced to take specific actions to counteract malicious behavior. This is also a violation of CR-Integrity, which our game $\GameCRI{Loki}{\adv}$ effectively captures.

These issues were missed in the original security analysis of Loki because CR-Integrity was entirely omitted from it.
\section{Attack on CR-Privacy}

We now examine the privacy requirement from K\"usters \emph{et al.}'s definition (CR-Privacy). Here, we demonstrate that the Loki voting system is susceptible to \emph{forced abstention} attacks, whereby a coercer can compel a voter to abstain, and the voter cannot deviate from their coercer's instructions without being detected (with non-negligible probability).

\subsection{Formalizing CR-Privacy}
\label{subsec: forced-abstention instantiate on loki}

\subsubsection{The CR-Privacy Game}

In essence, CR-Privacy mandates that an adversary cannot distinguish whether a coerced voter followed the protocol's evasion strategy or complied with the coercer's instructions. To reiterate, when a voter complies ($b=1$ in Figure~\ref{algo: game CR-P Gamma}), they reveal their true credentials (\emph{i.e.}, in Loki this is the correct list of indices from their cast ballot record corresponding to all ballots they have actually cast), along with their signing key, and refrain from casting further ballots. While a coercer could alternatively instruct the voter to cast specific ballots they prepared, this scenario is not formally considered within CR-Privacy. This is because the adversary could achieve the same outcome independently, using the provided credentials. Conversely, when a voter attempts to evade coercion ($b=0$ in Figure~\ref{algo: game CR-P Gamma}), they present the coercer with fake credentials (\emph{i.e.,} in Loki this is a list of indices that do not correspond to their ballots in their cast ballot record), and cast one before the end of the casting period (unless their intention was to abstain). In Loki, if the voter has already cast a ballot for their preferred option, they do not need to cast another after the point of active coercion. For e-voting scheme $\Gamma$, this yields the game $\GameCRP{\Gamma}{\cdot}$ specified in Figure~\ref{algo: game CR-P Gamma}. 

The final tally, combined with the adversary's knowledge of voter behaviour derived from $\DO$, $\DR^v$, and $\DT^v$, inherently grants the adversary a non-negligible advantage. For example, the coercer can compare the final tally with the expected outcome determined by the voting intention distribution, $\DO$. If the coercer cast a ballot for a highly unpopular candidate on behalf of the coerced voter, simply observing whether this candidate received any votes in the published tally reveals, with high probability, if the voter complied or employed the protocol's evasion strategy. Following a similar approach to~\cite{cortier2022jcj}, the CR-Privacy definition presented below addresses this by comparing the adversary's advantage in $\GameCRP{\Gamma}{\cdot}$ with that of an adversary against an idealized game, $\GameCRP{Ideal}{\cdot}$. In $\GameCRP{Ideal}{\cdot}$, illustrated in Figure 8 of Appendix~\ref{apx:ideal privacy game}, the adversary only learns the tally result.

\begin{figure}[!t]
\hrule
\vspace*{3pt}
$\GameCRP{\Gamma^{\DR^{vs},\DT^{vs}}}{\adv}$ 
\vspace*{3pt}
\noindent\hrule
\begin{algorithmic}[1]
\scriptsize
\Require{$1^{\lambda},\mathbf{param}(\mathbb{I}, \mathbb{O}, \DO, \DR^v, \DT^v, t_{end})$}
\State $clk:=0$ \algorithmiccomment{initialize the clock}
\State $\{(sk_T,pk_T),(sk_{VS},pk_{VS})\}\leftarrow \mathsf{Setup}(1^{\lambda},\mathbf{param})$
\State Initialize $\mathbb{BB}$ and $\PBB$ as $[\ ]$
\State Initialize the list $\mathcal{O}$ as $[\ ]$
\State $\{(upk_{id},usk_{id}),CR_{id},cr_{id}\}\leftarrow\mathsf{Register}(id)_{id\in\mathbb{I}}$
\State Send $(pk_T,pk_{VS},\{upk_{id}, CR_{id}\}_{id\in\mathbb{I}}, \PBB, \mathbf{param})$ to $\mathcal{A}$
\State Receive $\Icorr\in\mathbb{I}$ from $\mathcal{A}$ 
\algorithmiccomment{corruption}
\For{$id\in\mathbb{I}\backslash\Icorr$}
\State $o_{id} \leftarrow \DO(id)$
\State $\mathcal{O}:= \mathcal{O}||[(id,o_{id})]$
\If{$o_{id} \neq \Phi$} \algorithmiccomment{voter's intention is not to abstain}
\State $r^{id}_{v} \leftarrow\mathcal{D}_R^v(id)$, $\vec{t}^{id}_{v}\leftarrow\mathcal{D}_T^v(r_v^{id}, id)$
\Else
\State $r^{id}_v:= 0$, $\vec{t}^{id}_{v} := [\ ]$ \algorithmiccomment{distributions never map to 0}
\EndIf
\State $r^{id}_{vs} \leftarrow\mathcal{D}^{vs}_R(id)$, $\vec{t}^{id}_{vs}\leftarrow\mathcal{D}^{vs}_T(r_{vs}^{id}, id)$
\algorithmiccomment{third party voting distribution}
\EndFor
\State Send $(\{usk_{id},cr_{id}\}_{id\in\Icorr})$ to $\mathcal{A}$ 
\State Receive $(id_j,o)$ from $\mathcal{A}$
\algorithmiccomment{coerce voter $id_j$ whose intention is $o$}
\If{$id_j\notin\mathbb{I}\backslash \Icorr \lor o \notin \mathbb{O} $}
\State \Return 0
\EndIf
\State $b\leftarrow \{0,1\}$
\If{$b=1$}
\State Send $(usk_{id_j},cr_{id_j})$ to $\mathcal{A}$
\label{line: cr-p sends secrets to adv}
\Else
\State Send $(usk_{id_j},\mathsf{Fake}(id_j,\PBB, CR_{id_j}, cr_{id_j}))$ to $\mathcal{A}$ \JQdone{introduce $\mathsf{fake}$ in syntax}
\EndIf
\While{\JQ{$clk \leq t_{end}$}}
\State \JQ{Send $cr_{id_j}$ or $\mathsf{Fake}(id_j,\PBB,CR_{id_j},cr_{id_j})$ depending on $b$ if $\adv$ requests for credential}
\State $used:=\bot$ \algorithmiccomment{mark of moment-of-privacy}
\For {$id\in\mathbb{I}\backslash \Icorr$}
\State Send $(\PBB,clk)$ to $\mathcal{A}$ and receive list of ballots $\mathbb{L}$
\For{tuples $(id',\beta_{\mathcal{A}}) \in \mathbb{L}$}
\If{\JQ{no ballot added for $id'$ to $\mathbb{BB}$ at $clk$}}
\State $\PBB\leftarrow\mathsf{Publish}(\mathsf{Include}(id', sk_{VS}, \mathbb{BB}, \beta_{\mathcal{A}}))$
\EndIf
\If{$id'=id_j$}
\State $used := \top$ \algorithmiccomment{no moment-of-privacy}
\EndIf
\EndFor
\If{$clk\in\vec{t}^{id}_{v}$} \algorithmiccomment{voter $id$ casts ballot}
\If{$id=id_j \land used = \bot\land o \neq \Phi \land b = 0$}
\State \algorithmiccomment{$id_j$ choose to evade and she wants to cast}
\State $\beta\leftarrow \mathsf{Vote}(id_j,usk_{id_j},pk_T, pk_{VS},o,cr_{id_j})$
\State $\PBB\leftarrow\mathsf{Publish}(\mathsf{Include}(id_j, sk_{VS}, \mathbb{BB}, \beta))$
\ElsIf{$id\neq id_j$}
\State $\beta\leftarrow\mathsf{Vote}(id,usk_{id},pk_T,pk_{VS},o_{id},cr_{id})$
\State $\PBB\leftarrow\mathsf{Publish}(\mathsf{Include}(id, sk_{VS}, \mathbb{BB},\beta))$
\EndIf
\EndIf
\EndFor
\If{$clk\in\vec{t}^{id}_{vs}\land clk\notin\vec{t}^{id}_{v} \land \adv$ did not cast for $id$} 
\State $\PBB\leftarrow\mathsf{Publish}(\mathsf{Include}(id, sk_{VS}, \mathbb{BB}, \mathsf{Noise}(id, sk_{VS},pk_T,\PBB)))$
\EndIf
\State $clk++$ \algorithmiccomment{clock advances}
\EndWhile
\State $(W,\Pi)\leftarrow\mathsf{Tally}(\PBB,sk_{T})$ 
\State Send $(\PBB,W,\Pi)$ to $\adv$ and receive $b'$
\If{$b = b'$} 
\State\Return 1 \algorithmiccomment{$\mathcal{A}$ wins}
\Else
\State\Return 0 \algorithmiccomment{$\mathcal{A}$ fails}
\EndIf
\end{algorithmic}
\vspace*{3pt}\hrule\vspace*{3pt}
\caption{CR-Privacy game for protocol $\Gamma^{\DR^{vs},\DT^{vs}}$ and adversary.}
\label{algo: game CR-P Gamma}
\end{figure}

\begin{definition}[CR-Privacy]
\label{def: cr privacy}
Let $\mathbf{param} = (\mathbb{I}, \mathbb{O}, \DO, \DR^v, \DT^{v}, t_{end})$ be some voting parameters, and $\Gamma^{\DR^{vs},\DT^{vs}}$ be an e-voting scheme. $\Gamma^{\DR^{vs},\DT^{vs}}$ is said to satisfy \JQC{$\delta$-}CR-Privacy with respect to voting parameters $\mathbf{param}$, if for all PPT adversaries $\adv$, it holds that:
\begin{equation*}
    \begin{split}
        \Pr[\GameCRP{\Gamma^{\DR^{vs},\DT^{vs}}}{\adv}(1^{\lambda},\mathbf{param})=1] 
        \leq \delta.
    \end{split}
\end{equation*}

\end{definition}

\subsubsection{The Optimal Constant $\delta_{min}$}
Küsters \emph{et al.} define the constant $\delta_{min}$, the minimum level of coercion-resistance the ideal protocol provides. This value depends on the number of honest and not coerced voters ($n_h$), the number of voting options ($n_o$), and the distribution ($\DO$). Specifically, $\delta_{min}^{o,o_c}(n_h,n_o,\DO)$ represents the minimum coercion-resistance the ideal protocol achieves when a voter is coerced to vote for $o_c$ instead of their intended choice, $o$. Then, $\delta_{min}(n_h,n_o,\DO)$ is the maximum of $\delta_{min}^{o,o_c}(n_h,n_o,\DO)$ over all possible $o$ and $o_c$ in the set of options $\mathbb{O}$. 
For completeness, the detailed specification of the ideal protocol and the definitions of $\delta_{min}^{o,o_c}(n_h,n_o,\DO)$ and $\delta_{min}(n_h,n_o,\DO)$ are provided in Appendices~\ref{apx:ideal privacy game} and~\ref{apx: computation of delta min} respectively. We utilise their $\delta_{min}$-optimality theorem below in our subsequent attack.

\begin{theorem}[$\delta_{min}$-optimal~\cite{kusters2012game}]
\label{thm: delta min}
    For all adversaries $\adv$ corrupting and coercing at most $n$ voters, and for all ideal election parameters $(\mathbb{I}, \mathbb{O}, \DO)$:
    \[
    \delta_{min}(|\mathbb{I}|-n,|\mathbb{O}|,\DO) \le \Pr[\GameCRP{Ideal}{A}(1^\lambda, \mathbb{I}, \mathbb{O}, \DO)=1]
    \]
\end{theorem}

An e-voting scheme $\Gamma^{\DR^{vs},\DT^{vs}}$ is said to satisfy CR-Privacy if it satisfies  $\delta_{min}$-CR-Privacy. 

\subsection{Forced Abstention Attack Against Loki}
\label{subsec: forced-abstention attack on Loki}

In a forced abstention attack, the coercer intends to prevent the voter from casting any vote. In the ideal game, the adversary's ability to detect non-compliance is limited to information leaked by the election result and the voters' behaviour distributions, $\DO$, $\DR^v$, and $\DT^v$. However, an adversary in the real game possesses a greater advantage in detecting whether a coerced voter disobeyed and cast a ballot using the evasion strategy. Specifically, we demonstrate a forced abstention adversary, $\Afa$ (Figure~\ref{algo: A FA}), interacting with $\GameCRP{Loki}{\cdot}$. This adversary simply observes the length of the coerced voter $id_j$'s record, $\PBB^{id_j}$, and achieves a significantly greater winning advantage than any adversary in the ideal game. Indeed, if the coercer instructs their victim to abstain and at the end of the election the voter's cast ballot record contains only one ballot, the coercer can conclude that this ballot was a noise ballot cast by the server ($\DR^{vs}[0] = 0$), and thus that the voter complied with their instructions to abstain. We formalise this in our next theorem 
and see proofs of it in Appendix~\ref{apx: proof of thm forced abstention attack}.

\begin{figure}[!t]
\hrule
\vspace*{3pt}
$\mathcal{A}_{FA}$
\vspace*{3pt}
\noindent\hrule
\begin{algorithmic}[1]
\scriptsize
\Require{$\lambda$}

\UponReceiving{$(pk_T,pk_{VS},\{upk_{id}, L_{id}\}_{id\in\mathbb{I}}, \PBB, \mathbf{param})$}
\State send back $\Icorr = [\ ]$
\algorithmiccomment{does not corrupt anyone}
\EndUponReceiving

\UponReceiving{$(\{usk_{id},l_{id}\}_{id\in\Icorr})$}
\State $id_j\leftarrow_{\$}\mathbb{I}$
\State $o\leftarrow_{\$}\mathbb{O}\backslash\Phi$
\State send back $(id_j,o)$
\EndUponReceiving

\UponReceiving{$usk_j, l_{id_j}$}
\State do nothing
\EndUponReceiving

\State \JQ{do not request for updated $l_{id_j}$}

\WhileReceiving{$(\PBB,clk)$}
\State do nothing
\EndWhileReceiving
\State
\UponReceiving{$(\{\PBB\}_{id\in\mathbb{I}}, W,\Pi)$}
\If{$|\PBB^{id_j}|=1$}
\algorithmiccomment{not counting the twin re-encryption ballot}
\State send $b'=1$
\Else
\State send $b'\leftarrow_{\$}\{0,1\}$
\EndIf
\EndUponReceiving

\end{algorithmic}
\vspace*{3pt}\hrule\vspace*{3pt}
\caption{Forced absention attack algorithm $\mathcal{A}_{FA}$.}
\label{algo: A FA}
\end{figure}

\begin{theorem}
\label{thm: forced abstention attack}
Let $\mathbf{param} = (\mathbb{I}, \mathbb{O}, \DO, \DR^v, \DT^{v}, t_{end})$ be some election parameters such that $\DO(id_j)[\Phi] > 0$,
and $\DR^{v}(id_j)[1] = q < 1-\frac{4}{p}(\delta_{min}(|\mathbb{I}| - 1,|\mathbb{O}|,\DO)-\frac{1}{2})$ for some constants $p$ and $q$. 
Let distribution $\DR^{vs}$ be independent to $\DR^{v}$, $\DT^{vs}$ be independent to $\DT^{v}$, and assume that there exists a voter $id_j$ for which $\DR^{vs}(id_j)[1] = p > 0$. 
Then, there exists a PPT adversary $\Afa$, such that:

\begin{equation*}
    \begin{split}
        \Delta =\Big|
        &\Pr[\GameCRP{Loki^{\DR^{vs},\DT^{vs}}}{\Afa}(1^\lambda,\mathbf{param})=1] \\
        &- \delta_{min}(|\mathbb{I}|-1,|\mathbb{O}|,\DO)
        \Big|
        \geq p'
    \end{split}
\end{equation*}

where $p'>0$ is a constant.
\end{theorem}

\subsection{On the Pitfalls of Loki's Security Analysis}
\label{subsec: discussion on loki original argument}

Loki introduces an ad hoc coercion-resistance definition for its analysis that suffers from two critical shortcomings. 
First, the analysis assumes an unduly weak adversary (both lazy and $\DR$-agnostic). Theorem~\ref{thm: CR-I attack without DR} is aligned with, and formally captures, the threat model considered by the authors of Loki in\cite{giustolisi2023thwarting}. In particular, the brute-force attack is ineffective only if voters behave in a uniformly random manner, an assumption contradicted by empirical data such as that from Estonia. Theorem~\ref{thm: CR-I attack with DR} considers a more realistic model, accounting for real voters' revoting patterns. Under this setting, we demonstrate that even a single malicious ballot can yield an attack success rate of $20\%$. 
Second, although the original Loki paper constrains the noise distribution, it fails to account for how an adversary could exploit it. More fundamentally, the coercion-resistance definition abuses the twin-ballot mechanism (Algorithm~3, cast oracle, line~6 in~\cite{giustolisi2023thwarting}). It hardcodes in the definition that adversarially cast ballots are always deemed invalid and detected by the Voting Server, thereby rendering the noise distribution irrelevant in practice.
\section{On the Impossibility of Last Minute Coercion-Resistance without Pre-Agreed Secrets}
\label{sec: can we fix loki}

Loki, in fact, belongs to the broader class of signalling-based revote protocols. 
It introduces a trusted Voting Server (trusted solely for coercion-resistance) in order to simultaneously satisfy the following requirements:
\begin{itemize}
\item[(A1)] Request a coercion-free window at any time during the voting period, thereby addressing last-minute coercion.
\item[(A2)] Avoid long, unique, and pre-agreed secrets between voters and election authorities.
\end{itemize}

We formalize the syntax of a generic ballot-signalling mechanism and demonstrate that revoting protocols following a last-ballot-counts policy, when constructed using such a signalling mechanism, fail to provide adequate coercion-resistance while complying with (A1) and (A2) at the same time.

\subsection{Syntax of Ballot-Signalling Mechanism}
\label{subsec: signalling syntax}

We introduce five stateful algorithms—$\mathsf{InitSgn}$, $\mathsf{Sgn}$, $\tau_v$, $\mathsf{VrfSgn}$, and $\tau_{vs}$—which collectively function as a signalling scheme embedded into the e-voting protocol syntax. 
We assume that all public information associated with the e-voting protocol, including $\mathbf{param}$, is on the public bulletin board. Therefore, the five algorithms take $\PBB$ as parameter for the sake of inputting required public information.
The protocol $\mathsf{InitSgn}(id,1^{\lambda},\PBB)$ is invoked during the execution of the $\mathsf{Register}(\cdot)$ algorithm to initialize the signalling state on the voter's side and the Voting Server's side, respectively. Each state includes both a public and a private component: the private part acts as the private credential, while the public part serves as a public credential.

To cast a ballot via $\mathsf{Vote}(\cdot)$, given the current state $\Sigma^{t}_{id}$ and the current bulletin board $\PBB^t$, the voter first computes a signal $l_{id}^t \leftarrow \mathsf{Sgn}(id,\Sigma_{id}^t, \PBB^t; s_{v}^t)$, which represents a freely cast vote. The voter then updates their state using $\Sigma_{id}^{t+1} \leftarrow \tau_v(id, \Sigma_{id}^t, \PBB^t, l_{id}^{t})$, and generates the corresponding ballot embedding the signal $l_{id}^t$.

Upon receiving a ballot, the Voting Server invokes $\mathsf{Include}(\cdot)$, during which it verifies whether the ballot was freely cast by evaluating $\mathsf{VrfSgn}(id,\Sigma_{vs}^t, \PBB^t, l_{id}^t)$, where $l_{id}^t$ is the signal extracted from the ballot. If the verification fails, the ballot is flagged as having been cast under coercion. The Voting Server then updates its own internal state via $\Sigma_{vs}^{t+1} \leftarrow \tau_{vs}(id, \Sigma_{vs}^t, \PBB^t, l_{id}^t)$ and includes the received ballot along with any updates specified by $\mathsf{Include}(\cdot)$.
%

More precisely, we introduce the following algorithms:
\begin{itemize}[leftmargin=*]
    \item $\mathsf{InitSgn}(\mathbb{I},1^{\lambda},\PBB) \rightarrow \big( \big\{\Sigma_{id}^{0}=(\Sigma_{id,pb}^{0}, \Sigma_{id,pv}^{0})\big\}_{id\in\mathbb{I}},$ $\Sigma_{vs}^{0}=(\Sigma_{vs,pb}^{0}, \Sigma_{vs,pv}^{0}) \big)$: An interactive protocol run by all voters and the Voting Server with public and security parameters, which returns the initial state of every voter and the Voting Server, each having a public and a private part.
    \item $\mathsf{Sgn}(id,\Sigma_{id}^{t}, \PBB^t; s_{id}^t)\rightarrow l_{id}^t$: On input a voter identity $id$, their current state $\Sigma_{id}^t$, some public information from $\PBB^t$, and a random value $s_{id}^t$, it outputs a signal $l^t_{id}$.
    \item $\tau_v(id,\Sigma_{id}^{t}, \PBB^t, l_{id}^t)\rightarrow\Sigma_{id}^{t+1}$: The state transition function on voter's side takes the voter's identity, their current state $\Sigma_{id}^t$, some public information from $\PBB^t$, and the signal produced by $\mathsf{Sgn}(\cdot)$ as input and outputs a new state $\Sigma_{id}^{t+1}$. 
    \item $\mathsf{VrfSgn}(id,\Sigma_{vs}^{t}, \PBB^t, l_{id}^t)\rightarrow \top/\bot$: On input the voter's $id$, the Voting Server's current state $\Sigma_{vs}^t$, public information $\PBB^t$, and received signal $l_{id}^t$ from voter $id$, it verifies the received signal and returns accept ($\top$) or reject ($\bot$).
    \item $\tau_{vs}(id,\Sigma_{vs}^{t}, \PBB^t, l_{id}^t)\rightarrow\Sigma_{vs}^{t+1}$: The state transition function on the Voting Server's side takes a voter's identity, the current state of the Voting Server $\Sigma_{vs}^t$, some public information from $\PBB^t$, and a signal of $id$ accepted by $\mathsf{VrfSgn}(\cdot)$ as input and outputs a new state $\Sigma_{vs}^{t+1}$. 
\end{itemize}

The assumption (A2) that voters and Voting Server have no pre-agreed secrets is equivalent to requiring that their initial states' private components be empty, i.e. $\Sigma_{id,pv}^0 = \Sigma_{vs,pv}^0 = \emptyset$.

\subsection{Impossibility Result}
\label{subsec: on the impossibility}

We now present our no-go result, establishing that any class of e-voting protocols satisfying the following conditions cannot achieve coercion-resistance under assumptions (A1) and (A2):
\begin{itemize}
    \item[(B1)] They are constructed using the signalling mechanism defined in Section~\ref{subsec: signalling syntax}.
    \item[(B2)] They adopt a last-ballot-counts revote policy, i.e., in the case of multiple votes, the last valid ballot cast by a voter is selected. This represents the most commonly deployed revote policy.
    \item[(B3)] Any signal that was honestly generated by the signal-generation algorithm $\mathsf{Sgn}(\cdot)$ will be verified correctly. Formally, for all $id\in\mathbb{I}$, 
    \begin{equation*}
    \begin{split}
        \hspace{-0.8cm}\Pr\left[
        \begin{array}{l}
         (\{\Sigma_{id}^0\}_{id\in\mathbb{I}}, \Sigma_{vs}^0) \leftarrow \mathsf{InitSgn}(\mathbb{I},1^\lambda,\PBB);\\
         \mathsf{VrfSgn}(id,\Sigma_{vs}^0, \PBB^0, \mathsf{Sgn}(id,\Sigma_{id}^{0},\PBB^0))\rightarrow\top
        \end{array}
        \right] = 1.
    \end{split}
    \end{equation*} 
    \item[(B4)] The verification procedure $\mathsf{VerifyTally}(\cdot)$ of the bulletin board $\PBB$ against the published tally result $W$ outputs \emph{true} if the ballots recorded on $\PBB$ are honestly generated by the ballot-generation algorithm $\mathsf{Vote}(\cdot)$.
    \item[(B5)] For every reachable 
    Voting Server state $\Sigma^{t}_{vs}$ and $\mathbb{PBB}^t$ (via an execution of $\GameCRI{\Gamma}{\adv}$ or $\GameCRP{\Gamma}{\adv}$), there is a partition of $\Sigma^{t}_{vs}$ denoted by $\{\Sigma^{t}_{vs,id}\}_{id\in\mathbb{I}}$ and a partition of $\mathbb{PBB}^t$ denoted by $\{\mathbb{PBB}^t_{id}\}_{id\in\mathbb{I}}$ such that for every signal $\ell$ and $id\in\mathbb{I}$:
      \begin{equation*}
    \begin{split}
         \mathsf{VrfSgn}(id,\Sigma_{vs}^t, \PBB^t,\ell)=     \mathsf{VrfSgn}(id,\Sigma_{vs,id}^t, \PBB_{id}^t,\ell).
    \end{split}
    \end{equation*} 
    
    Intuitively, the above condition ensures non-interferences between different voters' signals and ballots. Hence, we can consider identity-based partitions of Voting Server's view of the execution so that the behavior of the signaling verification process can be determined solely on information indexed by the voter's identity.
\end{itemize}

We believe these conditions are reasonable and that they are satisfied by all known protocols designed to withstand coercion. For instance, in JCJ protocols, the signal emitted by voters is simply the encryption of their credential under the registrars' public key. The corresponding signal verification is the plaintext equivalence tests that the registrars run during tallying. Further, in JCJ protocols, the state of the voters and registrars does not need updating. 
~\cite{park2024zkvoting} introduce a new primitive, nullifiable commitment, where a real commitment key can have multiple shadow fake commitment keys, and the entity with the secret master key can verify if the commitment is generate from a real key or shadow fake keys. With this primitive, the signal is a commitment to options under real or fake commitment key and the signal verification is the commitment verification. ~\cite{achenbach2015improved} alters the credential in JCJ to a pair of signing keys and introduces authentic timestamps. The ballot signal is a encryption to the public signing key under the tallier's key and a timestamp. The verification is sorting out superseded ballots via encrypted plaintext equality tests (EPETs) and PETs. The difference is that the state of the voters changes as it includes timestamps. In contrast, for protocols like DeVoS~\cite{muller2024devos}, VoteAgain~\cite{lueks2020voteagain} that rely on revoting rather than a signaling mechanism, the signal is the empty string, and the signal verification algorithm always returns $\top$. 

\MAdone{What about [13] and [14] ?}
\JQdone{[13] is having an nullifiable commitment scheme and the commitment works as a signal. [14] they have a credential-like thing, which is a public key.} \MAdone{I don't know - I don't remember having looked how these two protocols work.}

\TZdone{We should provide some evidence of how broad this class is. Namely, explain why conditions (B2), (B3), and (B4) are very reasonable and if possible, explain how well-known e-voting designs fall into the scope of (B1).}

We now prove our impossibility result. It establishes that the above class of e-voting protocols cannot achieve coercion-resistance without either relying on pre-agreed secrets or assuming a last-minute coercion-free window. Specifically, we demonstrate that when the probability of voters casting ballots at both times $t_0 = 0$ and $t_1 = 1$ is non-negligible, an attack on CR-Integrity becomes feasible.

\begin{theorem}
\label{thm: incompatible assumptions}
Let $\mathbf{param} = (\mathbb{I}, \mathbb{O}, \DO, \DR^v, \DT^{v}, t_{end})$ be any election parameters such that
$\Pr[r_v^{id_j}\leftarrow\DR^{v}(id_j);t_0\in\DT^v(r_v^{id_j}, id_j)] \geq \alpha(\lambda)$ and $\Pr[r_v^{id_j}\leftarrow\DR^{v}(id_j);t_1\in\DT^v(r_v^{id_j}, id_j)] \geq \alpha'(\lambda)$, where $\alpha(\cdot)$ and $\alpha'(\cdot)$ are non-negligible functions. Let $\DR^{vs},\DT^{vs}$ be any distributions, and $\Gamma^{\DR^{vs},\DT^{vs}}$ be any revote protocol that satisfies conditions (B1), (B2), (B3), (B4) and (B5).
If both assumptions (A1) and (A2) hold, then there exists a PPT adversary $\adv$ such that $\Pr[\GameCRI{\Gamma^{\DR^{vs},\DT^{vs}}}{\adv}(1^{\lambda},\mathbf{param}) = 1]$ is non-negligible in $\lambda$.
\end{theorem}

The intuition behind this proof is to partition the possible class of protocols in two types based on their signalling mechanism. Either (i) the signalling scheme is easily overwritten: here, the coercer just casts their vote after the voter, thereby overwriting the voter's intent; or (ii) conversely, the signal is difficult to overwrite: here, the coercer casts their vote before the voter, leaving the coerced voter with the challenging task of circumventing the previously cast malicious signal. See Appendix~\ref{apx: impossibility result proof} for the complete proofs.

\section{CR-Loki: A Coercion-Resistant Variant of the Loki Protocol}
\label{subsec: one possible way}

In this section, we explore one variant of Loki in order to achieve coercion-resistance. Specifically, we replace the use of voters' voting history (or any function derived from it) as a coercion signal with randomly generated, unpredictable credentials. These credentials, similar to those employed in the JCJ protocol, are pre-agreed upon between the voter and the Voting Server. This change is designed to mitigate our integrity attack. Furthermore, we configure the server to ensure that a ballot and its corresponding twin ballot is appended to each voter's cast ballot record at every clock tick. This modification addresses our forced abstention attack. We refer to this variant of the Loki protocol as CR-Loki. 



\subsection{Overview of CR-Loki}
\label{sec: Loki overview}


This section provides an overview of the mechanisms implemented in CR-Loki to mitigate the aforementioned attacks. Due to space constraints, we focus on the differences between CR-Loki and Loki's original design. The full formal specification of CR-Loki is given in Figure~11 of Appendix~\ref{apx: formal description of cr-loki}.

{
\paragraph{Trust assumptions}
To facilitate the secure generation and pre-agreement of voter credentials, we adopt a trust model that deviates slightly from the original Loki scheme. The following trust assumptions from the original Loki protocol remain valid in our context: 
\begin{inparaenum}[(a)] 
\item the Voting Server is trusted to uphold coercion-resistance; 
\item voter authentication with the Voting Server is inalienable; 
\item each voter is able to cast at least one ballot free from coercion, hence respecting assumption (A1); and 
\item the bulletin board operates honestly. 
\end{inparaenum}
In addition to these, our model introduces two further assumptions that break assumption (A2), hence bypassing our impossibility result in Section~\ref{sec: can we fix loki}: first, the Voting Server must also be honest during the credential generation phase, and second, the adversary is assumed to be inactive during the registration phase—or, equivalently, that the registration process is untappable.}

\paragraph{Credentials as signalling mechanism}
Loki is vulnerable to an integrity attack because $l_{id}$ is guessable. Theorem \ref{thm: incompatible assumptions} establishes that without pre-agreed secrets, Loki cannot effectively resist coercion. Therefore, instead of relying on voters' voting history as a signalling mechanism, we propose utilizing JCJ-style credentials. During the registration phase, each eligible voter obtains their uniquely and randomly generated private credential $\{cr_{id}\leftarrow_{\$}\{0,1\}^{\lambda}\}_{id\in\mathbb{I}}$, \MAdone{Why define $l_{id}$ like that ? We are not exploiting any property of $\mathbb{Z}_p$ in CR-Loki. The credentials just need to be a random string of length $\lambda$. Maybe use $cr_{id}$ to avoid confusion with lists.} from the Voting Server. These credentials are kept secret by the voters. Simultaneously, the Voting Server computes the corresponding public credentials, $\{CR_{id}\leftarrow_{\$}\mathsf{Enc}_{CCA}(pk_{VS},cr_{id};r^{cr_{id}})\}_{id\in\mathbb{I}}$, and publishes them on $\PBB$. We require that the encryption scheme employed by the Voting Server be IND-CCA secure. 

{
\paragraph{Ballot generation}
Our intention is to respect Loki's design as much as possible. After changing to a static random credential $cr_{id}$ rather than an ever-changing list $l_{id}$, $ct_{l_{id}}$ becomes redundant. Therefore, CR-Loki ballots are of the form $\beta = ct_o \mathbin\Vert ct_{cr_{id}} \mathbin\Vert \pi$.
The encryption scheme used for the option ciphertext $ct_{o}$ is required to satisfy IND-RCCA security, whereas the encryption scheme used for the credential ciphertext $ct_{cr}$ satisfies IND-CCA security. Specifically,
\begin{align*}
    ct_{o_{id}}\leftarrow_{\$} \mathrm{Enc}_{RCCA}(pk_T,o_{id};r_o) \\
    ct_{cr_{id}}\leftarrow_{\$} \mathrm{Enc}_{CCA}(pk_{VS},cr_{id};r_{cr})
\end{align*}
The disjunctive NIZKP~\cite{cramer1994proofs} $\pi$ if retained for verifiability without revealing which kind of relation is proved.
Accordingly, we change the disjunctive relation to the disjunction of the following three relations, where $i-1$—treated as an implicit input—denotes the length of $\PBB^{id}$.
\begin{itemize}[leftmargin=.8em, labelsep=0.5em, itemindent=\dimexpr\labelwidth+\labelsep\relax]
    \item $R^{id}$ 
    : This is the relation underlying proof in ballots cast by voters. It contains the proof of plaintext knowledge of the option $o$, the proof of plaintext knowledge of credential $cr_{id}$, and proof of private key knowledge of $usk_{id}$. Formally,
    \begin{align*}
        (x, & w)\in R_i^{id}\ \mathrm{iff} \\
        & ct_{o_{id}}=\mathrm{Enc}_{RCCA}(pk_T,o_{id};r_o) \land o_{id}\in \mathbb{O} \land \\
        & ct_{cr_{id}} = \mathrm{Enc}_{CCA}(pk_{VS},cr_{id};r_{cr}) \land cr_{id} \in \{0,1\}^\lambda \land \\
        & (usk_{id},upk_{id}) =\mathsf{SKG}(1^{\lambda})
    \end{align*}
    where $x=(pk_T,pk_{VS},\mathbb{O},(ct_{o_{id}},ct_{cr_{id}}), upk_{id})$, and $w\leftarrow (usk_{id},(r_o,o_{id}),(r_{cr},cr_{id}))$.
    
    \item $R^{pred}$
    : This is the relation used in the proofs for ballots that are re-encryptions of the preceding ballot within the voter's cast ballot record, $\PBB^{id_j}$.
    These ballots, along with their associated proofs, are generated by the Voting Server (either as noise ballots or as twin ballots corresponding to actual votes) when the voter's previous ballot in their cast ballot record was cast using valid credentials. A proof for $R^{pred}$ includes: a proof of re-encryption of $ct_o$, which is the ciphertext of the last ballot in $\PBB^{id}$; a proof of plaintext knowledge of $cr_{id}$, where $cr_{id}$ is the decryption of $CR_{id}$ (which is publicly available on $\PBB^{id_j}$); a proof demonstrating the equivalence of the decryption of $CR_{id}$ and $ct'_{cr_{id}}$ within the last\footnote{Denote with the superscript $-1$.} ballot in $\PBB^{id}$. Formally,
    \begin{align*}
        (x, w) & \in R_i^{pred}\ \mathrm{iff} \\
        & ct_{o_{id}}=\mathrm{ReEnc}_{RCCA}(pk_T,ct_{o_{id}}^{-1};r_o) \land \\
        & ct_{cr_{id}}=\mathsf{Enc}_{CCA}(pk_{VS},cr_{id};r_{cr}) \land \\
        & cr_{id} = \mathsf{Dec}_{CCA}(sk_{VS},CR_{id}) \land \\
        & \mathsf{Dec}_{CCA}(sk_{VS},ct_{cr_{id}}^{-1})= \mathsf{Dec}_{CCA}(sk_{VS}, CR_{id}) \land \\
        & (pk_{VS}, sk_{VS}) = \mathsf{EKG}_{CCA}(1^{\lambda})
    \end{align*}
    where $x=(pk_T,pk_{VS},\mathbb{O},CR_{id},(ct_{o_{id}}^{-1},ct_{cr_{id}}^{-1}),$ $(ct_{o_{id}},ct_{cr_{id}}))$ and $w\leftarrow (sk_{VS},r_o,r_{cr})$.
    \item $R^{pred2}$ 
    : This is the relation used in the proofs for ballots that are re-encryptions of the second-to-last ballot in the voter's cast ballot record, $\PBB^{id}$ (i.e., skipping the last ballot). These ballots, along with their associated proofs, are generated by the Voting Server when the last ballot in $\PBB^{id_j}$ were cast with invalid/fake credentials, \emph{i.e.} under coercion, and thus skipped. A proof for $R^{pred2}$ includes: a proof of re-encryption of $ct_o$, which is the ciphertext of the second-to-last\footnote{Denote with the superscript $-2$.} ballot in $\PBB^{id}$; a proof of plaintext knowledge of $cr_{id}$, where $cr_{id}$ is the decryption of $CR_{id}$ (which is publicly available on $\PBB^{id_j}$); a proof demonstrating the non-equality of the decryption of $CR_{id}$ and $ct'_{cr_{id}}$ within the last ballot in $\PBB^{id}$. Formally,
    \begin{align*}
        (x,w) & \in R_i^{pred2}\ \mathrm{iff} \\
        &ct_{o_{id}}=\mathrm{ReEnc}_{RCCA}(pk_T,ct^{-2}_{o_{id}};r_o) \land \\
        & ct_{cr_{id}}=\mathsf{Enc}_{CCA}(pk_{VS},cr_{id};r_{cr}) \land \\
        & cr_{id} = \mathsf{Dec}_{CCA}(sk_{VS},CR_{id}) \land \\
        & \mathsf{Dec}_{CCA}(sk_{VS},ct^{-1}_{cr_{id}}) \neq \mathsf{Dec}_{CCA}(sk_{VS}, CR_{id}) \land \\ 
        & (pk_{VS}, sk_{VS}) = \mathsf{EKG}_{CCA}(1^{\lambda})
    \end{align*}
    where $x=(pk_T,pk_{VS},\mathbb{O},CR_{id},(ct^{-2}_{o_{id}},ct_{cr_{id}}^{-1}),$ $(ct_{o_{id}},ct_{cr_{id}}))$ and $w\leftarrow (sk_{VS},r_o,r_{cr})$.
\end{itemize}
}

\paragraph{Configuration of the Voting Server}
\label{sec: configure vs noise}
In the Loki system, the Voting Server's casting behaviour distributions, $\DR^{vs}$ and $\DT^{vs}$, are initially configured to mirror those of voters. However, this configuration leads to a privacy vulnerability when a significant number of voters intend to abstain. To address this, we modify $\DR^{vs}$ to a one-dimensional degenerate distribution and $\DT^{vs}$ to a $t_{end}$-dimensional degenerate distribution. Specifically, for all $id\in\mathbb{I}$, given $t_{end}$, the probability $\Pr_{r\leftarrow \DR^{vs}}[r=t_{end}]$ is set to 1, and the probability $\Pr_{\vec{t}\leftarrow \DT^{vs}}[\vec{t}=(0, \dots, t_{end}-1)]$ is also set to 1. In simpler terms, if the Voting Server does not receive a ballot from a voter at any given clock tick, it will automatically cast a noise ballot on their behalf to mask their inactivity.

\paragraph{Evasion strategy} 
When faced with coercion, a voter can present a fake credential $\tilde{cr}_{id}$ along with their private verification key. Fake credentials are generated by running $\tilde{cr}_{id}\leftarrow\mathsf{Fake}(id,\PBB, CR_{id},$ $cr_{id})$. The algorithm simply samples a random credential from $\{0,1\}^\lambda$ and  different to $cr_{id}$. During over-the-shoulder coercion, the voter follows the commands of the coercer but using their fake credential $\tilde{cr}_{id}$. During their moment-of-privacy, the voter uses their genuine credential $cr_{id}$ to cast their ballot for their intended option, or does nothing if they intended to abstain.

Ballots from voters and the Voting Server, as well as those cast with real and fake credentials, are indistinguishable. This indistinguishability prevents a voter from convincingly proving their vote to a coercer.

\subsection{Security Analysis of CR-Loki}

We show that CR-Loki is a coercion-resistant e-voting protocol (Theorem~\ref{thm: cr-loki is coercion resistance}). The proof proceeds in two steps, which we sketch here. For the complete proof, we refer the reader to Appendix~\ref{apx: coercion resistant proofs for cr-loki}.

\begin{theorem}
\label{thm: cr-loki is coercion resistance}
Let $(\mathsf{EKG}_{RCCA}, \mathsf{Enc}_{RCCA}, \mathsf{Dec}_{RCCA}$, $\mathsf{ReEnc}_{RCCA})$ be the IND-RCCA secure encryption scheme with perfect rerandomization of the talliers. Let $(\mathsf{EKG}_{CCA}$, $\mathsf{Enc}_{CCA} $, $\mathsf{Dec}_{CCA})$ be the IND-CCA secure encryption scheme of the Voting Server. Assuming UC-security of the key generation algorithm (KG) and of NIZKPs, CR-Loki$^{\DR^{vs}, \DT^{vs}}$ is coercion-resistant.
\end{theorem}

For completeness, we confirm that the alterations of CR-Loki to the original Loki design do not affect \emph{ballot privacy} and \emph{verifiability}. Similarly to the original paper, we adopt the ballot privacy definition by Bernhard \emph{et al.}~\cite{bernhard2012not} and the verifiability definition introduced in Cortier \emph{et al.}~\cite{cortier2014election}, to prove that CR-Loki satisfies privacy and verifiability. Details are in Appendix~\ref{apx: privacy analysis for cr-loki} and Appendix~\ref{apx: verifiability analysis for cr-loki}.

\subsection{Performance Comparison}

\begin{table*}[!t]
\centering
\begin{tabular}{|c|c|c|c|l|}
    \toprule 
    Protocol & Usability & $\DR^{vs}$ & $\mathsf{Exp}(n_{\beta})$ & Efficiency \\ \midrule
    CR-Loki & \multirow{2}{*}{random} & $\scriptstyle \Pr_{r\leftarrow \DR^{vs}}[r=t_{end}]=1$ & $\scriptstyle t_{end}\cdot n_v$ & $2t_{end}\cdot n_v$ re-encryptions and  $n_v$ decryptions \\ \addlinespace[0.2em]\cline{1-1}\cline{3-5}\addlinespace[0.2em]
    \multirow{4}{*}{JCJ*} & credentials 
    & \multirow{2}{*}{$\scriptstyle \Pr_{r\leftarrow \DR^{vs}}[r=t_{end}] = 1$} & \multirow{2}{*}{$\scriptstyle t_{end}\cdot n_v$} & $\scriptstyle 3(257+\lceil\log(t_{end}\cdot n_v)\rceil)(t_{end}+1) \cdot \log^2((t_{end}+1) \cdot n_v) \cdot n_v$  \\
    & & & & re-encryptions and $1.515n_v$ decryptions \\
    \addlinespace[0.2em]\cline{3-5}\addlinespace[0.2em]
     & & \multirow{2}{*}{$\scriptstyle \mathcal{D}_{uni}\sim\{0,\cdots,t_{end}\}$} & \multirow{2}{*}{$\scriptstyle (0.5807 + \frac{1}{2}t_{end})\cdot n_v$} &
    $\scriptscriptstyle 3(257+\lceil\log((0.5807+\frac{1}{2}t_{end})\cdot n_v)\rceil)(1.5807+\frac{1}{2}t_{end}) \cdot \log^2((1.5807+\frac{1}{2}t_{end})\cdot n_v) \cdot n_v$  \\ 
    & & & & re-encryptions and $1.515n_v$ decryptions \\
    \bottomrule
\end{tabular}
\caption{Comparison between CR-Loki, and JCJ*. We consider the distribution from Estonia's 2021 Local Election. We set the security parameter $\lambda=256$ consistent with the underlying ElGamal encryption. We set the number of talliers $n_T=3$.}
\label{tb: cr-loki vs jcj}
\end{table*}

To mitigate vulnerabilities in the original Loki design, we introduced JCJ-type credentials, negotiated prior to the casting phase. This adjustment, while enhancing security, impacts usability. Additionally, we configure the Voting Server to incorporate noise ballots into each voter's cast ballot record at every clock cycle, resulting in multiple re-encryptions on the Voting Server. This raises the question of efficiency, particularly in comparison to the JCJ protocol and its subsequent enhancements~\cite{araujo2016remote,grontas2019towards,spycher2011new,yin2023scalable}. Unfortunately, the original JCJ protocol, along with its optimizations, is vulnerable to the attack identified in~\cite{cortier2022jcj}. The only known coercion-resistant variant within the JCJ family that addresses this vulnerability is the fix proposed in the same work, which we denote as JCJ*. We investigated whether introducing a trusted Voting Server, which strengthens trust assumptions relative to JCJ*, offers any efficiency gain.

Let $n_{\beta}$ represent the number of ballots. CR-Loki requires $2n_{\beta}$ re-encryptions to maintain the bulletin board during the casting phase and $n_v$ decryptions for tallying. In contrast, JCJ* implements the tallying algorithm using secure multi-party computation (MPC), and 
the overall computational complexity of the tally phase is $\mathcal{O}((\lambda + m) n_T (n_{\beta} + n_v) \log^2(n_{\beta} + n_v))$, where $m = \lceil \log(n_{\beta}) \rceil + 1$ and $n_T$ denotes the number of talliers.
CR-Loki's computational complexity scales linearly with the number of ballots, whereas JCJ* scale quasi linearly~\cite{cortier2022jcj}. 

\JQ{We now examine the relation between the number of ballots, $n_{\beta}$, and the number of voters, $n_v$. In CR-Loki, and according to the calibration of the Voting Server's noise distributions $\DR^{vs}$ and $\DT^{vs}$, a fixed number of $n_v$ ballots is added to the board at each tick. Consequently, the total number of ballots is given by $n_{\beta} = t_{\mathrm{end}} \cdot n_v$, regardless of the voters' actual distribution.  
In contrast, the noise distribution in JCJ* is only informally described as ``the authorities add a random number of dummy ballots.'' To enable a concrete comparison, we instantiate the JCJ* noise distribution in two ways: (i) following their description, by setting $\DR^{vs}(id) = \mathcal{D}_{\mathrm{uni}} \sim \{0, \dots, t_{\mathrm{end}}\}$ for all $id \in \mathbb{I}$, and (ii) mirroring CR-Loki, by setting $\Pr_{r \leftarrow \DR^{vs}}[r = t_{\mathrm{end}}] = 1$. For the voters' distribution, we adopt the empirical data from the 2021 Estonian election. Using these assumptions, we compute the expected number of ballots in JCJ* under both noise calibrations. Table~\ref{tb: cr-loki vs jcj} summarizes this comparison between CR-Loki and JCJ*.}

\JQ{CR-Loki is more efficient compared to JCJ* when both employ the same noise distribution. Even if JCJ* uses a uniform random noise distribution, CR-Loki can still maintain this advantage by appropriately configuring its clock cycle length. This efficiency gain, however, relies on a stronger trust assumption on the Voting Server, than that placed on the JCJ* Registrars. Indeed,  verification of the signal in JCJ* doesn't reveal to the Registrars the voters' credentials, while the Voting Server needs to decrypt and check these in plaintext. This confirms the key trade-off: achieving greater tallying efficiency requires placing more trust in the authorities.}






\bibliographystyle{IEEEtran}
\bibliography{references}

\begin{appendices}
\renewcommand{\thesection}{\Alph{section}}

\section{Definitions}
\label{apx: definitions}

\begin{definition}[Non-negligible function]
\label{def: non-negligible function}
    A function $\alpha(\cdot)$ is non-negligible if there exists $c\in\mathbb{N}$ such that for infinitely many $n\in\mathbb{N}$,
    \begin{equation*}
    \begin{split}
        \alpha(n)\geq \frac{1}{n^c}
    \end{split}
    \end{equation*}
\end{definition}

\begin{definition}[Negligible Function]
\label{def: negligible function}
    A function $\mu(\cdot)$ is negligible if, for all $c\in\mathbb{N}$, there exists $n_c\in\mathbb{N}$ such that for all $n>n_c$,
    \begin{equation*}
        \begin{split}
            |\mu(n)| < \frac{1}{n^c}
        \end{split}
    \end{equation*}
\end{definition}

\begin{definition}[Super-polynomial Function]
\label{def: super-polynomial function}
    A function $\omega(\cdot)$ is called super-polynomial if, for every $c\in\mathbb{N}$, there exists $n_c\in\mathbb{N}$ such that for all $n>n_c$,
    \begin{equation*}
        \begin{split}
            |\omega(n)| > n^d
        \end{split}
    \end{equation*}
\end{definition}

\section{Brute-Force Attack Against Loki}

We present a formal proof demonstrating how a specific class of brute force attackers, illustrated in Figure~\ref{algo: A brute}, can compromise the CR-Integrity of the Loki protocol.

\subsection{Proof of Theorem \ref{thm: CR-I attack without DR}}
\label{apx: proof of thm CR-I attack without DR}

    Given $\kappa(\cdot) = \Theta(\log(\cdot))$, we have that 
    \[
    \exists k_1, k_2 > 0 \text{ such that } \kappa(\lambda) = k_1 \log(\lambda) + k_2 = \log \lambda^{k_1} + k_2.
    \]
    We let $c = k_1+k_2$, and consider the following non-negligible function: \JQdone{numerator is $\lambda$ or 1?}
    \[
    \alpha(\lambda)\ \triangleq\ \frac{1}{\lambda^{k_1}\cdot 2^{k_2}}.
    \]
    Let $\lambda, \delta \in\mathbb{N}$, and $t_{end} = \kappa(\lambda) + \delta$. Given that at most one vote (along with its twin, \emph{i.e.} its re-encryption from the voting server) is appended to a voter's cast ballot record $\PBB^{id}$, there are at most $\kappa(\lambda)$ ballots in $\PBB^{id}$ (not counting their twin re-encrypted ballot) at time $\kappa(\lambda)$. 
    Now, at clock tick $\kappa(\lambda)$, there is only one valid list of indices among the $F_{\mathcal{D}_{uni}}(\kappa(\lambda), id, t_{end}) = 2^{\kappa(\lambda)} = 2^{k_1\log(\lambda) + k_2} = \lambda^{k_1}\cdot 2^{k_2}$ computed by the naive adversary (the $\DR^v$-agnostic adversary). Given that at clock tick $\kappa(\lambda)$, there are only $\delta$ casting slots remaining, the naive adversary will cast at most $\delta$ guesses out of the $\lambda^{k_1}\cdot 2^{k_2}$ computed ones. Hence,

    \begin{equation*}
        \begin{split}
         & \Pr[\GameCRI{Loki^{\DR^{vs},\DT^{vs}}}{\AbruteOpt{\mathcal{D}_{uni}}}(1^\lambda, \mathbf{param}) = 1] \\
         =& \min(\frac{\delta}{\lambda^{k_1}\cdot 2^{k_2}}, 1)\\  
         =& \min(\delta\cdot \alpha(\lambda), 1). 
        \end{split}
    \end{equation*}
\qed

\subsection{Loki is Effective Under Unrealistic Settings}
\label{apx: proof of clm super ploy m has bf security}

Essentially, Theorem~\ref{thm: CR-I attack without DR} establishes that Loki can only be secure against \emph{lazy} adversaries - that is, adversaries that are passive until all voters' cast ballot records are long enough. Specifically, the Loki evasion strategy can be \emph{effective against naive brute-force adversaries} with $\kappa(\cdot) = \omega(\log(\cdot))$. We formalize this in our next result, proving that if the Voting Server casts a ballot at each clock tick, then the Loki evasion strategy is effective against lazy naive brute-force adversaries 

\begin{proposition}
\label{clm: super ploy m has bf security}
Let $\mathbf{param} = (\mathbb{I}, \mathbb{O}, \DO, \DR^v, \DT^{v}, t_{end})$ be any voting parameters and $\DR^{vs}, \DT^{vs}$ be any distributions such that $\forall id$, $\DR^{vs}(id)$ is the constant distribution $t_{end}$. Then, for $\kappa(\cdot)=\omega(\log({\cdot}))$ and any $c \in \mathbb{N}$, there is a negligible function $\mu(\cdot)$ such that
\[\Pr[\GameCRI{Loki^{\DR^{vs},\DT^{vs}}}{\AbruteOpt{\mathcal{D}_{uni}}}(1^{\lambda},\mathbf{param})=1]\ \leq\ \mu(\lambda).\]
\end{proposition}

Let $\lambda\in\mathbb{N}$ be the security parameter. At time $\kappa(\lambda)$, because the voting server casts a ballot at every clock tick when voters don't \JQ{($\forall id, \DR^{vs}(id)$ is the constant distribution $t_{end}$)}, the length of any voter $id$'s cast ballot record $\PBB^{id}$ is of length  $\kappa(\lambda)$. Now, according to the description of $\mathcal{A}_{brute}^{c,\kappa,\mathcal{D}_{uni}}$, we have that $\mathcal{A}_{brute}^{c,\kappa,\mathcal{D}_{uni}}$ makes $\lambda^c$ guesses   for the list of indices of the coerced voter against their cast ballot record of length $\kappa(\lambda)$. The attacker will then be able to cast ballots encoding a fraction $\delta\le \lambda^c$ of those depending on the number of remaining casting slots until $t_{end}$. 
    
    Since $\kappa(\lambda)=\omega(\log\lambda)$, we have that for every constant $c'$ there is a $\lambda'$ such that for every $\lambda>\lambda'$, it holds that $\kappa(\lambda)>(c'+c)\cdot\log\lambda=\log\lambda^{c'+c}$. Hence, we have that for every constant $c'$, and for sufficiently large $\lambda$, it holds that \MAdone{Fix layout.}
    \begin{equation*}
        \begin{split}
        &\Pr[\GameCRI{Loki^{\DR^{vs},\DT^{vs}}}{\AbruteOpt{\mathcal{D}_{uni}}}(1^{\lambda},\mathbf{param})=1] \\
        =& \frac{\delta}{2^{\kappa(\lambda)}} \leq \frac{\lambda^c}{2^{\kappa(\lambda)}}\leq\\
        \leq& \frac{\lambda^c}{2^{\log\lambda^{c'+c}}}=\frac{\lambda^c}{\lambda^{c'+c}} \\
        =&\frac{1}{\lambda^{c'}}.
        \end{split}
    \end{equation*}
    
    Thus a lazy brute-force adversary breaks CR-Integrity with negligible probability.
\qed

Note that this proposition does not mean that securing the Voting Server's distribution alone guarantees Loki's security. This statement only indicates that a \emph{lazy naive brute-force adversary} is ineffective against CR-Integrity. As we'll demonstrate in the following section, a more sophisticated adversary can leverage their knowledge of $\DR^v$ to optimize their attack in terms of the window of opportunity, and we will show how easily this can be accomplished with real-world voters.

\subsection{$\DR^v$ Affects Attack Difficulty}
\label{app:distribution affects difficulty}
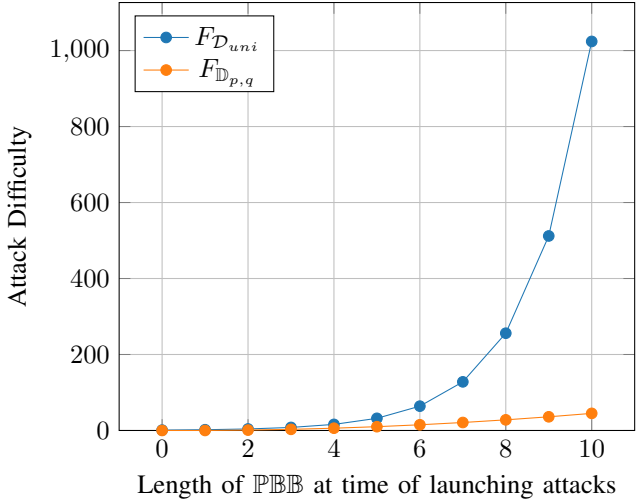
\begin{figure}
    \centering
    \pgfplotsset{compat=1.18}

\pgfplotstableread{
    X   Y_exp   Y_sum
    0   1       0
    1   2       0
    2   4       1
    3   8       3
    4   16      6
    5   32      10
    6   64      15
    7   128     21
    8   256     28
    9   512     36
    10  1024    45
}\mydata

\begin{tikzpicture}
    \pgfplotsset{
        style_y/.style={color=myblue, mark=*},
        style_z/.style={color=myorange, mark=*},
    }

    \begin{axis}[
        xlabel={Length of $\PBB$ at time of launching attacks},
        ylabel={Attack Difficulty},
        ymin=0,                   
        legend pos=north west,    
        grid=major,
    ]

        \addplot[style_y] table[x=X, y=Y_exp] from \mydata;
        \addlegendentry{$F_{\mathcal{D}_{uni}}$}

        \addplot[style_z] table[x=X, y=Y_sum] from \mydata;
        \addlegendimage{style_z}
        \addlegendentry{$F_{\mathbb{D}_{p,q}}$} 
    \end{axis}

\end{tikzpicture}
    \caption{The attack difficulty with respect to two type of voter re-voting distributions $\mathcal{D}_{uni}$ and $\mathbb{D}_{p,q}$.}
    \label{fig: F difficulty}
\end{figure}

In this section, we show that $\DR^v$ significantly affects the difficulty of launching brute-force attack. Specifically, we compare the uniform distribution $\mathcal{D}_{uni}$ with a class of distributions $\mathbb{D}_{p,q}$ that describes voter vote at most $q$ times with probability $p$. As show in Figure~\ref{fig: F difficulty}, the attack difficulty of $\mathbb{D}_{p,q}$ does not increase as fast as $\mathcal{D}_{uni}$, and the gap grows with the length of $\PBB$ at attacking time.

\subsection{Proof of Theorem \ref{thm: CR-I attack with DR}}
\label{apx: proof of thm CR-I attack with DR}

    We consider the following non-negligible function 
    \[
    \alpha(\lambda)\ \triangleq\ \frac{1}{(\kappa(\lambda)+1)^q}.
    \]
    Let $\lambda, \delta \in\mathbb{N}$ such that $\lambda>1$, and $t_{end} = \kappa(\lambda) + \delta$; 
    and let $c$ be any integer such that $2^c \ge \delta$.
    Since $\mathcal{D}^v_R \in \mathbb{D}_{p,q}$, it holds that
    \begin{equation*}
    \label{eq: law of prob}
    \begin{split}
    &\Pr[\GameCRI{Loki^{\DR^{vs},\DT^{vs}}}{\AbruteOpt{\mathcal{D}^*_{1,q}}}(1^\lambda, \mathbf{param})=1] \\
  \geq&\Pr[\GameCRI{Loki^{\DR^{vs},\DT^{vs}}}{\AbruteOpt{\mathcal{D}^*_{1,q}}}(1^\lambda, \mathbf{param})=1\land id_j\text{ votes}]\\
     \geq& \Pr[\GameCRI{Loki^{\DR^{vs},\DT^{vs}}}{\AbruteOpt{\mathcal{D}^*_{1,q}}}(1^\lambda, \mathbf{param})=1\land r^{id_j}_v \le q] \\
        +& \Pr[\GameCRI{Loki^{\DR^{vs},\DT^{vs}}}{\AbruteOpt{\mathcal{D}^*_{1,q}}}(1^\lambda, \mathbf{param})=1\land r^{id_j}_v > q] \\
    = & p\cdot \Pr[\GameCRI{Loki^{\DR^{vs},\DT^{vs}}}{\AbruteOpt{\mathcal{D}^*_{1,q}}}(1^\lambda, \mathbf{param})=1~|~r^{id_j}_v\le q] \\
    + &(1-p)\cdot \\
    & \Pr[\GameCRI{Loki^{\DR^{vs},\DT^{vs}}}{\AbruteOpt{\mathcal{D}^*_{1,q}}}(1^\lambda, \mathbf{param})=1~|~r^{id_j}_v> q]\\
    \geq&p\cdot \Pr[\GameCRI{Loki^{\DR^{vs},\DT^{vs}}}{\AbruteOpt{\mathcal{D}^*_{1,t}}}(1^\lambda, \mathbf{param})=1~|~r^{id_j}_v\le q]
    \end{split}
    \end{equation*}

    Now, given that at most one vote (along with its twin, \emph{i.e.}, its re-encryption from the voting server) is appended to a voter's cast ballot record $\PBB^{id}$, there are at most $\kappa(\lambda)$ ballots in $\PBB^{id}$ (not counting their twin re-encrypted ballot) at time $\kappa(\lambda)$. 

    Further, at clock tick $\kappa(\lambda)$, there is only one valid list of indices among the at most
    \begin{equation*}
    \begin{split}
     F_{\mathcal{D}^*_{1,q}}(\kappa(\lambda), id_j, t_{end}) &= \sum_{i=0}^{\min(\kappa(\lambda), q)} \binom{\kappa(\lambda)}{i} \leq\\
     &\leq(\kappa(\lambda) +1)^{\min(\kappa(\lambda), q)}\leq(\kappa(\lambda) +1)^q.
    \end{split}
\end{equation*}
 
\noindent computed ones by the adversary (the $\DR^v$-aware adversary). Given that at clock tick $\kappa(\lambda)$, there are only $\delta$ casting slots remaining, the adversary will cast at most $\delta$ guesses out of the less than $F_{\mathcal{D}^*_{1,q}}(\kappa(\lambda), id_j, t_{end})$ computed ones.
Since $\lambda^c\geq2^c\geq\delta$, and $F_{\mathcal{D}^*_{1,q}}(\kappa(\lambda), id_j, t_{end})\leq(\kappa(\lambda) +1)^q$,
    \begin{equation*}
    \begin{split}
        &\Pr[\GameCRI{Loki^{\DR^{vs},\DT^{vs}}}{\AbruteOpt{\mathcal{D}^*_{1,q}}}(1^\lambda, \mathbf{param})=1~|~r^{id_j}_v \le q] \\
        \ge & \min(\frac{\min(\lambda^c,\delta)}{F_{\mathcal{D}^*_{1,q}}(\kappa(\lambda), id_j, t_{end})}, 1) \\
        \ge & \min(\frac{\delta}{(\kappa(\lambda)+1)^q}, 1).
    \end{split}
    \end{equation*}
    \MAdone{$\min(1, \frac{\min(\delta, \lambda^c)}{(\kappa(\lambda)+1)^t})$}
    Finally, we can conclude that 
    \begin{equation*}
    \begin{split}
    &\Pr[\GameCRI{Loki^{\DR^{vs},\DT^{vs}}}{\AbruteOpt{\mathcal{D}^*_{1,q}}}(1^\lambda, \mathbf{param})=1] \\ 
    \ge & p\cdot\min( \frac{\delta}{(\kappa(\lambda)+1)^q}, 1) \\
    = & p\cdot\min(\delta\cdot \alpha(\lambda), 1).
    \end{split}
    \end{equation*}
\qed

\section{Estonia Case Study}
\label{apx: calculation details for estonia case}

In this appendix, a detailed, step-by-step computational breakdown of the case analysis involving Estonia, as discussed in Section~\ref{subsec: analysis with estonia}, is provided.

\subsection{Statistical Data on Estonian voters}
\label{apx: Estonia data}

We collected data on the distribution of voter voting times over the five recent Estonia election from Estonia State Electoral Office~\cite{estoniaElectionsInternetVoting} as show in Table~\ref{tb: estonia D_R}. 

\begin{table}[!t]
    \centering
    \resizebox{\linewidth}{!}{
    \begin{tabular}{|c|c|c|c|}
        \toprule 
        \multicolumn{1}{|c}{} & \multicolumn{3}{|c|}{\textbf{Voting times}} \\ \cline{2-4}
        & Abstention & 1 & $\geq 2$ \\ \midrule
        European Parliament elections 2024 & 62.36\% & 37.22\% & 0.43\% \\ \midrule
        Parliamentary elections 2023 & 36.47\% & 62.28\% & 1.25\% \\ \midrule
        Local elections 2021 & 45.21\% & 52.31\% & 2.38\% \\ \midrule
        European Parliament elections 2019 & 62.41\% & 37.3\% & 0.3\% \\ \midrule
        Parliamentary elections 2019 & 36.33\% & 62.94\% & 0.74\% \\
        \bottomrule
    \end{tabular}}
    \caption{Statistics of Estonian internet voting from the last five elections utilizing internet voting in Estonia. (Source: Estonia State Electoral Office)}
    \label{tb: estonia D_R}
\end{table}

\subsection{Calculation Details of $\DR^{LE21}$}
\label{apx: calculation of DR LE21}

By design, the distribution $\DR^{v}$ never maps to $0$. For the $\DR^v$ used in the 2021 Local Election, we condition on the data corresponding to individuals who actually voted. The details are outlined below.
\begin{equation*}
    \begin{split}
        \DR^{LE21}[1] &= \Pr[id \text{ voted once}~|~id \text{ voted}] \\
         &= \Pr[id \text{ voted once}\land id \text{ voted}] / \Pr[id \text{ voted}] \\
        &= \Pr[id \text{ voted once}] / \Pr[id \text{ voted}] \\
        &= 0.5231 / 0.5479 \\
        &= 0.9547 \\
        \DR^{LE21}[2+] &= 1 - \DR^{LE21}[1] \\
        &= 0.0453
    \end{split}
\end{equation*}

\subsection{Calculation Details of $\mathcal{D}^{LE21}_{|\PBB^{id}|}$}
\label{apx: calculation of PBB id length}

Leveraging the fact that, in Loki, the voting server employs the same $\DR$ as the voters (i.e. $\DR^v = \DR^{vs}$), we estimate the distribution of the length of each $\PBB^{id}$ under the hypothetical deployment of Loki in the 2021 Local Election.

The probability $\Pr[|\PBB^{id}| = 1]$ is lower bounded by the scenario in which the voter abstains and only the voting server casts a single ballot on their behalf.
\begin{align*}
    & \Pr[|\PBB^{id}| = 1] \\
    = & \Pr[|\PBB^{id}| = 1 \wedge id \text{ abstains}] + \Pr[|\PBB^{id}| = 1 \wedge id \text{ votes}] \\
    \geq & \Pr[|\PBB^{id}| = 1 \wedge id \text{ abstains}] \\
    = & \Pr[|\PBB^{id}| = 1~|~id \text{ abstains}] \cdot \Pr[id \text{ abstains}] \\
    = & \DR^{LE21}[1](id) \cdot \DO^{LE21}[\Phi](id) \\
    = & 0.9547 \cdot 0.4521 \\
    = & 0.4316
\end{align*}

Similarly, $\Pr[|\PBB^{id}| = 2]$ is lower bounded by the case where both the voter and the voting server each cast exactly one ballot.
\begin{align*}
    & \Pr[|\PBB^{id}| = 2] \\
    = & \Pr[|\PBB^{id}| = 2 \wedge id \text{ abstains}] + \Pr[|\PBB^{id}| = 2 \wedge id \text{ votes}] \\
    \geq & \Pr[|\PBB^{id}| = 2 \wedge id
    \text{ votes}] \\
    = & \Pr[|\PBB^{id}| = 2~|~id
    \text{ votes}] \times \Pr[id \text{ votes}] \\
    \geq & \DR^{LE21}[1](id) \times \DR^{LE21}[1](id) \times (1-\DO^{LE21}[\Phi](id)) \\
    = & 0.9547 \times 0.9547 \times 0.4521 \\
    = & 0.4120
\end{align*}

$\Pr[|\PBB^{id}| \geq 3]$ is simply the complement of the aforementioned probabilities.
\begin{equation*}
    \begin{split}
    & \Pr[|\PBB^{id}| \geq 3] \\
    = & 1 - (\Pr[|\PBB^{id}| =1] + \Pr[|\PBB^{id}| =2]) \\
    \leq & 1 - (0.4316 + 0.4120) \\
    = & 0.1564
    \end{split}
\end{equation*}

\subsection{Calculation Detail of Probability $\AbruteOpt{\mathcal{D}^*_{LE21}}$ wins}
\label{apx: calculation of prob. A brute wins in Estonia}

Given the high probability that a voter's cast ballot record contains no more than two ballots, the adversary focuses on ballot lengths that are considered to be "within control." For an adversary who submits coercive ballots three clock ticks prior to the end of the voting phase, we consider the following:
\begin{equation*}
    \begin{split}
        & \Pr[\GameCRI{Loki}{\AbruteOpt{\mathcal{D}^*_{LE21}}}(1^\lambda, \mathbf{param}) = 1] \\
        = & \Pr[\GameCRI{Loki^{\DR^{vs},\DT^{vs}}}{\AbruteOpt{\mathcal{D}^*_{LE21}}}(1^\lambda, \mathbf{param}) = 1 \wedge |\PBB^{id}| \leq 2] \\
        + & \Pr[\GameCRI{Loki^{\DR^{vs},\DT^{vs}}}{\AbruteOpt{\mathcal{D}^*_{LE21}}}(1^\lambda, \mathbf{param}) = 1 \wedge |\PBB^{id}| \geq 3] \\
        \geq & \Pr[\GameCRI{Loki^{\DR^{vs},\DT^{vs}}}{\AbruteOpt{\mathcal{D}^*_{LE21}}}(1^\lambda, \mathbf{param}) = 1 \wedge |\PBB^{id}| \leq 2] \\
        = & \Pr[\GameCRI{Loki^{\DR^{vs},\DT^{vs}}}{\AbruteOpt{\mathcal{D}^*_{LE21}}}(1^\lambda, \mathbf{param}) = 1~|~|\PBB^{id}| \leq 2] \\
        \times & \Pr[|\PBB^{id}| \leq 2] \\
        \geq & 0.9547 \times 0.8436 \\
        \geq & 0.8053
    \end{split}
\end{equation*}

For an adversary who submits coercive ballot two clock ticks prior to the end of the voting phase, he has two slots to try all his three coercive ballots, we have the probability 
\begin{equation*}
    \begin{split}
        & \Pr[\GameCRI{Loki^{\DR^{vs},\DT^{vs}}}{\AbruteOpt{\mathcal{D}^*_{LE21}}}(1^\lambda, \mathbf{param}) = 1] \\
        =& \frac{2}{3} \times 0.8053 = 0.5369
    \end{split}
\end{equation*}

For an adversary who submits a coercive ballot one clock ticks prior to the end of the voting phase, we have
\begin{equation*}
    \begin{split}
        & \Pr[\GameCRI{Loki^{\DR^{vs},\DT^{vs}}}{\AbruteOpt{\mathcal{D}^*_{LE21}}}(1^\lambda, \mathbf{param}) = 1] \\
        =& \frac{1}{3} \times 0.8053 = 0.2684
    \end{split}
\end{equation*}




\section{Forced Abstention Attack Against Loki}

We formally demonstrate how an attacker can enforce a forced abstention scenario, undermining voter participation in Loki.

\subsection{Ideal privacy - $\GameCRP{\Pideal}{\adv}$}
\label{apx:ideal privacy game}

Privacy is captured in comparison with the guarantees of an ideal protocol that only leaks the tally result. Specifically, similar to the approach in~[2], the CR-Privacy definition
presented above compares the adversary's
advantage in $\GameCRP{\Gamma}{\adv}$ with that of an adversary against
an idealized game, $\GameCRP{\Pideal}{\adv}$. In $\GameCRP{\Pideal}{\adv}$, illustrated in
Figure~\ref{apx:ideal privacy game} the adversary only learns the tally
result.

\begin{figure}[!t]
\hrule
\vspace*{3pt}
$\GameCRP{\Pideal}{\adv}$
\vspace*{3pt}
\noindent\hrule
\begin{algorithmic}[1]
\scriptsize
\Require{$1^{\lambda},\mathbf{param}(\mathbb{I}, \mathbb{O}, \DO, \DR^v, \DT^v, \DR^{vs}, \DT^{vs}, t_{end})$}
\State Initialize $\mathbb{B}$ as $[\ ]$
\State Send $\mathbf{param}$ to $\mathcal{A}$, and wait for $\Icorr$ and $\{(id, o_{id})_{id\in \Icorr}\}$ from $\adv$
\State \algorithmiccomment{corrupted voters and malicious options of them}
\State Receive $(id_j,o)$ and $(id_j, o_c)$ from $\mathcal{A}$ 
\algorithmiccomment{coerce $id_j$ whose intention is $o$ into $o_c$ \TZdone{I think that there is duplication w.r.t. line 3}\JQdone{$o_c$ is coerced option, $o$ is genuine option.}}
\If{$id_j\notin\mathbb{I}\backslash \Icorr \lor o \notin \mathbb{O} \lor o_c \notin \mathbb{O}$}
\State \Return 0
\EndIf
\State $b \leftarrow_{\$}\{0,1\}$
\For{$id\in \mathbb{I}\backslash \Icorr$}
\State $o_{id} \leftarrow \DO(id)$
\If{$id=id_j\land b=0\land o\neq\Phi$} \MAdone{$b \xleftarrow{\$}\{0,1\}$ ?}
\State $\mathbb{B} = \mathbb{B}\cup (id_j,o)$
\ElsIf{$id=id_j\land b=1 \land o_c\neq \Phi$}
\State $\mathbb{B} = \mathbb{B}\cup (id_j,o_c)$
\ElsIf{$id\neq id_j \land o_{id}\neq \Phi$}
\State $\mathbb{B} = \mathbb{B}\cup (id,o_{id})$ 
\EndIf
\EndFor
\State $\{\mathbb{B}\leftarrow\mathbb{B}\cup (id,o_{id})\}_{id\in \Icorr, o_{id}\in \mathbb{O}}$
\State $W\leftarrow \rho(\mathbb{B})$ \algorithmiccomment{$\rho$ is the result function of e-voting scheme\TZdone{Remind the reader what $\rho$ is}}
\State Send $W$ to $\mathcal{A}$ and receive $b'$
\If{$b'=b$}
\State\Return 1
\Else
\State\Return 0
\EndIf
\end{algorithmic}
\vspace*{3pt}\hrule\vspace*{3pt}
\caption{The CR-Privacy game of an ideal protocol $\Pideal$.}
\label{algo: game CR-P ideal}
\end{figure}

\subsection{Computation of $\delta_{min}$}
\label{apx: computation of delta min}
For the completeness, we append the computation methodology of $\delta_{min}$ in~\cite{kusters2012game}~\cite{kusters2011verifiability}, adapting with our notation.

The pure tally result is a tuple $W_p = (w_1,\dots,w_{n_o})$ such that $w_1+\cdots+w_{n_o}=n_h + 1$, where $w_i$, for $i\in\{1,\dots,n_o\}$, is the number of votes for the option $i$-th (including abstention). 
Denote $A^o_{W_p}$ the probability that honest voters plus the coerced voter yields the pure result $W_p$ given that $id_j$ votes for $o$. Say $o$ is the $i$-th option in $\mathbb{O}$. It is easy to see that
\begin{equation*}
    \begin{split}
        & A^o_{W_p} \\
        =& \binom{n_h}{w_0,\dots,w_{i-1},w_{i+1},\dots,w_{n_o}}\times \\
        &\Pr[\DO(id_j)=\Phi]^{w_0}\cdots\Pr[\DO(id_j)=\mathbb{O}(i-1)]^{w_{i-1}}\times \\
        &\Pr[\DO(id_j)=\mathbb{O}(i+1)]^{w_{i+1}}\cdots\Pr[\DO(id_j)=\mathbb{O}(n_o)]^{w_{n_o}} \\
        =& \frac{n_h!}{w_0!\cdots w_{n_o}!}
        \times \\
        &\Pr[\DO(id_j)=\Phi]^{w_0}\cdots
        \Pr[\DO(id_j)=\mathbb{O}(n_o)]^{w_{n_o}} \times\\
        & \frac{w_i}{\Pr[\DO(id_j)=\mathbb{O}(i)]^{w_{i}}}
    \end{split}
\end{equation*}
The best strategy of the coercer in the ideal protocol is to accept a run when $A^o_{W_p} \leq A^{o_c}_{W_p}$, where $o$ is the intentional option of the coerced voter and $o_c$ the coerced one. $\delta_{min}^{o,o_c}(n_h,n_o,\DO)$ is define as the sum of the difference between $A^o_{W_p}$ and $A^{o_c}_{W_p}$ in all runs that the attacker accepts. Formally,
\begin{equation*}
    \begin{split}
        \delta_{min}^{o,o_c}(n_h,n_o,\DO) = \sum\limits_{W_p\in M^*_{o,o_c}}(A^{o_c}_{W_p}-A^o_{W_p})
    \end{split}
\end{equation*}
where $M^*_{o,o_c}=\{W_p\in Res: A^{o}_{W_p}\leq A^{o_c}_{W_p} \}$. And
\begin{equation*}
    \delta_{min}(n_h,n_o,\DO) = \max\limits_{o\in\mathbb{O},o_c\in\mathbb{O}} \delta_{min}^{o,o_c}(n_h,n_o,\DO)
\end{equation*}

\subsection{Proof of Theorem~\ref{thm: forced abstention attack}}
\label{apx: proof of thm forced abstention attack}


To compute $\Pr[\GameCRP{Loki^{\DR^{vs},\DT^{vs}}}{\Afa}(1^{\lambda},\mathbf{param})=1]$ we consider three events: $|\PBB^{id_j}|>1$, $|\PBB^{id_j}|=1\land b=1$, and $|\PBB^{id_j}|=1\land b=1$.

\noindent\textit{Case 1:} When $|\PBB^{id_j}|=1\land b=1$ occurs, by construction $\Afa$ wins with probability 1:
\begin{align*}
    &\Pr[\GameCRP{Loki^{\DR^{vs},\DT^{vs}}}{\Afa}(1^{\lambda},\mathbf{param})=1~\big|~|\PBB^{id_j}|=1\land b=1] \\
    &\hspace{8cm} = 1.
\end{align*}
Furthermore, when $b=1$, having $|\PBB^{id_j}|=1$ means that the coerced voter didn't cast as per the coercer's instruction, and the only ballot in $\PBB^{id_j}$ comes from the voting server. When $b=1$, this is determined by the voting server's distribution $\DR^{vs}$:
\begin{equation*}
\begin{split}
    \Pr[|\PBB^{id_j}|=1 \land b=1] 
    =& \Pr[|\PBB^{id_j}|=1~\big|~b=1] \Pr[b=1] \\
    =& \frac{1}{2}\Pr[r_{vs}^{id_j}=1].
\end{split}
\end{equation*}
Thus the probability of $\Afa$ winning in case 1 is
\begin{equation*}
\begin{split}
&\Pr[\GameCRP{Loki^{\DR^{vs},\DT^{vs}}}{\Afa}(1^{\lambda},\mathbf{param})=1 \land |\PBB^{id_j}|=1\land b=1] \\
=& \frac{1}{2}\Pr[r_{vs}^{id_j}=1].
\end{split}
\end{equation*}

\noindent\textit{Case 2:} When $|\PBB^{id_j}|=1\land b=0$ occurs, by construction $\Afa$ always guesses the wrong bit, so it looses:
\begin{align*}
    &\Pr[\GameCRP{Loki^{\DR^{vs},\DT^{vs}}}{\Afa}(1^{\lambda},\mathbf{param})=1~\big|~|\PBB^{id_j}|=1\land b=0]\\
    &\hspace{8cm} = 0.
\end{align*}
And thus,
\begin{align*}
    & \Pr[\GameCRP{Loki^{\DR^{vs},\DT^{vs}}}{\Afa}(1^{\lambda},\mathbf{param})=1 \land |\PBB^{id_j}|=1\land b=0] \\
    &\hspace{8cm}= 0.
\end{align*}

\noindent \textit{Case 3:} When $|\PBB^{id_j}|>1$ occurs,
$\Afa$ randomly guesses the bit:
\begin{equation*}
\begin{split}
    &\Pr[\GameCRP{Loki^{\DR^{vs},\DT^{vs}}}{\Afa}(1^{\lambda},\mathbf{param})=1~\big|~|\PBB^{id_j}|>1] \\
    & = \frac{1}{2}.
\end{split}
\end{equation*}

In addition, the probability of $|\PBB^{id_j}|>1$ is equal to the sum of $\Pr[|\PBB^{id_j}|>1 \land b=1]$ and $\Pr[|\PBB^{id_j}|>1\land b=0]$.

When the coerced voter abstains to comply with the coercer's instruction ($b=1$), having $|\PBB^{id_j}|>1$ (not counting twin re-encryption ballots) means that the voting server must have cast more than just one noise ballots for the coerced voter $id_j$:
\begin{equation*}
\begin{split}
    \Pr[|\PBB^{id_j}|>1 \land b=1] 
    =& \Pr[|\PBB^{id_j}|>1 | b=1] \Pr[b=1] \\
    &= \frac{1}{2}\Pr[r_{vs}^{id_j}\geq2] \\
    &= \frac{1}{2}-\frac{1}{2}\Pr[r_{vs}^{id_j}=1].
\end{split}
\end{equation*}
And, when the coerced voter evades and casts a ballot ignoring the coercer's instructions ($b=0$), having $|\PBB^{id_j}|=1$ means that the voter's ballot and the voting server's ballot must have collision at some time $t$. Indeed, by definition the voting server casts at least one noise ballot for each voter. Thus, because of the independence \JQ{between $\DR^{vs}$ and $\DR^{v}$ as well as $\DT^{vs}$ and $\DT^{v}$}, and of the assumption that $\Pr[o_{id_j} \not= \Phi] < 1$
\JQ{
\begin{align*}
    &\Pr[|\PBB^{id_j}|=1|b=0] \\
    =& \Pr[r_{vs}^{id_j}=1 \land r_{v}^{id_j}=1 \land \vec{t}^{id_j}_{v} = \vec{t}^{id_j}_{vs}] \\
    =& \Pr[\vec{t}^{id_j}_{v} = \vec{t}^{id_j}_{vs} | r_{vs}^{id_j}=1 \land r_{v}^{id_j}=1]\Pr[r_{vs}^{id_j}=1 \land r_{v}^{id_j}=1] \\
    =& \sum_{t=0}^{t_{end}}\Pr[\vec{t}_v^{id_j}=[t] \land \vec{t}_{vs}^{id_j}=[t]]\Pr[r_{vs}^{id_j}=1 \land r_{v}^{id_j}=1] \\
    =& \sum_{t=0}^{t_{end}}\Pr[\vec{t}_v^{id_j}=[t]]\Pr[\vec{t}_{vs}^{id_j}=[t]]\Pr[r_{vs}^{id_j}=1]\Pr[r_{v}^{id_j}=1]\\
    \le& \Pr[r_{vs}^{id_j}=1]\Pr[r_{v}^{id_j}=1].\\
\end{align*}
}
Thus, we can conclude that 
\begin{equation*}
\begin{split}
    \Pr[|\PBB^{id_j}|>1\land b=0] =& \Pr[|\PBB^{id_j}|>1 | b=0] \Pr[b=0] \\
    =& \frac{1}{2}(1- \Pr[|\PBB^{id_j}|=1 | b=0]) \\
    >& \frac{1}{2}(1- \Pr[r_{vs}^{id_j}=1]\Pr[r_{v}^{id_j}=1]).
\end{split}
\end{equation*}
And that 
\begin{equation*}
\begin{split}
    &\Pr[|\PBB^{id_j}|>1] \\
    =& \Pr[|\PBB^{id_j}|>1 \land b=1] + \Pr[|\PBB^{id_j}|>1\land b=0] \\
    >& 1 - \frac{1}{2}\Pr[r^{id_j}_{vs}=1] - \frac{1}{2}\Pr[r^{id_j}_{vs}=1]\Pr[r^{id_j}_{v}=1]
\end{split}
\end{equation*}

By putting everything together we can now conclude the third case as follows:
\begin{equation*}
\begin{split}
    &\Pr[\GameCRP{Loki^{\DR^{vs},\DT^{vs}}}{\Afa}(1^{\lambda},\mathbf{param})=1 \land |\PBB^{id_j}|>1] \\
    =& \Pr[\GameCRP{Loki^{\DR^{vs},\DT^{vs}}}{\Afa}(1^{\lambda},\mathbf{param})=1~\big|~|\PBB^{id_j}|>1] \\
    &\hspace{5.5cm}\cdot\Pr[|\PBB^{id_j}|>1] \\
    >& \frac{1}{2}(1 - \frac{1}{2}\Pr[r^{id_j}_{vs}=1] - \frac{1}{2}\Pr[r^{id_j}_{vs}=1]\Pr[r^{id_j}_{v}=1]) \\
    =& \frac{1}{2} - \frac{1}{4}\Pr[r^{id_j}_{vs}=1] - \frac{1}{4}\Pr[r^{id_j}_{vs}=1]\Pr[r^{id_j}_{v}=1]. \\
\end{split}
\end{equation*}
Adding the probabilities of the above three mutually exclusive events, we obtain the probability of the forced abstention adversary of winning the CR-Privacy game:
\begin{equation}
\label{eq: Pr FA wins}
\begin{split}    
    & \Pr[\GameCRP{Loki^{\DR^{vs},\DT^{vs}}}{\Afa}(1^{\lambda},\mathbf{param})=1] \\
    >& \frac{1}{2} + \frac{1}{4}\Pr[r_{vs}^{id_j}=1]\big(1-\Pr[r_{v}^{id_j}=1]).
\end{split}
\end{equation}

Finally, given our assumptions on the distributions $\DR^{vs}$ and $\DR^v$, we have that $\Pr[r_{vs}^{id_j}=1] = p > 0$ and 
$\Pr[r_{v}^{id_j}=1]  < 1-\frac{4}{p}(\delta_{min}(|\mathbb{I}| - 1,|\mathbb{O}|,\DO)-\frac{1}{2})$. Replacing these in Equation~\ref{eq: Pr FA wins} implies that the success probability of $\Afa$ is strictly greater than $\delta_{min}(|\mathbb{I}| - 1,|\mathbb{O}|,\DO)$. And thanks to ~Theorem \ref{thm: delta min}, we can conclude that for all PPT adversaries $\adv'$
\begin{equation*}
    \begin{split}
    &\Pr[\GameCRP{Loki^{\DR^{vs},\DT^{vs}}}{\Afa}(1^{\lambda},\mathbf{param})=1] \\
        >& \delta_{min}(|\mathbb{I}| - 1,|\mathbb{O}|,\DO) \\
        \geq& \Pr[\GameCRP{Ideal}{\adv'}(1^{\lambda},(\mathbb{I}, \mathbb{O}, \DO))=1]
    \end{split}
\end{equation*}
Observe that none of the terms of Equation~\ref{eq: Pr FA wins} are a function of the security parameter $\lambda$, therefore $\Delta$ is greater than a positive constant.\qed

\section{Proofs for The Impossibility Result}
\label{apx: impossibility result proof}

We provide a rigorous impossibility proof showing that protocols fall into the category of (B1), (B2), (B3), and (B4) cannot achieve full coercion-resistance.

\subsection{Proof of Theorem~\ref{thm: incompatible assumptions}}
\label{apx: proofs of thm no-go result}


For integrity to hold, the ballot of voter casting at time $t_0$ should be accepted by the Voting Server. In other words, by condition (B3):
\begin{equation*}
\label{eq: accept init signal}
\begin{split}
    \Pr\left[
    \begin{array}{l}
  (\{\Sigma_{id}^0\}_{id\in\mathbb{I}}, \Sigma_{vs}^0) \leftarrow \mathsf{InitSgn}(\mathbb{I},1^\lambda,\PBB) : \\
     \mathsf{VrfSgn}(id,\Sigma_{vs}^0, \PBB^0, \mathsf{Sgn}(id,\Sigma_{id}^{0},\PBB^0))\rightarrow\top
    \end{array}
    \right] = 1
\end{split}
\end{equation*}

We consider two cases for the signalling algorithms covering all scenarios:

\noindent\textit{C1 -} for some non-negligible function $\alpha_{C1}(\cdot)$:
\begin{equation*} 
    \begin{split} 
    & \Pr \left[ 
    \begin{array}{l} 
    (\{\Sigma_{id}^0\}_{id\in\mathbb{I}}, \Sigma_{vs}^0) \leftarrow \mathsf{InitSgn}(\mathbb{I},1^\lambda,\PBB) :\\ l_{id}^0\leftarrow\mathsf{Sgn}(id, \Sigma_{id}^{0},\PBB^0); \\ \Sigma_{vs}^1\leftarrow\tau_{vs}(id,\Sigma_{vs}^0,\PBB^0, l_{id}^0): \\ 
    \mathsf{VrfSgn}(id, \Sigma_{vs}^1, \PBB^1, \mathsf{Sgn}(\Sigma^{0}_{id},\PBB^0))\rightarrow\top 
    \end{array} \right] \\ 
    &\hspace{7cm} \geq \alpha_{C1} (\lambda) 
    \end{split} 
\end{equation*}

\noindent\textit{C2 -} for some non-negligible function $\alpha_{C2}(\cdot)$:
\begin{equation*} 
    \begin{split} 
    & \Pr \left[ 
    \begin{array}{l} 
    (\{\Sigma_{id}^0\}_{id\in\mathbb{I}}, \Sigma_{vs}^0) \leftarrow \mathsf{InitSgn}(\mathbb{I},1^\lambda,\PBB) ;\\ l_{id}^0\leftarrow\mathsf{Sgn}(id, \Sigma_{id}^{0},\PBB^0); \\ \Sigma_{vs}^1\leftarrow\tau_{vs}(id,\Sigma_{vs}^0,\PBB^0, l_{id}^0): \\ 
    \mathsf{VrfSgn}(id, \Sigma_{vs}^1, \PBB^1, \mathsf{Sgn}(\Sigma^{0}_{id},\PBB^0))\rightarrow\bot 
    \end{array} \right] \\ 
    &\hspace{7cm} \geq \alpha_{C2} (\lambda) 
    \end{split} 
\end{equation*}

Under the assumption (A2), the private part of initial states both for voters and the Voting Server are $\emptyset$. This means that, only with the public part of the voter's initial state, the adversary is able to reproduce the complete initial state by setting the private part as $\emptyset$.


\textbf{Case C1.} We construct adversary $\mathcal{A}_{C1}$ (Figure \ref{algo: A C1}) that allows $t_0$ as the coerced voter's only moment of privacy. At all clock cycles $t \geq t_1$ the adversary casts a ballot on behalf of the coerced voter.
\begin{align*}
    &\Pr[flag=\top]=\Pr[r_v^{id_j}\leftarrow\DR^{v}(id_j);t_0\in\DT^v(r_v^{id_j}, id_j)] \\
    &\hspace{7cm} \geq \alpha(\lambda)
\end{align*}
Besides, given that ballot cast by $\adv_{C1}$ at $t_1$ are generated by $\mathsf{Vote}(\cdot)$, and according to condition (B4), we have the following for some negligible function $\mu(\cdot)$:
\begin{equation*}
    \Pr[\mathsf{VerifyTally}(\PBB^{t_{end}}, W,\Pi,pk_T)]\geq 1- \mu(\lambda)
\end{equation*}
Since the protocol adopts the last-ballot revote policy (condition (B2)), the probability that the coerced option $o_c$ is included in the tally result $W$ instead of the true intention $o$ of the coerced voter $id_j$, solely depends on the probability of the ballot cast by the coercer at time $t_1$ being accepted by the Voting Server. 
As explained, $\adv_{C1}$ could reconstruct $\Sigma_{id}^0$ because of assumption (A2), which in case C1 will be accepted in the current Voting Server state against the current $\PBB^{t}$ (by condition (B5)) with probability $\alpha_{1}(\lambda)$.
That is:
\begin{equation*}
    \Pr[(id_j, o)\in\mathcal{O} \land W = \rho(\mathcal{O}[(id_j, o_c)/(id_j, o)])] \geq \alpha_{C1}(\lambda).
\end{equation*}

Putting everything together we have that:
\begin{equation*}
    \begin{split}
    &\Pr[\GameCRI{Loki}{\adv_{C1}}(1^{\lambda},\mathbf{param}) = 1] \\
    \geq& \Pr[\neg\gamma_{ee}(\mathbb{I},\emptyset,\rho,\mathcal{O},W) \land \mathsf{VerifyTally}(\PBB^{t_{end}}, W,\Pi,pk_T) \land \\
    & flag=\top \land (id_j, o)\in\mathcal{O} \land W = \rho(\mathcal{O}[(id_j, o_c)/(id_j, o)])] \\
    =& 1\cdot \Pr[flag=\top \land \mathsf{VerifyTally}(\PBB^{t_{end}}, W,\Pi,pk_T) \land \\
    &\hspace{2cm}(id_j, o)\in\mathcal{O} \land W = \rho(\mathcal{O}[(id_j, o_c)/(id_j, o)])] \\
    \geq& \alpha(\lambda)\cdot\Pr[\mathsf{VerifyTally}(\PBB^{t_{end}}, W,\Pi,pk_T) \land (id_j, o)\in\mathcal{O} \land \\
    &\hspace{4cm}W = \rho(\mathcal{O}[(id_j, o_c)/(id_j, o)])] \\
    \geq& \alpha(\lambda)\cdot(1-\mu(\lambda))\cdot \\
    &\hspace{1.5cm} \Pr[(id_j, o)\in\mathcal{O} \land W = \rho(\mathcal{O}[(id_j, o_c)/(id_j, o)])] \\
    \geq& \alpha(\lambda)\cdot(1-\mu(\lambda))\cdot(1-\alpha_{C1}(\lambda)) \\
    \geq& \alpha_{1}(\lambda)
    \end{split}
\end{equation*}
where $\alpha_{1}(\cdot)$ is a non-negligible function.

\begin{figure}[!t]
\hrule
\vspace*{3pt}
$\mathcal{A}_{C1}$
\vspace*{3pt}
\noindent\hrule
\begin{algorithmic}[1]
\scriptsize
\Require{$\lambda$}
\UponReceiving{$(pk_T,pk_{VS},\{upk_{id}, L_{id}\}_{id\in\mathbb{I}}, \PBB, \mathbf{param})$}
\State return $[\ ]$ as $\Icorr$
\EndUponReceiving

\UponReceiving{$(\{usk_{id},l_{id}\}_{id\in\Icorr})$}
\State $(id_j,o)\leftarrow_{\$}(\mathbb{I}, \mathbb{O}\backslash\Phi)$
\algorithmiccomment{select coerced voter $id_j$ and their intention $o$}
\State send back $(id_j,o)$
\EndUponReceiving

\UponReceiving{$usk_{id_j}$}
\State $o_c \leftarrow_{\$}\mathbb{O}\backslash o$
\algorithmiccomment{choose a malicious option that differs $o$}
\EndUponReceiving

\While{receiving $(\PBB,clk)$}
\If{$clk=t_0$}
\State do nothing
\Else
\State $\tilde{s}_{id}\leftarrow_{\$}\{0,1\}^{\lambda}$
\State $\Sigma_{id,pb}^0\leftarrow\PBB$
\algorithmiccomment{extract public state from the board}
\State $\Sigma_{id}^0 :=(\emptyset, \Sigma_{id,pb}^0)$
\State $\beta_{\mathcal{A}}\leftarrow\mathsf{Vote}(id_j,usk_{id_j},pk_T,pk_{VS},o_c,\mathsf{Sgn}(\Sigma_{id}^0,\PBB; \tilde{s}_{id}))$
\State Send $(id_j,\beta_{\mathcal{A}})$
\EndIf
\EndWhile

\end{algorithmic}
\vspace*{3pt}\hrule\vspace*{3pt}
\caption{Attack algorithm of $\mathcal{A}_{C1}$.}
\label{algo: A C1}
\end{figure}

\textbf{Case C2.}  We construct adversary $\mathcal{A}_{C2}$ (Figure \ref{algo: A C2}) that allows $t_1$ as the coerced voter's only moment of privacy. At all clock cycles $t \not= t_1$ the adversary casts a ballot on behalf of the coerced voter. The proof for case C2 is similar to that of case C1.
\begin{align*}
    &\Pr[flag=\top]=\Pr[r_v^{id_j}\leftarrow\DR^{v}(id_j);t_1\in\DT^v(r_v^{id_j},id_j)] \\
    &\hspace{7cm} \geq \alpha'(\lambda)
\end{align*}
For the same reason as $\adv_{C1}$, by condition (B4), we have the following for some negligible function $\mu(\cdot)$:
\begin{equation*}
    \Pr[\mathsf{VerifyTally}(\PBB^{t_{end}}, W,\Pi,pk_T)]\geq 1- \mu(\lambda)
\end{equation*}
And the probability that the coerced option $o_c$ is included in the tally result $W$ instead of the true intention $o$ of the coerced voter $id_j$, solely depends on the probability of the ballot cast by the coerced voter at time $t_1$ being rejected by the Voting Server. 
For the same reason, $\adv_{C2}$ can reconstruct $\Sigma_{id}^0$ if it casts first, which allows the experiment proceeds to where the voter cast at $t_1$. In case C2, it will be rejected in the current Voting Server state against the current $\PBB^t$ (by condition (B5)) with probability $\alpha_{2}(\lambda)$. That is:
\begin{equation*}
    \Pr[(id_j, o)\in\mathcal{O} \land W = \rho(\mathcal{O}[(id_j, o_c)/(id_j, o)])] \geq \alpha_{C2}(\lambda). 
\end{equation*}

Putting everything together we have that:
\begin{equation*}
    \begin{split}
    &\Pr[\GameCRI{Loki}{\adv_{C2}}(1^{\lambda},\mathbf{param}) = 1] \\
    \geq& \Pr[\neg\gamma_{ee}(\mathbb{I},\emptyset,\rho,\mathcal{O},W) \land \mathsf{VerifyTally}(\PBB^{t_{end}}, W,\Pi,pk_T) \land \\
    &flag=\top \land (id_j, o)\in\mathcal{O} \land W = \rho(\mathcal{O}[(id_j, o_c)/(id_j, o)])] \\
    =& 1\cdot \Pr[flag=\top \land \mathsf{VerifyTally}(\PBB^{t_{end}}, W,\Pi,pk_T) \land \\
    &\hspace{2cm}(id_j, o)\in\mathcal{O} \land W = \rho(\mathcal{O}[(id_j, o_c)/(id_j, o)])] \\
    \geq& \alpha'(\lambda)\cdot\Pr[\mathsf{VerifyTally}(\PBB^{t_{end}}, W,\Pi,pk_T) \land (id_j, o)\in\mathcal{O} \land \\
    &\hspace{4cm}W = \rho(\mathcal{O}[(id_j, o_c)/(id_j, o)])] \\
    \geq& \alpha'(\lambda)\cdot(1-\mu(\lambda))\cdot \\
    &\hspace{1cm} \Pr[(id_j, o)\in\mathcal{O} \land W = \rho(\mathcal{O}[(id_j, o_c)/(id_j, o)])] \\
    \geq& \alpha'(\lambda)\cdot(1-\mu(\lambda))\cdot\alpha_{C2}(\lambda) \\
    \geq& \alpha_{2}(\lambda)
    \end{split}
\end{equation*}
where $\alpha_{2}(\cdot)$ is a non-negligible function.\qed

\begin{figure}[!t]
\hrule
\vspace*{3pt}
$\mathcal{A}_{C2}$
\vspace*{3pt}
\noindent\hrule
\begin{algorithmic}[1]
\scriptsize
\Require{$\lambda$}
\UponReceiving{$(pk_T,pk_{VS},\{upk_{id}, L_{id}\}_{id\in\mathbb{I}}, \PBB, \mathbf{param})$}
\State return $[\ ]$ as $\Icorr$
\EndUponReceiving

\UponReceiving{$(\{usk_{id},l_{id}\}_{id\in\Icorr})$}
\State $(id_j,o)\leftarrow_{\$}(\mathbb{I}, \mathbb{O}\backslash\Phi)$
\algorithmiccomment{select coerced voter $id_j$ and their intention $o$}
\State send back $(id_j,o)$
\EndUponReceiving

\UponReceiving{$usk_{id_j}$}
\State $o_c \leftarrow_{\$}\mathbb{O}\backslash o$
\algorithmiccomment{choose a malicious option that differs $o$}
\EndUponReceiving

\While{receiving $(\PBB,clk)$}
\If{$clk=t_0$}
\State $\tilde{s}_{id}\leftarrow_{\$}\{0,1\}^{\lambda}$
\State $\Sigma_{id,pb}^0\leftarrow\PBB$
\algorithmiccomment{extract public state from the board}
\State $\Sigma_{id}^0 :=(\emptyset, \Sigma_{id,pb}^0)$
\State $\beta_{\mathcal{A}}\leftarrow\mathsf{Vote}(id_j,usk_{id_j},pk_T,pk_{VS},o_c,\mathsf{Sgn}(\Sigma_{id}^{0},\PBB; \tilde{s}_{id}))$
\State Send $(id_j,\beta_{\mathcal{A}})$
\ElsIf{$clk=t_1$}
\State do nothing
\Else
\State $\tilde{s}_{id}\leftarrow_{\$}\{0,1\}^{\lambda}$
\State $\Sigma_{id,pb}^0\leftarrow\PBB$
\algorithmiccomment{extract public state from the board}
\State $\Sigma_{id}^0 :=(\emptyset, \Sigma_{id,pb}^0)$
\State $\beta_{\mathcal{A}}\leftarrow\mathsf{Vote}(id_j,usk_{id_j},pk_T,pk_{VS},o_c,\mathsf{Sgn}(\Sigma_{id}^{0},\PBB; \tilde{s}_{id}))$
\State Send $(id_j,\beta_{\mathcal{A}})$
\EndIf
\EndWhile
\end{algorithmic}
\vspace*{3pt}\hrule\vspace*{3pt}
\caption{Attack algorithm of $\mathcal{A}_{C2}$.}
\label{algo: A C2}
\end{figure}

In reality, an attacker doesn't even need to specifically target the coercion-free window ($t_0$ or $t_1$). If the signalling algorithms are functioning correctly, the attacker can simply cast a vote after the coerced voter, effectively overriding their choice. Alternatively, if the attacker votes before the coerced voter's privacy window, the voting server will reject the voter's subsequent attempts.

\section{Formal Specification of CR-Loki}
\label{apx: formal description of cr-loki}

This appendix includes a complete specification of the CR-Loki protocol and the underlying cryptographic primitives used.

\subsection{Cryptographic Primitives}
\label{apx: crypto primitives of cr-loki}

\subsubsection{Encryption Scheme}

The encryption scheme of the talliers and the Voting Server are different. 

For the talliers, it is required to be IND-RCCA secure, which can be achieved by ElGamal encryption~\cite{cramer1998secure} along with NIZKPs. The message space is $m\in\mathbb{M}$ and the randomness space is $r\leftarrow_{\$} \mathbb{Z}$, consisting of the following algorithms:
\begin{description}
    \item[$\mathsf{EKG}_{RCCA}(1^\lambda)$ -] On input of security parameter $1^\lambda$, it outputs a pair of decryption and encryption keys $(sk,pk)$.
    \item[$\mathsf{Enc}_{RCCA}(pk,m;r)$ -] Given a public key $pk$, a message $m\in\mathbb{M}$, and some randomness $r\leftarrow_{\$} \mathbb{Z}$, it outputs a ciphertext $ct$.
    \item[$\mathsf{Dec}_{RCCA}(sk,ct)$ -] Given a secret key $sk$, it outputs the message $m$ of $ct$.
    \item[$\mathsf{ReEnc}_{RCCA}(pk,ct;r)$ -] Given a public key $pk$, a ciphertext $ct$, using randomness $r\leftarrow_{\$}\mathbb{Z}$, outputs the re-encryption of $ct$. 
    For every plaintext $m$ and every honestly generated ciphertext $ct\leftarrow_{\$}\mathsf{Enc}_{RCCA}(pk,m;r)$, the distribution of $\mathsf{ReEnc}_{RCCA}(pk,ct;r)$ is identical to that of $\mathsf{Enc}_{RCCA}(pk,m;r)$. For every purported ciphertext $ct$ and every $ct'\leftarrow_{\$}\mathsf{ReEnc}_{RCCA}(pk,ct;r)$, we must have that $\mathsf{Dec}_{RCCA}(sk,ct)=\mathsf{Dec}_{RCCA}(sk,ct')$.
\end{description}

For the Voting Server, it is a IND-CCA secure encryption scheme, message space $m\in\mathbb{M'}$ and randomness space $r\leftarrow_{\$} \mathbb{Z'}$, consisting of the following algorithms:
\begin{description}
    \item[$\mathsf{EKG}_{CCA}(1^\lambda)$ -] On input of security parameter $1^\lambda$, it outputs a pair of decryption and encryption keys $(sk,pk)$.
    \item[$\mathsf{Enc}_{CCA}(pk,m;r)$ -] Given a public key $pk$, a message $m\in\mathbb{M'}$, and some randomness $r\leftarrow_{\$} \mathbb{Z'}$, outputs a ciphertext $ct$.
    \item[$\mathsf{Dec}_{CCA}(sk,ct)$ -] Given a secret key $sk$, it outputs the message $m$ of $ct$.
\end{description}

\subsubsection{Verifying Key Pair}
$\mathsf{SKG}(1^\lambda)$, which, on input of security parameter $1^\lambda$, outputs a pair of private and public keys $(usk,upk)$.

\subsubsection{Zero-knowledge Proofs}
The disjunctive NIZKP~\cite{groth2006perfect} is retained for verifiability without revealing which kind of relation is trying to proved. 
The disjunctive proof is defined as
$$
(x=(x_1,\cdots,x_n),\omega)\in R\ \mathrm{iff}\ (x_1,\omega)\in R_1\lor\cdots (x_n,\omega)\in R_n
$$

The proving algorithm $\mathsf{DisjProof}(x,w)$ is: by inputting public statement $x$ and private witness $\omega$ of a disjunctive relation $R=R_1\lor\cdots\lor R_n$, outputs a NIZKP $\pi$ that proves the knowledge of $w_j$ w.r.t $x_j$ of relation $R_j$ and simulates the proof of knowledge of other relations. 

The verifying algorithm $\mathsf{Verify}(x,\pi)$, which on input public statement $x$ and NIZKP $\pi$ checks the proof $\pi$ w.r.t $x$ of relation $R$ and outputs $\top$ if all checks succeeds, otherwise return $\bot$.



\subsection{Algorithms Defining CR-Loki}
\label{apx: algorithms definiting cr-loki}

In this section, we present formal descriptions (Figure~\ref{algo: cr-loki algorithms}) of algorithms $(\mathsf{Setup}$, $\mathsf{Register}$, $\mathsf{Vote}$, $\mathsf{Noise}$, $\mathsf{Validate}$, $\mathsf{Include}$, $\mathsf{Publish}$, $\mathsf{Fake}$, $\mathsf{Tally}$, $\mathsf{VerifyTally}$ and $\mathsf{VerifyVote})$ that define CR-Loki.

\lstset{
    basicstyle=\footnotesize,
    numbers=left,
    numberstyle=\tiny\color{gray},
    stepnumber=1,
    numbersep=5pt,
    numberblanklines = true,
    breaklines=true,
    breakatwhitespace=false,
    frame=single,
    rulecolor=\color{lightgray},
    mathescape=true,
    escapeinside={/*}{*/}
}

\begin{figure*}

\begin{lstlisting}
$\mathsf{Setup}(\lambda,\mathbf{param})\rightarrow((pk_T,sk_T),(pk_{VS},sk_{VS}))$: On input of the security parameter $\lambda$, and public parameters $\mathbf{param}$ ($\mathbb{I}, \mathbb{O}, \DO, \DR^v, \DT^v, \DR^{vs}, \DT^{vs}, t_{end}$), compute $(pk_T,sk_T)\leftarrow_{\$}\mathsf{EKG}_{RCCA}(1^\lambda)$ and $(pk_{VS},sk_{VS})\leftarrow_{\$}\mathsf{EKG}_{CCA}(1^\lambda)$. 
$\mathsf{Register}(id) \rightarrow ((usk_{id},upk_{id}),CR_{id},cr_{id})$: On input voter identity $id\in \mathbb{I}$, do the following.
    1)Generate $(usk_{id},upk_{id}) \leftarrow_{\$} \mathsf{SKG}(1^{\lambda})$
    2)Compute $cr_{id}\leftarrow_{\$}\{0,1\}^\lambda$, and set $CR_{id}\leftarrow_{\$}\mathsf{Enc}_{CCA}(pk_{VS},cr_{id};r)$.
    3)Compute $ct_{o}^{0}\leftarrow_{\$} \mathrm{Enc}_{RCCA}(pk_T,0;r_o)$, $ct_{cr}^{0}\leftarrow_{\$} \mathrm{Enc}_{CCA}(pk_{VS},0;r_{cr})$ and set $\beta^0_{id} = (id,usk_{id},(ct^{0}_{o},ct^{0}_{cr},r_{o},r_{cr}))$/*\label{line: embeded rand}*/
    4)Return $((usk_{id},upk_{id}),CR_{id},cr_{id})$ and publish $CR_{id}$ and $\beta^0_{id}$ on the $\PBB^{id}$
$\mathsf{Vote}(id, usk_{id},pk_T,pk_{VS},o_{id},cr_{id})\rightarrow\beta$: On input a candidate $o_{id}\in\mathbb{O}$, a credential $cr_{id}$, keys $(usk_{id},pk_T,pk_{VS})$, and implict input $\PBB^{id}$ of length $i-1$ do the following.
    1)Compute $ct_{o_{id}}\leftarrow_{\$} \mathrm{Enc}_{RCCA}(pk_T,o_{id};r_o)$, $ct_{cr_{id}}\leftarrow_{\$} \mathrm{Enc}_{CCA}(pk_{VS},cr_{id};r_{cr})$
    2)Run $\pi \leftarrow_{\$} \mathrm{DisjProof}(x,w)$, where $x=(pk_T,pk_{VS},\mathbb{O},(ct_{o_{id}},ct_{cr_{id}}), upk_{id})$ and $\omega\leftarrow (usk_{id},(r_o,o_{id}),(r_{cr},cr_{id}))$ s.t. $(x,w)\in R_i^{id}\ \mathrm{iff}$
            $ct_{o_{id}}=\mathrm{Enc}_{RCCA}(pk_T,o_{id};r_o) \land o_{id}\in \mathbb{O} \land ct_{cr_{id}} = \mathrm{Enc}_{CCA}(pk_{VS},cr_{id};r_{cr}) \land cr_{id} \in \{0,1\}^\lambda \land$ /*\label{line: R id-1}*/
            $(usk_{id},upk_{id}) =\mathsf{SKG}(1^{\lambda})$ /*\label{line: R id-2}*/
    where $R_i = R^{id}_i \lor R^{pred}_i \lor R^{pred2}_i$, $i \geq 1$
    3)Set $\bar{ct} = (ct_{o_{id}},ct_{cr_{id}},\pi)$. Return the ballot $\beta = (id,upk_{id},\bar{ct})$.
$\mathsf{Noise}(id,sk_{VS}, pk_{T}, \mathbb{PBB}^{id})\rightarrow \beta$: On input the public bulletin board $\PBB^{id}$ of length $i-1$, obtains $\beta^{-1}=(id,upk_{id},\bar{ct}^{-1})$ (the last ballot on $\mathbb{BB}^{id}$). Then, it does the following.
    1)Compute $ct_{o_{id}}\leftarrow_{\$} \mathsf{ReEnc}_{RCCA}(pk_T,ct^{-1}_{o_{id}};r_o)$, $ct_{cr_{id}}\leftarrow_{\$} \mathsf{Enc}_{CCA}(pk_{VS},cr_{id};r_{cr})$, where $cr_{id} = \mathsf{Dec}_{CCA}(sk_{VS},CR_{id})$.
    2)Runs $\pi \leftarrow_{\$} \mathrm{DisjProof}(x,w)$, where $x=(pk_T,pk_{VS},\mathbb{O},CR_{id},(ct_{o_{id}}^{-1},ct_{cr_{id}}^{-1}),(ct_{o_{id}},ct_{cr_{id}}))$ and $\omega\leftarrow (sk_{VS},r_o,r_{cr})$ s.t. $(x,w)\in R_i^{pred}\ \mathrm{iff}$
            $ct_{o_{id}}=\mathrm{ReEnc}_{RCCA}(pk_T,ct_{o_{id}}^{-1};r_o) \land ct_{cr_{id}}=\mathsf{Enc}_{CCA}(pk_{VS},cr_{id};r_{cr}) \land cr_{id} = \mathsf{Dec}_{CCA}(sk_{VS},CR_{id})\land$
            $\mathsf{Dec}_{CCA}(sk_{VS},ct_{cr_{id}}^{-1})= \mathsf{Dec}_{CCA}(sk_{VS}, CR_{id})\land (pk_{VS}, sk_{VS}) = \mathsf{EKG}_{CCA}(1^{\lambda})$
    where $R_i = R^{id}_i \lor R^{pred}_i \lor R^{pred2}_i$, $i \geq 1$
    3)Set $\bar{ct} = (ct_{o_{id}},ct_{cr_{id}},\pi)$. Return the ballot $\beta = (id,upk_{id},\bar{ct})$.
$\mathsf{GenTwin}(id,sk_{VS}, pk_{T}, \mathbb{BB}^{id})\rightarrow \beta$: On input ballot box $\mathbb{BB}^{id}$ of length $i-1$, obtains $CR_{id}$, $\beta^{-1}=(id,upk_{id},\bar{ct}^{-1})$ and $\beta^{-2}=(id,upk_{id},\bar{ct}^{-2})$ (the last and second last ballot on $\mathbb{BB}^{id}$), where $\bar{ct}^{-1}=(ct_{o_{id}}^{-1},ct_{cr_{id}}^{-1},\pi^{-1})$ and $\bar{ct}^{-2}=(ct_{o_{id}}^{-2},ct_{cr_{id}}^{-2},\pi^{-2})$. Then, do the following.
    1)If $\mathsf{Dec}_{CCA}(sk_{VS},ct_{cr_{id}}^{-1})=\mathsf{Dec}_{CCA}(sk_{VS}, CR_{id})$, compute $ct_{o_{id}}\leftarrow_{\$} \mathsf{ReEnc}_{RCCA}(pk_T,ct^{-1}_{o_{id}};r_o)$, $ct_{cr_{id}}\leftarrow_{\$} \mathsf{Enc}_{CCA}(pk_{VS},cr_{id};r_{cr})$, where $cr_{id} = \mathsf{Dec}_{CCA}(sk_{VS},CR_{id})$ and run $\pi \leftarrow_{\$} \mathrm{DisjProof}(x,w)$, where $x=(pk_T,pk_{VS},\mathbb{O},CR_{id},(ct_{o_{id}}^{-1},ct_{cr_{id}}^{-1}),(ct_{o_{id}},ct_{cr_{id}}))$ and $\omega\leftarrow (sk_{VS},r_o,r_{cr})$ s.t. $(x,w)\in R_i^{pred}\ \mathrm{iff}$
            $ct_{o_{id}}=\mathrm{ReEnc}_{RCCA}(pk_T,ct_{o_{id}}^{-1};r_o) \land ct_{cr_{id}}=\mathsf{Enc}_{CCA}(pk_{VS},cr_{id};r_{cr}) \land cr_{id} = \mathsf{Dec}_{CCA}(sk_{VS},CR_{id})\land$ /*\label{line: R pred-1}*/
            $\mathsf{Dec}_{CCA}(sk_{VS},ct_{cr_{id}}^{-1})= \mathsf{Dec}_{CCA}(sk_{VS}, CR_{id})\land (pk_{VS}, sk_{VS}) = \mathsf{EKG}_{CCA}(1^{\lambda})$ /*\label{line: R pred-2}*/
    2)Else compute $ct_{o_{id}}\leftarrow_{\$} \mathsf{ReEnc}_{RCCA}(pk_T,ct^{-2}_{o_{id}};r_o)$, $ct_{cr_{id}}\leftarrow_{\$} \mathsf{Enc}_{CCA}(pk_{VS},cr_{id};r_{cr})$, where $cr_{id} = \mathsf{Dec}_{CCA}(sk_{VS},CR_{id})$ and run $\pi \leftarrow_{\$} \mathrm{DisjProof}(x,w)$, where $x=(pk_T,pk_{VS},\mathbb{O},CR_{id},(ct^{-2}_{o_{id}},ct_{cr_{id}}^{-1}),(ct_{o_{id}},ct_{cr_{id}}))$ and $\omega\leftarrow (sk_{VS},r_o,r_{cr})$ such that $(x,w)\in R_i^{pred2}\ \mathrm{iff}$
            $ct_{o_{id}}=\mathrm{ReEnc}_{RCCA}(pk_T,ct^{-2}_{o_{id}};r_o) \land 
            ct_{cr_{id}}=\mathsf{Enc}_{CCA}(pk_{VS},cr_{id};r_{cr}) \land 
            cr_{id} = \mathsf{Dec}_{CCA}(sk_{VS},CR_{id})\land$ /*\label{line: R pred2-1}*/
            $\mathsf{Dec}_{CCA}(sk_{VS},ct^{-1}_{cr_{id}}) \neq \mathsf{Dec}_{CCA}(sk_{VS}, CR_{id})\land (pk_{VS}, sk_{VS}) = \mathsf{EKG}_{CCA}(1^{\lambda})$ /*\label{line: R pred2-2}*/
    where $R_i = R^{id}_i \lor R^{pred}_i \lor R^{pred2}_i$, $i \geq 1$
    3)Set $\bar{ct} = (ct_{o_{id}},ct_{cr_{id}},\pi)$. Return the ballot $\beta = (id,upk_{id},\bar{ct})$.
$\mathsf{Validate}(id,\PBB^{id},\beta)\rightarrow\top\backslash\bot$: On input the public bulletion board $\PBB^{id}$ and a ballot $\beta$, it checks that $(id\in \mathbb{I}) \land (\beta \text{ does not already appended to } \PBB^{id}) \land (\top\leftarrow\mathsf{Verify}(x,\pi))$. If any of three checks fail, it returns $\bot$. Otherwise $\top$.
$\mathsf{Include}(id, sk_{VS}, \mathbb{BB}^{id}, \beta) \rightarrow \mathbb{BB}^{id}$: On input the ballot box $\mathbb{BB}^{id}$ and a ballot $\beta$, it does the following:
    1)It checks that $\mathsf{Validate}(id,\PBB^{id},\beta)=\top$ and updates $\mathbb{BB}^{id}\leftarrow \mathbb{BB}^{id} \cup \beta$. /*\label{line: validate true}*/
    2) It generates a twin ballot $\beta' \leftarrow \mathsf{GenTwin}(id,sk_{VS}, pk_{T}, \mathbb{BB}^{id})$ and updates $\mathbb{BB}^{id}\leftarrow \mathbb{BB}^{id} \cup \beta'$ to append the twin ballot as well.
$\mathsf{Publish}(\mathbb{BB}^{id})\rightarrow\PBB^{id}$: On input $\mathbb{BB}^{id}$, it synchronizes $\PBB^{id}\leftarrow\mathbb{BB}^{id}$.
$\mathsf{Fake}(id, \PBB, CR_{id}, cr_{id})\rightarrow\tilde{cr}_{id}$: On input voter identity $id$, the public bulletin board $\PBB$, the public part of the credential $CR_{id}$, and the real private credential $cr_{id}$, and implicit input of security parameter $\lambda$, it does the following:
    1)If it is the first time to run for voter $id$, it generates a fake credential $\tilde{cr}_{id}\leftarrow_{\$}\{0,1\}^{\lambda}\backslash cr_{id}$, record it as a tuple of $(id, \tilde{cr}_{id})$, and returns $\tilde{cr}_{id}$.
    2)Else, it searches the recorded tuple with the first element equals to $id$ and returns the second element of the tuple as $\tilde{cr}_{id}$.
$\mathsf{Tally}(\{\PBB^{id}\}_{id\in\mathbb{I}},sk_T)\rightarrow(W,\Pi)$: On input $\PBB^{id}$ of all eligible voters, and the decryption key $sk_T$, let $\{ct_x\}_{x=1}^{n_v}$ be the encrypted option ($ct_{o_{id}}$) of the last ballots from each sub-chain ($\PBB^{id}$), and let $ct_x=\{ct_j\}_{j=1}^{n_o}$ where $ct_j$ denotes the encrypted option for the selection $o_j\in\mathbb{O}$.
    1)Compute $T_j =\prod_{x=1}^{n_o} ct_x^j$. The tally $t_j$ for selection option $o_{j}$ is produced by decrypting $T_j$ with $sk_T$.
    2)Compute the result $W=(w_1,\dots,w_{n_o})$ and $\Pi$, a Fiat-Shamir proof of correct decryption. Return $(W,\Pi)$
$\mathsf{VerifyTally}(\PBB, W, \Pi)\rightarrow\top\backslash\bot$: On input $\PBB$, the tally result $W$ and proof $\Pi$, verifies the correctness of $(W,\Pi)$ and returns the check outcome.
$\mathsf{VerifyVote}(id,usk_{id},upk_{id},\PBB,\beta)\rightarrow\top\backslash\bot$: On input voter identity $id$, signing key pair $usk_{id},upk_{id}$, $\PBB$ and a ballot $\beta$, it checks $usk_{id}$ corresponds to the one in ballot and that $\mathsf{Validate}(id,\PBB^{id},\beta) = \top$. If any checks fail, it returns $\bot$ and $\top$ otherwise. 
\end{lstlisting}

\caption{The algorithms defining CR-Loki.}
\label{algo: cr-loki algorithms}
\end{figure*}

\section{Security Proofs For CR-Loki}

\subsection{Coercion-resistant Proofs for CR-Loki}
\label{apx: coercion resistant proofs for cr-loki}

In this section, we present the full proof of coercion-resistance of CR-Loki. The proof is conducted for a centralized tallier (as often done in the literature), but it can be generalised to distributed tallying with distributed key generation and threshold decryption.

\subsubsection{Simulation of Disjunctive NIZKPs}
\label{apx: simulation disjunctive proof}

Similar to the original Loki protocol, the Non-Interactive Zero-Knowledge Proofs (NIZKPs) within ballots are disjunctive, demonstrating that one of the relations, $R^{id}$, $R^{pred}$, or $R^{pred2}$, holds. These disjunctive proofs are constructed by generating a genuine NIZKP $\pi$ for one of the relations and simulating the other two. Whenever a proof for any one relation can be constructed, the corresponding disjunctive proof can also be generated. 
For simplicity, we only explicitly describe how the proof for the ``actually-satisfied'' relation is built. It's important to note that the relation being proven is indistinguishable from the proof itself.

\subsubsection{Proof of CR-Integrity}
\label{sec: proof of CR-Integrity}

Let $\mathbf{param}$ be any election parameters with the Voting Server's distributions as specified in Section~\ref{sec: Loki overview}. We prove this by by reduction to the IND-CCA security of the Voting Server's encryption scheme. Assume that there exists an adversary $\adv$ that wins $\GameCRI{CR-Loki}{\adv}$ with advantage $\alpha_\adv(\lambda)$ for some non-negligible function $\alpha_\adv(\cdot)$. We are then able to  construct an adversary $\mathcal{B}$ which wins the IND-CCA game with advantage $\frac{1}{2}+\alpha_\mathcal{B}(\lambda)$ for some non-negligible function $\alpha_\mathcal{B}(\cdot)$.

$\mathcal{B}$ invokes $\adv$ with the security parameter $\lambda$ and public parameters $\mathbf{param}$. $\mathcal{B}$ gets $pk$ from $\mathcal{G}ame_{IND-CCA}$. 
$\mathcal{B}$ generates two credential $cr^0$ and $cr^1$, and sends $(cr^0,cr^1)$ to $\mathcal{G}ame_{IND-CCA}$ as the challenge, to which it receives back $CR^b = \mathsf{Enc}_{CCA}(pk, cr^b; r)$.
$\mathcal{B}$ runs the setup phase in $\GameCRI{CR-Loki}{\adv}$ (lines 1 - 4) but replacing the public key of the voting server with $pk$ ($pk_{VS}\leftarrow pk$) and abandoning $sk_{VS}$. 
$\mathcal{B}$ runs the registration phase in $\GameCRI{CR-Loki}{\adv}$ (line 5) but setting the coerced voter's $id_j$ public credential to $CR_{id_j} \leftarrow CR^b$. \MAdone{In CR-Integrity game, and CR-Privacy game, registration needs to happen after coercion.}
Next, $\mathcal{B}$ sends $(pk_T,pk_{VS},\{upk_{id},CR_{id}\}_{id\in\mathbb{I}}, \{\PBB^{id}\}_{id\in\mathbb{I}}$, $\mathbf{param})$ to $\adv$ and receives the set of corrupted voters $\Icorr$. It samples the true intention from $\DO$, the number of casts from $\DR^v$, and the time of cast from $\DT^v$ for all non-corrupted voters. It returns to $\adv$ $(usk_{id}, cr_{id})$ for $id\in\Icorr$ and receives $(id_j,o)$ as the coerced voter and their true intention, which overrides the option sampled from $\DO$. 

Recall that a ballot $\beta$ is a tuple $(ct_{o_{id}}, ct_{cr_{id}}, \pi)$. For ease of notation, we use negative superscripts to denote a ballot's position from the end of the bulletin board $\PBB$. For example, $\beta^{-1} = (ct_{o_{id}}^{-1}, ct_{cr_{id}}^{-1}, \pi^{-1})$ denotes the most recent (\emph{i.e.}, last) ballot in $\PBB^{id}$. \MAdone{Why -1 and not 0?}\JQdone{By default, 0 is the newly created ballot}
For any query to the decryption or encryption oracle of the IND-CCA game, $\mathcal{B}$ simulates the corresponding proof of correct encryption or decryption. To avoid redundancy, we do not explicitly mention these simulated proofs in the subsequent description of $\mathcal{B}$.

At each clock cycle $clk$, $\mathcal{B}$ performs the following operations:
\begin{itemize}
    \item For every tuple $(id,\beta_{\adv})$ in $\mathbb{L}$, it verifies that $id$ is either a corrupted or coerced voter, ensures the validity of the accompanying proofs, and confirms that $ct_{cr_{id}}$ is not a replay. Subsequently, it appends $\beta_{\adv}$ to $\PBB^{id}$ and queries the decryption oracle of $\mathcal{G}ame_{IND-CCA}$ for $ct_{cr_{id}}$ to obtain the plaintext $\tilde{cr}_{id}$. A corresponding twin ballot is then constructed and appended to $\PBB^{id}$. Specifically, if $id$ is a corrupted voter, it checks whether $\tilde{cr}_{id}$ matches $cr_{id}$, indicating that the ballot was formed using their valid credential. If $id$ is a coerced voter, it checks if $\tilde{cr}_{id} \in \{cr^0, cr^1\}$. If yes, it re-encrypts $ct_{o_{id}}^{-1}$ (\emph{i.e.}, the ciphertext from $\beta_{\adv}$); otherwise, it re-encrypts $ct_{o_{id}}^{-2}$. Both re-encryptions are accompanied by the appropriate simulated proofs.
    
    \item For ballots cast by honest voters or submitted on their behalf by the voting server, it faithfully replicates the behaviour of $\GameCRI{CR-Loki}{\cdot}$ at lines 45 and 48, respectively.
    
    \item For the moment-of-privacy of the coerced voter or for noise ballots added by the voting server for $id_j$, whether the ballot is cast directly by the voter or on their behalf by the voting server, $\mathcal{B}$ queries the encryption oracle of $\G_{IND-CCA}$ for a fresh encryption of $cr^b$ in both cases. The distinction lies in the origin of $ct_o$: if the coerced voter $id_j$ casts the ballot, $\mathcal{B}$ computes the encryption of the adversary-specified genuine vote option $o$ as $ct_o$ (along with proofs); if the ballot is a noise one generated by the voting server, it re-encrypts $ct_o^{-1}$ (along with the corresponding proofs). The resulting ballot is appended to $\PBB^{id_j}$, thereby concluding the simulation of the original ballot. The behaviour for the twin ballot is identical in both scenarios: $\mathcal{B}$ again queries the encryption oracle for a fresh encryption $ct_{cr^b}$ of $cr^b$, simulates a proof of equal decryption between $ct_{cr^b}$ and $CR_{id_j}$, simulates a proof of correct decryption for $CR_{id_j}$, and re-encrypts $ct_o^{-1}$ with accompanying proofs. The twin ballot is then appended to $\PBB^{id_j}$.
\end{itemize}

For $\mathbb{L}$ received at the last moment, it does the same as above.
Finally, at time $t_{end}$, it takes the last ballot of every voter $id$ on $\PBB^{id}$, verifies the proofs and decrypts $ct_{o_{id}}$. $\mathcal{B}$ computes $W = \rho(\{id,o_{id}\}_{id\in\mathbb{I}\backslash\Icorr})$ and sends $W$ to $\adv$ with the corresponding proof of correct tallying.

At this point, if there exists a response $\tilde{cr}_{id_j}$ to a decryption oracle query to $\mathcal{G}ame_{IND-CCA}$ equal to $cr^0$, then $\mathcal{B}$ outputs $0$. If there exists a response $\tilde{cr}_{id_j}$ to a decryption oracle query to $\mathcal{G}ame_{IND-CCA}$ equal to $cr^1$, then $\mathcal{B}$ outputs $1$. Otherwise $\mathcal{B}$ outputs a random guess.

A complete and formal specification of $\mathcal{B}$ is provided in Figure~\ref{algo: cr-loki algorithms}.

We denote $E^b_{atk}$, where $b\in\{0,1\}$, the event where a response $\tilde{cr}_{id_j}$ to a decryption oracle query to $\mathcal{G}ame_{IND-CCA}$ returned $cr^b$; and we let event $E_{atk} \triangleq E^0_{atk} \lor E^1_{atk}$. We are now going to compute $\mathcal{B}$'s winning advantage against $\mathcal{G}ame_{IND-CCA}$.

First by definition of $\mathcal{B}$ we have that
\begin{equation}\label{eq: not Eatk}
    \begin{split}
    &\Pr[\mathcal{G}ame_{IND-CCA}^\mathcal{B}(1^\lambda) = 1 \land \neg E_{atk}] \\
    =&  \Pr[\mathcal{G}ame_{IND-CCA}^\mathcal{B}(1^\lambda) = 1~|~\neg E_{atk}]\cdot \Pr[\neg E_{atk}] \\
    =& \frac{1}{2}\Pr[\neg E_{atk}],
    \end{split}
\end{equation}
as well as
\begin{equation}\label{eq: Eatk}
    \begin{split}
        &\Pr[\mathcal{G}ame_{IND-CCA}^\mathcal{B}(1^\lambda) = 1 \land E_{atk}] \\
        =& \Pr[\mathcal{G}ame_{IND-CCA}^\mathcal{B}(1^\lambda) = 1 \land E^0_{atk} \land b=0] \\
        +& \Pr[\mathcal{G}ame_{IND-CCA}^\mathcal{B}(1^\lambda) = 1 \land E^1_{atk} \land b=0] \\
        +& \Pr[\mathcal{G}ame_{IND-CCA}^\mathcal{B}(1^\lambda) = 1 \land E^0_{atk} \land b=1] \\
        +& \Pr[\mathcal{G}ame_{IND-CCA}^\mathcal{B}(1^\lambda) = 1 \land E^1_{atk} \land b=1] \\
        =& \Pr[\mathcal{G}ame_{IND-CCA}^\mathcal{B}(1^\lambda) = 1~|~E^0_{atk} \land b=0]\cdot\\&\cdot\Pr[E^0_{atk} \land b=0] \\
        +& \Pr[\mathcal{G}ame_{IND-CCA}^\mathcal{B}(1^\lambda) = 1~|~E^1_{atk} \land b=0]\cdot\\&\cdot\Pr[E^1_{atk} \land b=0] \\
        +& \Pr[\mathcal{G}ame_{IND-CCA}^\mathcal{B}(1^\lambda) = 1~|~E^0_{atk} \land b=1]\cdot\\&\cdot\Pr[E^0_{atk} \land b=1] \\
        +& \Pr[\mathcal{G}ame_{IND-CCA}^\mathcal{B}(1^\lambda) = 1~|~E^1_{atk} \land b=1]\cdot\\&\cdot\Pr[E^1_{atk} \land b=1] \\
        =& \Pr[E^0_{atk} \land b=0] + \Pr[E^1_{atk} \land b=1] \\
        =& \Pr[b=0~|~E^0_{atk}]\cdot\Pr[E^0_{atk}] + \Pr[b=1~|~E^1_{atk}]\cdot\Pr[E^1_{atk}] \\
        =& (1-\frac{q(\lambda)}{2^\lambda})\cdot \Pr[E_{atk}]
    \end{split}
\end{equation}
for some polynomial function $q(\cdot)$ that captures the maximum number of ballots (and thus guesses of $id_j$'s credential) $\adv$ might have cast over the election period. Combining Equations~\ref{eq: not Eatk} and~\ref{eq: Eatk} we obtain
\begin{equation*}
    \begin{split}
    &\Pr[\mathcal{G}ame_{IND-CCA}^\mathcal{B}(1^\lambda) = 1] \\
    =& \frac{1}{2} + (\frac{1}{2} - \frac{q(\lambda)}{2^\lambda})\cdot\Pr[E_{atk}]
    \end{split}
\end{equation*}
Now, the probability of event $E_{atk}$ occurring is greater than the advantage of $\adv$ derives from the Voting Server chaining of ballots on each cast ballot record, and predicates $R^{pred}$ and $R^{pred2}$ \MAdone{The following would deserve a more detailed proof as this relies on the specific protocol but no time just now}:
\begin{equation*}
    \begin{split}
        &\Pr[E_{atk}] \\
        \geq& \Pr[E_{atk} \land \GameCRI{CR-Loki}{\adv}(1^\lambda,\mathbf{param})=1] \\
        =& \Pr[E_{atk}~|~\GameCRI{CR-Loki}{\adv}(1^\lambda,\mathbf{param})=1]\cdot \\
        &\Pr[\GameCRI{CR-Loki}{\adv}(1^\lambda,\mathbf{param})=1] \\
        =& \Pr[\GameCRI{CR-Loki}{\adv}(1^\lambda,\mathbf{param})=1] \\
        =& \alpha_\adv(\lambda)
    \end{split}
\end{equation*}
Finally, combining all the above we conclude that
\begin{equation*}
    \begin{split}
    \Pr[\mathcal{G}ame_{IND-CCA}^\mathcal{B}(1^\lambda) = 1]\geq& \frac{1}{2} + (\frac{1}{2} - \frac{q(\lambda)}{2^\lambda})\alpha_\adv(\lambda) \\
    =&\frac{1}{2} + \alpha_\mathcal{B}(\lambda)
    \end{split}
\end{equation*}
for non-negligible function $\alpha_\mathcal{B}(\lambda) = (\frac{1}{2} - \frac{q(\lambda)}{2^\lambda})\cdot\alpha_\adv(\lambda)$. This contradicts the assumption of IND-CCA security of the Voting Server's encryption scheme, and concludes the proof of CR-Integrity.\qed

\begin{figure*}
\hrule
\vspace*{3pt}
\textbf{$\mathcal{B}$ in IND-CCA game for CR-Integrity Proof}
\vspace*{3pt}
\noindent\hrule
\begin{algorithmic}[1]
\scriptsize
\State receive $pk$ from IND-CCA
\State $\{(pk_T,sk_T),(pk_{VS},sk_{VS})\}\leftarrow\mathsf{Setup}(1^\lambda,\mathbf{param})$
\State replace $pk_{VS}$ with $pk$ and drops $sk_{VS}$
\State initialize $\PBB$ and $\mathcal{O}$ as $[\quad]$
\State receive $\Icorr$ and $id_j$ from $\adv$
\State $\{(upk_{id},usk_{id}),CR_{id},cr_{id}\}\leftarrow\mathsf{Register}(id)_{id\in\mathbb{I}}$
\State $cr^0,cr^1\leftarrow_{\$}\{0,1\}^{\lambda}$
\State query encryption oracle and receive a ciphertext $CR^b$ to replace $CR_{id_j}$
\State send $(pk_T,pk,\{upk_{id},CR_{id}\}_{id\in\mathbb{I}},\{\PBB^{id}\}_{id\in\mathbb{I}},\mathbf{param})$ and $(\{usk_{id}, cr_{id}\}_{id\in\Icorr})$ to $\adv$
\State create the true intention set $\mathcal{O}$, number of voting $r_v^{id}$, and voting time vector $\vec{t}_{v}^{id}$ for non-corrupted voters by sampling from distributions.
\State receive $(id_j,o)$ (overrides the previous sampling if validity check passes) from $\adv$ and send $usk_{id_j}$ back.

\While{$clk<t_{end}$}
\State receive $\mathbb{L}$ from $\adv$

\For{$(id,\beta_{\adv}) \in \mathbb{L}$ such that $id\in\Icorr\cup\{id_j\}$}
\label{line: start for for beta adv}
\State check the validity of proofs
\State check $ct_{cr_{id}}$ of $\beta_{\adv}$ is not a replay
\State append $\beta_{\adv}$ to $\PBB^{id}$
\State query decryption oracle of $ct^{-1}_{cr_{id}}$ to IND-CCA and receive $\tilde{cr}_{id}$.

\If{$id\in\Icorr$}
\State $ct_{cr_{id}} \leftarrow_{\$} \mathsf{Enc}(pk, cr_{id};r_{cr})$ and compute $\pi_{\mathsf{Enc}(cr_{id})}$
\If{$\tilde{cr}_{id} = cr_{id}$}
\State $ct_{o_{id}} \leftarrow_{\$} \mathsf{ReEnc} (pk_{T}, ct_{o_{id}}^{-1};r_{o})$, compute $\pi_{\mathsf{ReEnc}(ct_{o_{id}}^{-1})}$, and simulate proofs $\mathsf{sim}_{\mathsf{Dec}(ct^{-1}_{cr_{id}})= \mathsf{Dec}(CR_{id})}$ and $\mathsf{sim}_{\mathsf{Dec}(CR_{id})=cr_{id}}$.
\Else
\State $ct_{o_{id}} \leftarrow_{\$} \mathsf{ReEnc}(pk_T, ct_{o_{id}}^{-2};r_o)$, compute $\pi_{\mathsf{ReEnc}(ct_{o_{id}}^{-2})}$, and simulate proofs $\mathsf{sim}_{\mathsf{Dec}(ct^{-1}_{cr_{id}})\neq \mathsf{Dec}(CR_{id})}$ and $\mathsf{sim}_{\mathsf{Dec}(CR_{id})=cr_{id}}$.
\EndIf
\Else\algorithmiccomment{malicious ballot for the coerced voter $id_k$}
\State query encryption oracle to receive $ct_{cr_{b}}$ from IND-CCA and simulate the proof $\mathsf{sim}_{\mathsf{Enc}(cr_{b})}$
\State $ct_{cr_{id_j}} \leftarrow ct_{cr_{b}}$
\If{$\tilde{cr}_{id_j} \in \{cr^{0}, cr^{1}\}$}
\algorithmiccomment{$\tilde{cr}_{id_j}$ is received at line 18}
\State \textbf{output $b'\leftarrow b$ to IND-CCA game}
\algorithmiccomment{$E_{atk}$ occurs}
\State $ct_{o_{id_j}} \leftarrow_{\$} \mathsf{ReEnc}(pk_T, ct_{o_{id_j}}^{-1};r_{o})$ and compute $\pi_{\mathsf{ReEnc}(ct_{o_{id_j}}^{-1})}$
\Else
\State $ct_{o_{id_j}} \leftarrow_{\$} \mathsf{ReEnc}(pk_T, ct_{o_{id_j}}^{-2};r_{o})$ and compute $\pi_{\mathsf{ReEnc}(ct_{o_{id_j}}^{-2})}$
\EndIf
\EndIf
\State form the twin ballot $\beta'_{\adv}$ and append it to the corresponding $\PBB$
\EndFor
\label{line: end for for beta adv}

\For{honest voters cast or the voting server casts noise for honest voters at $clk$}
\State compute as in $\GameCRI{CR-Loki}{\cdot}$
\EndFor

\For{the coerced voter casts or the voting server casts noise for $id_j$ at $clk$}
\State query encryption oracle to receive $ct_{cr_{b}}$ from IND-CCA and simulate the proof $\mathsf{sim}_{\mathsf{Enc}(cr_{b})}$
\State $ct_{cr_{id_j}} \leftarrow ct_{cr_{b}}$
\If{the $id_j$ casts}
\State $ct_{o_{id_j}} \leftarrow_{\$} \mathsf{Enc}_{RCCA}(pk_{T}, o;r_o)$, compute $\pi_{\mathsf{Enc}(o)}$ and $\pi_{\mathsf{KG}(usk_{id_j})}$
\Else
\algorithmiccomment{the voting server noise for $id_j$}
\State $ct_{o_{id_j}} \leftarrow_{\$}\mathsf{ReEnc}(pk_T, ct_{o_{id_j}}^{-1};r_o)$, compute $\pi_{\mathsf{ReEnc}(ct_{o_{id_j}}^{-1})}$ and simulate proofs $\mathsf{sim}_{\mathsf{Dec}(ct^{-1}_{cr_{id_j}})= \mathsf{Dec}(CR_{id_j})}$ and $\mathsf{sim}_{\mathsf{Dec}(CR_{id_j})=cr_{id_j}}$.
\EndIf
\State form the original ballot $\beta_{id_j}$ and append it to $\PBB^{id_j}$
\State query encryption oracle to receive $ct'_{cr_{b}}$ from IND-CCA again and simulate the proof $\mathsf{sim}_{\mathsf{Enc}(cr_{id_j})}$.
\State $ct'_{cr_{id_j}}\leftarrow ct'_{cr_{b}}$
\State simulate proofs $\mathsf{sim}_{\mathsf{Dec}(ct^{-1}_{cr_{id_j}})= \mathsf{Dec}(CR_{id_j})}$ and $\mathsf{sim}_{\mathsf{Dec}(CR_{id_j})=cr_{id_j}}$.
\State $ct'_{o_{id_j}} \leftarrow_{\$}\mathsf{ReEnc}(pk_{T}, ct_{o_{id_j}}^{-1};r'_o)$ and compute $\pi_{\mathsf{ReEnc}(ct_{o_{id_j}}^{-1})}$
\State form the twin ballot $\beta'_{id_j}$ and append it to $\PBB^{id_j}$
\EndFor

\EndWhile

\State for $\mathbb{L}$ received at $t_{end}$, repeat line \ref{line: start for for beta adv} - \ref{line: end for for beta adv}.
\If{the coerced voter never casts}
\State it sends $0$ to $\adv$ and output $b'\leftarrow_{\$}\{0,1\}$ to IND-CCA game
\EndIf

\For{last ballot on $\PBB$ of $id\in\mathbb{I}\backslash\Icorr$}
\algorithmiccomment{re-encryption ballots}
\State $o_{id}\leftarrow\mathsf{Dec}_{RCCA}(sk_{T}, ct_{o_{id}})$
\EndFor
\State $W\leftarrow\rho(\{id,o_{id}\}_{id\in\mathbb{I}\backslash\Icorr})$
\State send $W$ to $\adv$

\If{it has not output $b'$ to IND-CCA}
\State output $b'\leftarrow_{\$}\{0,1\}$
\EndIf

\end{algorithmic}
\vspace*{3pt}\hrule\vspace*{3pt}
\caption{$\mathcal{B}$ plays in IND-CCA game and simulates $\GameCRI{CR-Loki}{\adv}$ for $\adv$}
\label{algo: B in CCA game for CR-Loki CR-Integrity}
\end{figure*}

\subsubsection{Proof of CR-Privacy}
\label{sec: proof of CR-Privacy}
To prove that our protocol satisfies CR-Privacy, we create a sequence of ten games labelled as Game 0-10 such that
\begin{enumerate}
    \item Game 0 against PPT adversary $\adv$ matches $\GameCRP{CR-Loki}{\adv}$;
    \item Game 10 against PPT adversary $\adv'$ matches $\GameCRP{\Pideal}{\adv'}$. 
\end{enumerate}

Next, we prove that for any PPT adversary $\mathcal{A}_i$ against Game $i$, we can construct a PPT adversary $\mathcal{A}_{i+1}$ against Game $i+1$ such that the absolute difference between the probabilities that $\mathcal{A}_{i+1}$ wins Game $i+1$ and $\mathcal{A}_i$ wins Game $i$ is negligible. Namely, let $S_i$ denote the probability that $\mathcal{A}_i$ wins Game $i$. We show that for $i=0,\ldots,9$, there exists a negligible function $\mu_i(\cdot)$ such that
\[|S_{i+1}-S_i|\leq\mu_i(\lambda).\]

\textbf{Game 1:} In this game, the proof of correct tallying is not published. Due to the simulatability property of NIZKPs, the adversary $\adv_1$ can generate a proof that verifies correctly for $\adv_0$. \TZdone{I added more detail on this as follows.} In more detail, assume a hybrid game labelled as Game $Sim$, where the  NIZKPs of final tallying are replaced by simulated proofs. By the zero-knowledge property of the non-interactive proofs, we get that the following is negligible:
\[\Big|\Pr[\mathcal{G}ame_{Sim}^{CR-Loki,\adv_0}(1^{\lambda},\mathbf{param}) = 1]-S_0\Big|.\]
Now, we construct an adversary $\adv_1$ against Game $1$ that operates as $\adv_0$ except that it generates the simulated proofs locally itself. Clearly, it holds that 
\[S_1=\Pr[\mathcal{G}ame_{Sim}^{CR-Loki,\adv_0}(1^{\lambda},\mathbf{param}) = 1].\]
By the above, we have that $|S_1-S_0|$ is negligible.\smallskip

\textbf{Game 2:} In this game, we internally initialize a set $B$ to record the intentions received from the eligible voters. For each ballot in the list $\mathbb{L}$ submitted by $\mathcal{A}_2$ at each clock cycle $clk$, the ballot is decrypted to obtain $\tilde{cr}_{id}$ and the corresponding intention $\tilde{o}$. If $\tilde{cr}_{id} = cr_{id}$ (i.e., the malicious ballot carries the correct credential), we update $B = B \cup \{(id, \tilde{o})\}$, allowing overwrites. After $B$ is updated using the final list $\mathbb{L}$ at time $t_{end}$, we append the intention set $\mathcal{O}$  to $B$ corresponding to the honest voters (excluding the coerced voter $id_j$). If $b = 0$ (indicating the coerced voter chose to evade), we overwrite the entry for $id_j$ in $B$ with $(id_j, o)$. We observe that $B$ corresponds to the set $\mathbb{B}$ in $\GameCRP{Ideal}{\cdot}$.

Notably, Game 2 only involves internal modifications to the execution and does not alter any interactions with the adversary. Therefore, we can construct an adversary $\mathcal{A}_2$ that achieves a perfect simulation of $\mathcal{A}_1$ by simply forwarding all communications. The latter implies that $S_2=S_1$.\smallskip

\textbf{Game 3:} In this game, instead of computing the tally result $W$ from $\PBB$, we apply the result function $\rho$ on $B$. We argue that $|S_3-S_2|$ is negligible by building an adversary $\mathcal{A}_3$ that behaves as $\mathcal{A}_2$.

When $b=0$ (the coerced voter evaded), CR-Integrity guarantees that the probability of the tally result of honest voters plus the coerced voter (denote $W_{ncor}$) deviates from their true intentions ($\rho(\{o_{id}\}_{id\in\mathbb{I}\backslash \Icorr})$) is smaller or equal than $\mu(\lambda)$ for some negligible function $\mu(\cdot)$.
Thanks to the separation of individual $\PBB^{id}$, $W = W_{ncor} \mathbin\Vert W_{cor}$, where $W_{cor} = \rho(\{\tilde{o}_{id}\}_{id\in \Icorr})$. Therefore, $\Pr[W\neq \rho(B)]\leq \mu(\lambda)$. \TZdone{What is $\epsilon_{PA0}$?\JQdone{It is typo of CCA} I do not see it anywhere in CR-Integrity proof.}
When $b=1$, $W = \rho(\{\tilde{o}_{id}\}_{id\in\Icorr\cup \{id_j\}}\cup\{o_{id}\}_{id\in\mathbb{I}\backslash\{\Icorr\cup\{id_j\}\}})$, which is exactly $\rho(B)$. Hence we can upper bound $S_3$ as follows, 
\begin{equation*}
\begin{split}
S_3=&\Pr[\mathcal{G}ame_{\mbox{Game 3}}^{\adv_3}(1^{\lambda},\mathbf{param}) = 1]\\
=&\Pr[W=\rho(B)]\cdot\\
&\cdot\Pr[\mathcal{G}ame_{\mbox{Game 3}}^{\adv_3}(1^{\lambda},\mathbf{param}) = 1|W=\rho(B)]+\\
&+\Pr[W\neq\rho(B)]\cdot\\
&\cdot\Pr[\mathcal{G}ame_{\mbox{Game 3}}^{\adv_3}(1^{\lambda},\mathbf{param}) = 1|W\neq\rho(B)]\leq\\
\leq&1\cdot S_2+\mu(\lambda)\cdot1\\
=&S_2+\mu(\lambda)
\end{split}
\end{equation*}
In addition, we can lower bound $S_3$ as follows,
\begin{equation*}
\begin{split}
S_3\geq&\\
\geq&\Pr[W=\rho(B)]\cdot\\
&\cdot\Pr[\mathcal{G}ame_{\mbox{Game 3}}^{CR-Loki,\adv_3}(1^{\lambda},\mathbf{param}) = 1|W=\rho(B)]\geq\\
\geq&(1-\mu(\lambda))\cdot S_2\geq\\
\geq&S_2-\mu(\lambda)
\end{split}
\end{equation*}
The above two inequalities imply that 
$$
|S_3-S_2|\leq \mu(\lambda)
$$

{
\textbf{Game 4:} In this game, the adversary does not submit encrypted ballots.

The adversary $\adv_4$ internally runs the NIZKP simulator $\mathcal{S}_{NIZKP}$. For each malicious encrypted ballot $\tilde{\beta}=(id,upk_{id},\bar{ct})$ submitted by $\adv_3$, $\adv_4$ runs $\mathsf{Validate}(id,\PBB^{id},\tilde{\beta})$. For ballots pass the validation, it parses $\bar{ct}$ as $(ct_{o_{id}},ct_{cr_{id}},\pi)$. It then sends $(upk_{id}, ct_{o_{id}}, ct_{cr_{id}}, \pi)$ to $\mathcal{S}_{NIZKP}$, which returns a witness $((r_{\tilde{o}}, \tilde{o}_{id}), (r_{\tilde{cr}}, \tilde{cr}_{id}))$ for the proof $\pi$. Finally, $\adv_4$ submits the corresponding plaintext ballot $(id, \tilde{o}_{id} \mathbin\Vert \tilde{cr}_{id})$ to Game 4 and appends the validated malicious ballot to the bulletin board in order to simulate the view of Game 3 for $\adv_3$.

By the extractability of the UC secure NIZKPs, the view of $\adv_3$ in this simulation is identical to its view in Game 3. Therefore, $\adv_4$ provides a perfect simulation of Game 3 for $\adv_3$.}


\textbf{Game 5:} In this game, $ct_o$ (option ciphertext) of all ballots on $\PBB$, except for the adversarial ballots $\beta_{\adv}$ received from the adversary, are replaced with dummy option encryption. This includes ballots from honest voters, twin ballots and noise ballots from the Voting Server, and the genuine ballots of the coerced voter (if present), and we call the set of those ballots the \emph{replace set} (\emph{i.e.} $\PBB\backslash\{\beta_{\adv}\}$). Now, each ballot in the replace set takes the form:
\[
\mathsf{Enc}_{RCCA}(pk_T, 0^{|o|}; r_{o}) \mathbin\Vert ct_{cr_{id}} \mathbin\Vert \pi^*,
\]

We argue that $\mathcal{A}_5$, which behaves identically to $\mathcal{A}_4$, achieves a negligible distinguishing advantage, i.e., $|S_5 - S_4|$ is negligible. The disjunctive NIZKPs ensure that the proof $\pi^*$ does not reveal which specific relation it is attesting to—namely, whether it corresponds to a correct ballot encryption, a correct re-encryption of the previous ballot, or a correct re-encryption of the pre-previous ballot—nor does it leak any information about the associated witnesses. Consequently, any advantage obtained by the adversary $\adv_5$ arises solely from the underlying encryption scheme. 

Let $\zeta(\lambda)$ (a poly-nominal function of $\lambda$) denote the number of ballots replaced. 
We give another succession of hops $H_0,\cdots, H_{\zeta(\lambda)}$, such that in game $H_{i}$, last $i-1$ ballots within the replace set ($\beta\in\PBB\backslash\{\beta_{\adv}\}$) are replaced as describe above. In this way, Game 4 is $H_0$ and Game 5 is $H_{\zeta(\lambda)}$. Denote $S_{H_i}$ the probability of $\adv_5$ winning $H_i$. 
Let $i$ be some index. We construct adversary $\mathcal{B}$ for IND-RCCA to simulate $H_{i}$. 

$\mathcal{B}$ invokes $\adv_{5}$ with the security parameter $\lambda$ and public parameter $\mathbf{param}$. $\mathcal{B}$ gets $pk$ from $\G_{IND-RCCA}$. $\mathcal{B}$ runs the setup phase but replacing the public key of the tallier with $pk$ ($pk_T\leftarrow pk$) and abandoning its secrete key $sk_{T}$. $\mathcal{B}$ receives corrupted voters $\Icorr$ and the coerced voter $id_j$ from $\adv_{5}$, and runs the registration phase to generate public credentials of eligible voters. Next, it sends $(pk,pk_{VS},\{upk_{id},CR_{id}\}_{id\in\mathbb{I}}, \{\PBB^{id}\}_{id\in\mathbb{I}},\mathbf{param})$ to $\adv_{5}$ and samples $\mathcal{O}$, $\{r^{id}_{v},\vec{t}_{v}^{id}\}_{id\in\mathbb{I}\backslash\Icorr}$ from the distributions, with the true intention of the coerced voter being overridden. 
Then, it draws a random bit $b$. 

At each clock cycle $clk$, for every tuple $(id,\beta_{\adv})$ in $\mathbb{L}$, it verifies that $id$ is either a corrupted or coerced voter and ensures the validity of the accompanying proofs. Then, it appends $\beta_{\adv}$ to $\PBB^{id}$. For ballots that are not from $\adv_{5}$, it first check if the ballot is to be replaced before, at, or after hob $i$ ($H_{i}$), and does the following:
\begin{itemize}
        \item If the ballot should be replaced after $H_i$, $\mathcal{B}$ honestly generate them as Game 5. Specifically, if the ballot is produced by the voting server, it re-encrypts a ciphertext on the corresponding bulletin board depending on the credential check, and computes proofs. Otherwise, for ballots of coerced and honest voters, it queries the encryption oracle for $o$ or $o_{id}$, simulates the correct encryption proof and compute a correct key generation proof.
        \item If the ballot should be replaced before $H_{i}$, $\mathcal{B}$ just encrypts with dummy for $ct_{o}$ to maintain hob continuity. Proofs are either generated or simulated depending on ballot type.
        \item If the ballot is ought to be replace at the current hob, $\mathcal{B}$ sends the challenge message $m_0= o_{id}$ and $m_1 = 0^{|o|}$ to IND-RCCA to receive $ct_{o_b}$. Note that because of the perfect rerandomization, the re-encryption of ciphertexts and the encryption of intentions are indistinguishable. Even if $ct_o$ of this ballot was a re-encryption, the encryption of $o_{id}$ is a perfect simulation of $H_{i}$.
    \end{itemize}
$\mathcal{B}$ always faithfully creates the credential ciphertext and the proof of correct credential encryption to form a complete ballot and append it to the corresponding bulletin board.

The tally result is given to $\adv_5$ by directly applying the result function $\rho$ on $\mathcal{O}$ along with adversarial options sent by $\adv_5$. Upon receiving $b'$ from $\adv_5$, if $b'=b$, $\mathcal{B}$ returns $0$ to IND-RCCA, and $1$ otherwise.

A complete and formal description of $\mathcal{B}$ is given in Figure~\ref{algo: C in RCCA game for CR-Loki CR-Privacy Game 5}.

When IND-RCCA encrypts $o_{id}$ (resp. $0^{|o|}$), $\mathcal{B}$ plays a perfect simulation of $H_i$ (resp. $H_{i+1}$), so that the probability of $b=b'$ is $S_{H_i}$ (resp. $S_{H_{i+1}}$). Given that the advantage of $\mathcal{B}$ wins IND-RCCA is negligible, we have $|\frac{1}{2}(S_{H_i}-S_{H_{i+1}}) - \frac{1}{2}| \leq \epsilon_{RCCA}(\lambda)$. By the triangular inequality,
$$
|S_5-S_4|\leq 2\zeta(\lambda)\epsilon_{RCCA}(\lambda)
$$

\begin{figure*}
\hrule
\vspace*{3pt}
\textbf{$\mathcal{B}$ in IND-RCCA game for Game 5 of CR-Privacy Proof}
\vspace*{3pt}
\noindent\hrule
\begin{algorithmic}[1]
\scriptsize
\State receive $pk$ from IND-RCCA
\State $\{(pk_T,sk_T),(pk_{VS},sk_{VS})\}\leftarrow\mathsf{Setup}(1^\lambda,\mathbf{param})$
\State replace $pk_{T}$ with $pk$ and drops $sk_{T}$
\State initialize $\PBB$ and $\mathcal{O}$ as $[\quad]$
\State receive $\Icorr$ and $id_j$ from $\adv$
\State $\{(upk_{id},usk_{id}),CR_{id},cr_{id}\}\leftarrow\mathsf{Register}(id)_{id\in\mathbb{I}}$
\State send $(pk,pk_{VS},\{upk_{id},CR_{id}\}_{id\in\mathbb{I}},\{\PBB^{id}\}_{id\in\mathbb{I}},\mathbf{param})$ and $(\{usk_{id}, cr_{id}\}_{id\in\Icorr})$ to $\adv$
\State create the true intention set $\mathcal{O}$, number of voting $r_v^{id}$, and voting time vector $\vec{t}_{v}^{id}$ for non-corrupted voters by sampling from distributions.
\State receive $(id_j,o)$ (overrides the previous sampling if validity check passes) from $\adv$ and send $usk_{id_j}$ back.
\State $b\leftarrow_{\$}\{0,1\}$

\While{$clk<t_{end}$}
\State receive $\mathbb{L}$ from $\adv$

\For{$(id,\beta_{\adv}) \in \mathbb{L}$ such that $id\in\Icorr\cup\{id_j\}$}
\State check the validity of proofs
\State append $\beta_{\adv}$ to $\PBB^{id}$

\EndFor

\For{twin ballots, the voting server noise ballots, and coerced and honest voters' genuine ballots}
\algorithmiccomment{ballots in the replace set}

\If{the ballot is before $i$-th to-be-replaced ballot}
\algorithmiccomment{the ballot is not replaced and should been generated truthfully}

\If{the ballot is either a twin ballot or a voting server noise ballot}
\State $\tilde{cr}_{id} \leftarrow\mathsf{Dec}_{CCA}(ct_{cr_{id}}^{-1},pk_{VS})$
\algorithmiccomment{decrypt the credential of last ballot}
\If{$\tilde{cr}_{id} = cr_{id}$}
\State $ct_{o_{id}}\leftarrow_{\$}\mathsf{ReEnc}(pk, ct_{o_{id}}^{-1};r_o)$, compute $\pi_{\mathsf{ReEnc}(ct_{o_{id}}^{-1})}$, $\pi_{\mathsf{Dec}(ct^{-1}_{cr_{id}})= \mathsf{Dec}(CR_{id})}$ and $\pi_{\mathsf{Dec}(CR_{id})=cr_{id}}$.
\Else
\State $ct_{o_{id}}\leftarrow_{\$}\mathsf{ReEnc}(pk, ct_{o_{id}}^{-2};r_o)$, compute $\pi_{\mathsf{ReEnc}(ct_{o_{id}}^{-2})}$, $\pi_{\mathsf{Dec}(ct^{-1}_{cr_{id}})\neq \mathsf{Dec}(CR_{id})}$ and $\pi_{\mathsf{Dec}(CR_{id})=cr_{id}}$.
\EndIf
\Else\algorithmiccomment{coerced and honest voters' genuine ballots}
\State query encryption oracle for $o_{id}$ (\emph{i.e.} $(id,o_{id})\in \mathcal{O}\cup \{(id_j,o)\}$) to receive $ct_{o_{id}}$, simulate $\mathsf{sim}_{\mathsf{Enc}(o_{id})}$, and compute $\pi_{\mathsf{KG}(usk_{id})}$.
\EndIf

\Else
\If{the ballot is after $i$-th to-be-replaced ballot}
\algorithmiccomment{the ballot has been replaced}
\State query encryption oracle for $0^{|o|}$ to receive $ct_{o_{id}}$
\Else\algorithmiccomment{the $i$-th is being replaced}
\State challenge the encryption oracle with $o_{id}$ (either from $\mathcal{O}$ or $\beta_{\adv}$) and $0^{|o|}$ to receive $ct_{o_{b}}$
\State \algorithmiccomment{even if the $ct_o$ of the ballot is a re-encryption, because of the indistinguishability of $\mathsf{Enc_{RCCA}}$ and $\mathsf{ReEnc}$, still can challenge with two encryptions}
\EndIf
\If{the ballot is either a twin ballot or a voting server noise ballot}
\algorithmiccomment{the proof generation depends on ballot type}
\State $\tilde{cr}_{id} \leftarrow\mathsf{Dec}_{CCA}(ct_{cr_{id}}^{-1},pk_{VS})$
\If{$\tilde{cr}_{id} = cr_{id}$}
\State simulate $\mathsf{sim}_{\mathsf{ReEnc}(ct_{o_{id}}^{-1})}$, compute $\pi_{\mathsf{Dec}(ct^{-1}_{cr_{id}})= \mathsf{Dec}(CR_{id})}$ and $\pi_{\mathsf{Dec}(CR_{id})=cr_{id}}$.
\Else
\State simulate $\mathsf{sim}_{\mathsf{ReEnc}(ct_{o_{id}}^{-2})}$, compute $\pi_{\mathsf{Dec}(ct^{-1}_{cr_{id}})\neq \mathsf{Dec}(CR_{id})}$ and $\pi_{\mathsf{Dec}(CR_{id})=cr_{id}}$.
\EndIf
\Else
\State simulate $\mathsf{sim}_{\mathsf{Enc}(o_{id})}$ and compute $\pi_{\mathsf{KG}(usk_{id})}$.
\EndIf
\EndIf

\State $ct_{cr_{id}} \leftarrow_{\$} \mathsf{Enc}_{CCA}(pk_{VS}, cr_{id};r_{cr})$ and compute $\pi_{\mathsf{Enc}(cr_{id})}$
\State form the ballot as $(ct_{o_{id}},ct_{cr_{id}},\pi)$ where $\pi$ is a disjunctive proof and append it to $\PBB^{id}$.
\EndFor

\EndWhile

\State for $\mathbb{L}$ received at $t_{end}$, repeat above.
\If{$b=1$}
\State $W\leftarrow\rho(\mathcal{O}_{id\in\mathbb{I}\backslash\Icorr\cup\{id_j\}}\cup(id,o_{id})_{id\in\Icorr})$
\Else
\State $W\leftarrow\rho(\mathcal{O}_{id\in\mathbb{I}\backslash\Icorr\cup\{id_j\}}\cup(id_j,o)\cup(id,o_{id})_{id\in\Icorr})$
\EndIf
\State send $W$ to $\adv$ and receive $b'$

\If{$b' = b$}
\State return $0$ to IND-RCCA
\Else
\State return $1$ to IND-RCCA
\EndIf

\end{algorithmic}
\vspace*{3pt}\hrule\vspace*{3pt}
\caption{$\mathcal{B}$ plays in IND-RCCA game and simulates $\G^{CR-Loki,\adv_{5}}_{H_{i}}$ for $\adv_{5}$}
\label{algo: C in RCCA game for CR-Loki CR-Privacy Game 5}
\end{figure*}


\textbf{Game 6:} In this game, $ct_{cr}$ (credential ciphertext) of ballots in the replace set are replaced with dummy credential encryption. Ballots in the replace set is now in a form of
\[
\mathsf{Enc}_{RCCA}(pk_T, 0^{|o|};r_o) \mathbin\Vert \mathsf{Enc}_{CCA}(pk_{VS}, 0^{|cr|};r_{cr}) \mathbin\Vert \pi^*
\]
where $\pi^* = \pi_{\mathsf{Enc}(0^{|o|})} \mathbin\Vert \pi_{\mathsf{Enc}(0^{|cr_{id}|})} \mathbin\Vert \pi_{\mathsf{KG}(usk_{id})}$, which all can computed. Clearly, $\pi^*$ verifies true. 

We argue that $\mathcal{A}_6$, which behaves identically to $\mathcal{A}_5$, achieves a negligible distinguishing advantage, i.e., $|S_6 - S_5|$ is negligible. 

Similarly, any advantage obtained by the adversary $\adv_6$ arises solely from the underlying encryption scheme. Follow the same proving strategy, let $\zeta(\lambda)$ (a poly-nominal function of $\lambda$) denote the number of ballots replaced, and give another succession of hops $H_0,\cdots, H_{\zeta(\lambda)}$, such that in game $H_{i}$, last $i-1$ ballots within the replace set ($\beta\in\PBB\backslash\{\beta_{\adv}\}$) are replaced as pure dummy ballots. In this way, Game 5 is $H_0$ and Game 6 is $H_{\zeta(\lambda)}$. Denote $S_{H_i}$ the probability of $\adv_6$ winning $H_i$. 
Let $i$ be some index. We construct adversary $\mathcal{B}$ for IND-CCA to simulate $H_{i}$. Generally, $\mathcal{B}$ plays the challenge message to IND-CCA when it simulates the ballot that is to-be-replaced at the current hob. How does $\mathcal{B}$ behave is alike to what is in Game 5. For the sake of concise, we don't repeat the narrative description. A complete and formal description is given in Figure~\ref{algo: C in CCA game for CR-Loki CR-Privacy Game 6}.

Given that the advantage of $\mathcal{B}$ wins IND-CCA is negligible, we have $|\frac{1}{2}(S_{H_i}-S_{H_{i+1}}) - \frac{1}{2}| \leq \epsilon_{CCA}(\lambda)$. By the triangular inequality,
$$
|S_6-S_5|\leq 2\zeta(\lambda)\epsilon_{CCA}(\lambda)
$$

\begin{figure*}
\hrule
\vspace*{3pt}
\textbf{$\mathcal{B}$ in IND-CCA game for Game 6 of CR-Privacy Proof}
\vspace*{3pt}
\noindent\hrule
\begin{algorithmic}[1]
\scriptsize
\State receive $pk$ from IND-CCA
\State $\{(pk_T,sk_T),(pk_{VS},sk_{VS})\}\leftarrow\mathsf{Setup}(1^\lambda,\mathbf{param})$
\State replace $pk_{VS}$ with $pk$ and drops $sk_{VS}$
\State initialize $\PBB$ and $\mathcal{O}$ as $[\quad]$
\State receive $\Icorr$ and $id_j$ from $\adv$
\State $\{(upk_{id},usk_{id}),CR_{id},cr_{id}\}\leftarrow\mathsf{Register}(id)_{id\in\mathbb{I}}$
\State send $(pk_T,pk,\{upk_{id},CR_{id}\}_{id\in\mathbb{I}},\{\PBB^{id}\}_{id\in\mathbb{I}},\mathbf{param})$ and $(\{usk_{id}, cr_{id}\}_{id\in\Icorr})$ to $\adv$
\State create the true intention set $\mathcal{O}$, number of voting $r_v^{id}$, and voting time vector $\vec{t}_{v}^{id}$ for non-corrupted voters by sampling from distributions.
\State receive $(id_j,o)$ (overrides the previous sampling if validity check passes) from $\adv$ and send $usk_{id_j}$ back.
\State $b\leftarrow_{\$}\{0,1\}$

\While{$clk<t_{end}$}
\State receive $\mathbb{L}$ from $\adv$

\For{$(id,\beta_{\adv}) \in \mathbb{L}$ such that $id\in\Icorr\cup\{id_j\}$}
\label{line: start for for beta adv}
\State check the validity of proofs
\State check $ct_{cr_{id}}$ of $\beta_{\adv}$ is not a replay
\State append $\beta_{\adv}$ to $\PBB^{id}$
\EndFor

\For{twin ballots, the voting server noise ballots, and coerced and honest voters' genuine ballots}
\algorithmiccomment{ballots in the replace set}

\State $ct_{o_{id}}\leftarrow_{\$}\mathsf{Enc}_{RCCA}(pk_T,0^{|o|};r_{o})$
\If{the ballot is before $i$-th to-be-replaced ballot}
\algorithmiccomment{the ballot is not replaced and should encrypt with real credentials}
\State query encryption oracle for $cr_{id}$ to receive $ct_{cr_{id}}$
\State simulate $\mathsf{sim}_{\mathsf{Enc}(cr_{id})}$
\If{the ballot is either a twin ballot or a voting server noise ballot}
\algorithmiccomment{the proof generation depends on ballot type}
\State query decryption oracle for $ct_{cr_{id}}^{-1}$ to receive $\tilde{cr}_{id}$
\If{$\tilde{cr}_{id} = cr_{id}$}
\State simulate $\mathsf{sim}_{\mathsf{ReEnc}(ct_{o_{id}}^{-1})}$, $\mathsf{sim}_{\mathsf{Dec}(ct^{-1}_{cr_{id}})= \mathsf{Dec}(CR_{id})}$ and $\mathsf{sim}_{\mathsf{Dec}(CR_{id})=cr_{id}}$.
\Else
\State simulate $\mathsf{sim}_{\mathsf{ReEnc}(ct_{o_{id}}^{-2})}$, $\mathsf{sim}_{\mathsf{Dec}(ct^{-1}_{cr_{id}})\neq \mathsf{Dec}(CR_{id})}$ and $\mathsf{sim}_{\mathsf{Dec}(CR_{id})=cr_{id}}$.
\EndIf
\Else
\State compute $\pi_{\mathsf{Enc}(0^{|o|})}$ and $\pi_{\mathsf{KG}(usk_{id})}$.
\EndIf
\Else
\If{the ballot is after $i$-th to-be-replaced ballot}
\algorithmiccomment{the ballot has been replaced}
\State query encryption oracle for $0^{|cr|}$ to receive $ct_{cr_{id}}$
\Else
\algorithmiccomment{the $i$-th is being replaced}
\State challenge the encryption oracle with $cr_{id}$ and $0^{|cr|}$ to receive $ct_{cr_b}$
\EndIf
\State simulate $\mathsf{sim}_{\mathsf{Enc}(0^{|cr|})}$
\State compute $\pi_{\mathsf{Enc}(0^{|o|})}$ and $\pi_{\mathsf{KG}(usk_{id})}$
\EndIf
\State form the ballot as $(ct_{o_{id}},ct_{cr_{id}},\pi)$ where $\pi$ is a disjunctive proof and append it to $\PBB^{id}$.
\EndFor

\EndWhile

\State for $\mathbb{L}$ received at $t_{end}$, repeat above.

\If{$b=1$}
\State $W\leftarrow\rho(\mathcal{O}_{id\in\mathbb{I}\backslash\Icorr\cup\{id_j\}}\cup(id,o_{id})_{id\in\Icorr})$
\Else
\State $W\leftarrow\rho(\mathcal{O}_{id\in\mathbb{I}\backslash\Icorr\cup\{id_j\}}\cup(id_j,o)\cup(id,o_{id})_{id\in\Icorr})$
\EndIf
\State send $W$ to $\adv$ and receive $b'$

\If{$b' = b$}
\State return $0$ to IND-CCA
\Else
\State return $1$ to IND-CCA
\EndIf

\end{algorithmic}
\vspace*{3pt}\hrule\vspace*{3pt}
\caption{$\mathcal{B}$ plays in IND-CCA game and simulates $\G^{CR-Loki,\adv_6}_{H_i}$ for $\adv_6$}
\label{algo: C in CCA game for CR-Loki CR-Privacy Game 6}
\end{figure*}

\textbf{Game 7:} In this game, the public bulletin board $\PBB$ is removed from the game.

To simulate this setting, we construct an adversary $\mathcal{A}_7$ that internally reconstructs $\PBB$ while interacting with Game 7. For each eligible voter $id$, it initializes a corresponding $\PBB^{id}$. At each clock cycle $clk$, whenever $\mathcal{A}_7$ sends a message $(id, \tilde{o}_{id} \mathbin\Vert \tilde{cr}_{id})$ to Game 7, it also generates a corresponding encrypted ballot:
\begin{align*}
    \beta_{\adv} = (id, &\mathsf{Enc}_{RCCA}(pk_T, \tilde{o}_{id}; r_o) \mathbin\Vert \mathsf{Enc}_{CCA}(pk_{VS}, \tilde{cr}_{id}; r_{cr}) \mathbin\Vert \\&\pi_{\mathsf{Enc}(\tilde{o}_{id})} \mathbin\Vert \pi_{\mathsf{Enc}(\tilde{cr}_{id})} \mathbin\Vert \mathsf{sim}_{\mathsf{KG}(usk_{id})})
\end{align*}
where $\mathsf{Sim}_{\mathsf{KG}(usk_{id})}$ is the simulation of $\pi_{\mathsf{KG}(usk_{id})}$, and the other components are computed honestly. This ballot is then appended to the appropriate $\PBB^{id}$. \JQ{Additionally, the size of $\PBB$ needs to remain consistent with previous game at all ticks. Due to the configuration of the Voting Server such that $\Pr_{\vec{t}\leftarrow_{\$}\DT^{vs}}[\vec{t} = (0,\cdots, t_{end} - 1)]$, regardless of voters' casting, there is always a ballot appended to the $\PBB$ (not including twin ballots). $\adv_7$ fill the $\PBB$ with a dummy ballot for each voters if it does not send a ballot message to Game 7. In this way, it keeps the growing pace of $\PBB$ synchronized with the previous game.}

$\mathcal{A}_7$ provides a simulation of $\mathcal{A}_6$ that is perfect apart from including $\mathsf{sim}_{\mathsf{KG}(usk_{id})}$ instead of $\pi_{\mathsf{KG}(usk_{id})}$.
Let $\zeta_{\beta_\adv}(\lambda)$ (a polynomial function of $\lambda$) denote the total number of adversarial ballots sent to the game.
If $\epsilon_{ZK(\mathsf{KG})}$ is the (negligible) zero-knowledge error for the proof of key generation of $usk_{id}$, then we have that
\[|S_7-S_6|\leq \zeta_{\beta_\adv}(\lambda)\epsilon_{ZK(\mathsf{KG})}(\lambda), \]
Thus, $|S_5-S_4|$ is negligible.
\TZdone{The simulation is not perfect as $\pi_{\mathsf{KG}(usk_{id})}$ is simulated. It is better to say that: $\mathcal{A}_5$ provides a simulation of $\mathcal{A}_4$ that is perfect apart from including $\mathsf{sim}_{\mathsf{KG}(usk_{id})}$ instead of $\pi_{\mathsf{KG}(usk_{id})}$. If $\epsilon_{ZK,i}$ is the (negligible) zero-knowledge error for the proof of $id_i$, then we have that
\[|S_5-S_4|\leq\sum_{i=1}^{n_v}\epsilon_{ZK,i},\]
where $n_v$ is the number of eligible voters. Thus, $|S_5-S_4|$ is negligible.
}

\textbf{Game 8:} In this game, the adversary can no longer submit $\mathbb{L}$ during the cast of honest voters, but all at the end. Since there is no $\PBB$ in Game 7, $\mathcal{A}_7$ does not learn anything from adaptive insertion. What $\mathcal{A}_8$ does to have a perfect simulation for $\mathcal{A}_7$ is storing malicious ballots (with overwrite) and send the batch at $t_{end}$. Therefore, $S_8 = S_7$.

\textbf{Game 9:} In this game, all steps related to $(pk_T, pk_{VS},\{upk_{id},$ $CR_{id}\}_{id\in\mathbb{I}})$ are removed. 

Let us say $\mathcal{H}_1$ is a hybrid game where every $CR_{id}$ is replace by $\mathsf{Enc}_{CCA}(pk_T,0^{|cr_{id}|};r_{cr})$. Because $CR_{id}$ is the ciphertext of an IND-CCA secure encryption scheme, we have that
$$
|S_{\mathcal{H}_1} - S_8| \le n_{v}\epsilon_{CCA}(\lambda)
$$
\MAdone{I am not clear on why we define this next hybrid  game like that. We can now omit the publication of the $CR_{id}$ as the adversary can simulate it, and finally we can just not sample or publish $pk_T, pk_{VS}, upk_{id}$ since nothing else published after that.}\JQdone{this step is just to break simulation of $CR_{id}$ and simulation of public keys, as they use different errors.}

From the hybrid game $\mathcal{H}_1$, we can create another hybrid game $\mathcal{H}_2$, where $pk_{VS}$, $pk_{T}$ and $\{usk\}_{id\in\mathbb{I}}$ are replaced by simulated public keys. By the assumption of UC-secure key generation, we get that
$$
|S_{\mathcal{H}_2} - S_{\mathcal{H}_1}| \le (n_v+2)\epsilon_{KE}(\lambda)
$$
where $\epsilon_{KE}$ is one (negligible) key generation error.

Now, we construct an adversary $\adv_9$ against Game 9 that operates as $\adv_8$ except that it generates dummy credential encryption as $CR_{id}$ for $id\in\mathbb{I}$ and the simulated public keys locally itself. Clearly, it holds that $S_9 = S_{\mathcal{H}_2}$. By above, we have that
$$
|S_9-S_8|\le n_{v}\epsilon_{CCA}(\lambda) + (n_v+2)\epsilon_{KE}(\lambda)
$$

\textbf{Game 10:} The final game is the ideal game, where any steps that is not used in Game 9 are removed, including sample from distribution for non-corrupted voters and sending the credential of the coerced voter.
$\mathcal{A}_{10}$ might generate them itself internally, thus 
runs a perfect simulation for $\adv_9$. Clearly, $S_{10}=S_9$.\qed

\subsection{Ballot privacy}
\label{apx: privacy analysis for cr-loki}

In this appendix, we adapt the privacy definition introduced in Bernhard \emph{et al.}~\cite{bernhard2012not}, which is the same definition original Loki uses, and prove CR-Loki achieves ballot privacy against dishonest Voting Server. \TZdone{If this is the same definition as the one used in the original Loki paper, then we should mention this. In addition, I cannot see if the adversary can corrupt voters. Are all ballots generated via the algorithm $\mathsf{Vote}(\cdot)$? If not, isn't this definition only against passive adversaries?}\JQdone{Yes, both the privacy and verifiability definition are from the original Loki with the alteration of our syntax. The corrupted voters's ballot are casted via $\mathcal{O}cast$.}

\paragraph{Privacy Definition}
The privacy game maintains two bulletin boards and one of those is visible to the adversary $\adv$ depending on $b$. The adversary $\adv$ has access to four oracles: $\mathcal{O}\mathrm{voteLR}, \mathcal{O}\mathrm{cast}, \mathcal{O}\mathrm{board}$, and $\mathcal{O}\mathrm{tally}$.

The ballots on the bulletin board are generated by $\mathcal{O}\mathrm{voteLR}$ and $\mathcal{O}\mathrm{cast}$. $\mathcal{O}\mathrm{voteLR}$ simulates the potential vote of honest voters who are under observation. Ballot $\beta_{id}^{b}$ is appended to the corresponding bulletin board. The game captures a malicious Voting Server by giving the private key $sk_{VS}$ to the adversary $\adv$. Therefore regardless of the malicious twin ballots or noise ballots, the game allows the adversary to first observe the board via $\mathcal{O}\mathsf{board}$ and send malicious ballots via $\mathcal{O}\mathsf{cast}$. For $\mathcal{O}\mathsf{cast}$ oracle, the same ballot (naturally a same vote option) is appended to both bulletin board. The result is calculated on the last ballots by $\mathcal{O}\mathrm{tally}$. 

The adversary $\adv$ outputs a guess $b'$ at the end of the game. We say that the adversary wins the game if $b'=b$.

\begin{definition}[Privacy]
    \label{def: ballot privacy}
    et $\Gamma^{\DR^{vs},\DT^{vs}}$ be an e-voting scheme and $\mathbf{param}=(\mathbb{I},\mathbb{O}, \DO, \DR^{v},\DT^{v})$ be some voting parameters. $\Gamma^{\DR^{vs},\DT^{vs}}$ is said to satisfy Privacy if there exists a simulation algorithm $\mathsf{SimTally}$ such that for any PPT adversaries $\adv$
    \begin{align*}
        | &\Pr[\mathcal{G}ame^{\adv}_{Pri,0}(1^\lambda,\mathbf{param}) = 1] - \\&\Pr[\mathcal{G}ame^{\adv}_{Pri,1}(1^\lambda,\mathbf{param}) = 1]|
    \end{align*}
    is a negligible function in the security parameter $\lambda$.
\end{definition}

\begin{figure}
\hrule
\vspace*{3pt}
$\mathcal{G}ame_{Pri,b}^{\adv}$
\vspace*{3pt}
\noindent\hrule
\begin{algorithmic}[1]
\scriptsize
\Require{$\lambda,\mathbf{param}$}

\State $(pk_T,sk_T),(pk_{VS},sk_{VS})\leftarrow\mathsf{Setup}(1^{\lambda},\mathbf{param})$
\State $\{(upk_{id},usk_{id}),CR_{id},cr_{id}\}\leftarrow\mathsf{Register}(id)_{id\in\mathbb{I}}$
\State Initialize $\PBB_0,\PBB_1$
\State $b'\leftarrow\mathcal{A}^{\mathcal{O}}(pk_{T},(pk_{VS},sk_{VS}),\{upk_{id},CR_{id}\}_{id\in\mathbb{I}},\mathbf{param})$
\State \Return $b'$
\State

\State $\mathcal{O}\mathrm{voteLR}(id,o_0,o_1)$
\If{$o_0 \notin \mathbb{O}$ or $o_1 \notin \mathbb{O}$ or $id \notin \mathbb{I}$}
\State \Return $\bot$
\EndIf
\State $\beta_{o_0}\leftarrow\mathsf{Vote}(id,usk_{id},pk_T,pk_{VS},o_0,cr_{id})$
\State $\beta_{o_1}\leftarrow\mathsf{Vote}(id,usk_{id},pk_T,pk_{VS},o_1,cr_{id})$
\If{$\mathsf{Validate}(id,\PBB_b,\beta_{o_b}) = \bot$}
\State \Return $\bot$
\EndIf
\State $\PBB_0\leftarrow\PBB_0\cup\beta_{o_0}$
\State $\PBB_1\leftarrow\PBB_1\cup\beta_{o_1}$
\State

\State $\mathcal{O}\mathrm{cast}(id,\beta_{\adv})$
\If{$\mathsf{Validate}(id,\PBB_b,\beta_{\adv}) = \bot$}
\State \Return $\bot$
\EndIf
\State $\PBB_0\leftarrow\PBB_0\cup\beta_{\adv}$
\State $\PBB_1\leftarrow\PBB_1\cup\beta_{\adv}$
\State

\State $\mathcal{O}\mathrm{board}()$
\State \Return $\PBB_b$
\State

\State $\mathcal{O}\mathrm{tally}()$
\State $(W,\Pi_0)\leftarrow\mathsf{Tally}(\PBB_0,sk_T)$
\State $\Pi_1\leftarrow\mathsf{SimTally}(\PBB_1, W)$
\State \Return $(W,\Pi_b)$

\end{algorithmic}
\vspace*{3pt}\hrule\vspace*{3pt}
\caption{The privacy game in which the adversary $\mathcal{A}$ has access to the oracles $\mathcal{O}=\{\mathcal{O}\mathrm{voteLR}, \mathcal{O}\mathrm{cast}, \mathcal{O}\mathrm{board},\mathcal{O}\mathrm{tally}\}$}
\label{alg: privacy game}
\end{figure}

\paragraph{CR-Loki Achieves Ballot Privacy}
\label{apx: cr-loki is private}

\begin{theorem}
\label{thm: cr-loki is private}
Under the assumption that the tallier's encryption scheme is IND-RCCA secure and UC-security of non-interactive zero-knowledge proofs (NIZKPs), CR-Loki achieves ballot privacy.
\end{theorem}

We define a sequence of games to show that our scheme provides ballot privacy. In these games, the adversary $\adv$ interacts with a challenger in the privacy game, starting with $b= 0$ and ending up with $b=1$. We show that for any PPT adversary $\adv_{i}$ against Game $i$, we can construct a PPT adversary $\adv_{i+1}$ against Game $i+1$ such that the difference of probability $\adv_i$ and $\adv_{i+1}$ wins is negligible in the security parameter $\lambda$. 
\JQ{Recall that both ballot proofs and tally proofs are NIZKPs which can be simulated under the programmable random oracle. Accordingly, $\mathsf{SimTally}(\cdot)$ exists as the algorithm of simulator of NIZKPs.}
\TZdone{The security definition asks us to find an algorithm $\mathsf{SimTally}$ that achieves indistinguishability. I could not find the description of this algorithm in the proof.} 

\textbf{Game 0:} This game is $\mathcal{G}ame_{Pri,0}^{\adv}$ where $\adv_0$ sees $\PBB_0$ and the output of $\mathcal{O}\mathrm{tally}$ is computed on $\PBB_0$.

\textbf{Game 1:} In this game, all proofs made by honest parties, including the talliers, are simulated. It means that the output tally proof $\Pi_0$ and the proofs in ballots generated by $\mathcal{O}\mathrm{voteLR}$ are now simulated. Under the zero-knowledge property of assumption NIZKPs, the advantage of $\adv_1$ distinguishing real proofs from simulated ones is negligible. 

\textbf{Game 2:} In this game, ballots cast by $\mathcal{O}\mathrm{voteLR}$ on $\PBB_0$ are replaced by $\beta_{o_1}$. In another words, in this game, $\PBB_0$ and $\PBB_1$ always have ballots of a same option appended no matter through $\mathcal{O}\mathrm{voteLR}$ or $\mathcal{O}\mathrm{cast}$. \TZdone{I do not understand this point. Is the ballot always the same? Or do we mean that in each invocation we append the ballot that corresponds to the option indexed by 1?}

Let $\zeta(\lambda)$ denotes the number of ballots cast by $\mathcal{O}\mathrm{voteLR}$ on $\PBB_0$ (and $\PBB_1$). We give another succession of hops $H_0,\cdots,H_{\zeta(\lambda)}$ where at each hop $H_i$ the last $i-1$ ballots cast by $\mathcal{O}\mathrm{voteLR}$ on $\PBB_0$ are replaced by $\beta_{o_1}$, and the rest of ballots remain $\beta_{o_0}$. In this way, Game 1 is $H_0$ and Game 2 is $H_{\zeta(\lambda)}$. Let $i$ be some index. We construct an adversary $\mathcal{B}$ for IND-RCCA to simulate $H_i$. $\mathcal{B}$ runs $\mathsf{Setup}(\cdot)$ and $\mathsf{Register}(\cdot)$ for every eligible voter, and sends public information as well as the private key of the Voting Server to invoke $\adv_2$. 
\begin{itemize}
    \item Upon receiving enquiry of oracles $\mathcal{O}\mathrm{cast}$ and $\mathcal{O}\mathrm{board}$, it honestly behaves like $\mathcal{G}ame_{Pri,b}^{\adv}$. 
    \item Upon receiving the $i$th enquiry of $\mathcal{O}\mathrm{voteLR}$, it sends $(o_0,o_1)$ to the IND-RCCA challenger and receives a ciphertext $ct_o^b$. It then creates the credential ciphertext $ct_{cr}\leftarrow\mathsf{Enc}_{RCCA}(pk_{VS},cr_{id};r)$ \TZdone{$\mathsf{Enc}_{RCCA}(pk_{VS},cr_{id};r)$ maybe?}, simulates proofs for relation $R^{id}$ (with simulated proof of plaintext knowledge $\mathsf{sim}_{\mathsf{Enc}(o_b)}$ and $\mathsf{sim}_{\mathsf{Enc}(cr_{id})}$, and proof of private key knowledge $\mathsf{sim}_{\mathsf{KG}(usk_{id})}$), and forms a ballot $\beta_{o_b}$ to append both on $\PBB_0$ and $\PBB_1$. For enquiries before $i$-th ballot, it behaves like the oracle programmed, and for enquires after $i$-th ballot, it appends $\beta_{o_1}$ to both bulletin boards. It also records the option plaintext from query parameters for future tallying.
    \item Upon receiving enquiry of $\mathcal{O}\mathrm{tally}$, $\mathcal{B}$ does the following. For the last ballots of each voter $id$ on $\PBB_0$ (recall that the ballot format of CR-Loki is $(id,upk_{id},(ct_o,ct_{cr},\pi))$), if the ballot is cast via $\mathcal{O}\mathrm{voteLR}$, it refers to the previous record plaintext option. Otherwise, it queries the decryption oracle of IND-RCCA game for option plaintext of $\beta_{\adv}$. Note that the decryption oracle will return $\bot$ if the malicious ballots encrypt one of the options in the challenge message pair. 
    To solve this problem, $\mathcal{B}$ modifies the option ciphertext of $\beta_{\adv}$ before querying it to the decryption oracle. Recall that the tallier's encryption scheme consists of ElGamal encryption and NIZKP, which means the form of option ciphertext is $ct_o=(c_1,c_2)$. It modifies $ct_o$ of all malicious ballots $\beta_\adv$ except for those in the name of $id$ whom is the $i$th enquiry of $\mathcal{O}\mathrm{voteLR}$ for. The modified $ct_o'$ is $ct_o' = (c_1,c_2\cdot pk_T^{H(upk_{id})})$ and the IND-RCCA decryption oracle returns $o_{\adv}\cdot pk_T^{H(usk_{id})}$ as the plaintext where $\mathcal{B}$ extracts $o_{\adv}$ from it for returning the tallying result to $\adv_2$.
    With $\{(id,o_{id})\}_{id\in\mathbb{I}}$ (includes adversary options), it applies the result function $\rho$ to have $W$ and simulates the tally proof $\Pi_0$. Upon receiving $b'$ from $\adv_2$, it returns the coin back to IND-RCCA game.
\end{itemize}
When IND-RCCA encrypts $o_0$ (resp. $o_1$), $\mathcal{B}$ plays a perfect simulation of $H_i$ (resp. $H_{i+1}$). Therefore the probability that $\adv_2$ distinguishes $H_i$ from $H_{i+1}$ equals to the probability of $\mathcal{B}$ winning the IND-RCCA game. Overall, the difference of probability of $\adv_2$ and $\adv_1$ wins is smaller or equal than $\zeta(\lambda)\cdot\epsilon_{RCCA}$.

After Game 2, the view of the adversary of the bulletin board $\PBB_0$ is switched to $\PBB_1$, therefore Game 2 is $\mathcal{G}ame_{Pri,1}^{\adv}$. We show that the advantage of $\adv$ in distinguishing the transition from $\mathcal{G}ame_{Pri,0}^{\adv}$ to $\mathcal{G}ame_{Pri,1}^{\adv}$ is negligible in the security parameter.\qed

\subsection{Verifiability}
\label{apx: verifiability analysis for cr-loki}

In this appendix, we adapt the verifiability definition introduced in Cortier \emph{et al.}~\cite{cortier2014election}, which is used by the original Loki, and prove that CR-Loki achieves verifiability with honest registration and dishonest Voting Server and talliers. \TZdone{Again, is this adaptation similar to the one in the original Loki paper?}

\paragraph{Verifiability Definition}
Let denote $\mathbb{U}$ a tuple of secret and public information of the registered voter and $\mathbb{U}_{corr}$ is that of corrupted voters.
Let $\mathbb{HB}$ contain triples $(id, o_{id},\beta)$ that have been output by $\mathcal{O}\mathrm{vote}$ (if voter $id$ voted multiple times, only the last ballot is retained), while $\mathbb{CK}$ contain all tuples $(id, o_{id},\beta)\in\mathbb{HB}$ such that $\mathsf{VerifyVote}(id,usk_{id},upk_{id},\PBB,\beta) =\top$. In another words, $\mathbb{CK}$ corresponds to voters that have checked that their ballots will be counted. The adversary will have access to three oracles, namely
\begin{description}[style=unboxed,leftmargin=.5cm]
    \item[$\mathcal{O}\mathrm{register}(id)$:] it invokes algorithm $\mathsf{Register}(\cdot)$ to register the given voter, record the output in set $\mathbb{U}$, and return the adversary the public information in the registration.
    \item[$\mathcal{O}\mathrm{corrupt}(id)$:] it allows the adversary to corrupt an registered voter and obtain all the private information of the corrupted voter.
    \item[$\mathcal{O}\mathrm{vote}(id,o_{id})$:] it generates a ballot for the uncorrupted registered voter for a legitimate option $o_{id}$ and updates $\mathbb{HB}$ for recording.
\end{description}
which are parts of $\mathcal{G}ame^{\adv}_{Ver}(1^{\lambda},\mathbf{param})$ as in Figure~\ref{alg: verifiability game}.
\begin{figure}
\hrule
\vspace*{3pt}
$\mathcal{G}ame_{Ver}^{\adv}$
\vspace*{3pt}
\noindent\hrule
\begin{algorithmic}[1]
\scriptsize
\Require{$\lambda,\mathbf{param}$}

\State $(pk_T,sk_T),(pk_{VS},sk_{VS})\leftarrow\mathsf{Setup}(1^{\lambda},\mathbf{param})$
\State $(\PBB, W,\Pi)\leftarrow\mathcal{A}^{\mathcal{O}}((sk_{VS},pk_{VS}),(sk_{T},pk_T),\mathbf{param})$
\If{$\mathsf{VerifyTally}(\PBB, W,\Pi) = \bot$ }
\State\Return $0$
\EndIf
\State $\mathbb{CK} =\{(id_i^k, o_{i}^k,*)\}_{i=1}^{n_k}$
\If{$\exists\{(id_i^{A},o^{A}_{i})\}_{i=1}^{n_A}$ such that $0\leq n_A \leq |\mathbb{U}_{corr}|$, $\exists\{(id_i^{l},o^{l}_i,*)\}_{i=1}^{n_{lz}} \in \mathbb{HB}\setminus \mathbb{CK}$ such that $W = \rho(\{id^k_i,o_{i}^{k}\}_{i=1}^{n_k}) + \rho(\{(id_i^{A},o^{A}_{i})\}_{i=1}^{n_A}) + \rho(\{(id_i^{l},o^{l}_i)\}_{i=1}^{n_{lz}})$}
\State\Return $0$
\Else
\State\Return $1$
\EndIf
\State

\State $\mathcal{O}\mathrm{register}(id)$
\If{$id \notin \mathbb{U}$}
\State $\{(upk_{id},usk_{id}),CR_{id},cr_{id}\}\leftarrow\mathsf{Register}(id)$
\State $\mathbb{U} \leftarrow \mathbb{U} \cup \{id,(upk_{id},usk_{id}),CR_{id},cr_{id}\}$
\State\Return $(id, upk_{id},CR_{id})$
\Else
\State \Return $\bot$
\EndIf
\State

\State $\mathcal{O}\mathrm{corrupt}(id)$
\If{$\{id,(*,*),*,*\}\in\mathbb{U}$ and $\{id,(*,*),*,*\}\notin\mathbb{U}_{corr}$}
\State\Return $\mathbb{U}_{corr}\leftarrow\mathbb{U}_{corr}\cup\{id,(upk_{id},usk_{id}),CR_{id},cr_{id}\}$
\Else
\State\Return $\bot$
\EndIf
\State

\State $\mathcal{O}\mathrm{vote}(id,o_{id})$
\If{$\{id,(*,*),*,*\}\notin\mathbb{U}$ or $\{id,(*,*),*,*\}\in \mathbb{U}_{corr}$ or $o_{id}\notin \mathbb{O}$}
\State\Return $\bot$
\Else
\State $\beta\leftarrow\mathsf{Vote}(id,usk_{id},pk_T,pk_{VS},o_{id},cr_{id})$
\State Replaces any previous entry $(id,*,*)$ in $\mathbb{HB}$ with $(id,o_{id},\beta)$
\State\Return $\beta$
\EndIf

\end{algorithmic}
\vspace*{3pt}\hrule\vspace*{3pt}
\caption{The verifiability game in which the adversary $\mathcal{A}$ has access to the oracles $\mathcal{O}=\{\mathcal{O}\mathrm{register}, \mathcal{O}\mathrm{corrupt}, \mathcal{O}\mathrm{vote}\}$}
\label{alg: verifiability game}
\end{figure}

\begin{definition}[Verifiability]
\label{def: verifiability}
    Let $\Gamma^{\DR^{vs},\DT^{vs}}$ be an e-voting scheme and $\mathbf{param}=(\mathbb{I},\mathbb{O}, \DO, \DR^{v},\DT^{v})$ be some voting parameters. $\Gamma^{\DR^{vs},\DT^{vs}}$ is said to satisfy Verifiability with respect to voting parameters $\mathbf{param}$ if for all PPT adversaries $\mathcal{A}$, there exists a negligible function $\mu(\cdot)$ such that:
    $$
    \Pr[\mathcal{G}ame_{Ver}^{\adv}(1^{\lambda},\mathbf{param}) = 1] \leq \mu(\lambda)
    $$
\end{definition}

\paragraph{CR-Loki is Strong Verifiable}
\label{apx: cr-loki is strong verifiable}

\begin{theorem}
\label{thm: cr-loki is strong verifiable}
    Under the assumption of a trusted registration authority and the UC-security of NIZKPs, CR-Loki is strongly verifiable.
\end{theorem}

Let the adversary $\adv$ output a set of ballots, the result $W$ and the corresponding proof $\Pi$ for the final tally. Suppose that the last ballot for each voter form a set i.e. $\mathbb{T} = \{\beta_1,\dots,\beta_{n_v}\}$. $\mathbb{T}$, $W$ and $\Pi$ are all on the public bulletin board. The soundness of proof $\Pi$ verifies that the result $W$ is obtained from the correct decryption of $\{ct_{o_{id}}\}_{id = 1}^{n_v}$ where $ct_{o_{id}}\in\beta_{id}$. We can conclude that $\mathsf{VerifyTally}(W,\Pi)$ only returns $\top$ when $W$ is the correct result of $\mathbb{T}$ on $\PBB$. \TZdone{I do not understand why this holds. Some more justification is needed.}\JQdone{Because of the previous sentence. The correct decryption of $\{ct_{o_{id}}\}_{id = 1}^{n_v}$ where $ct_{o_{id}}\in\beta_{id}$ is actually $\mathbb{T}$.}

As ballots appended to both $\mathbb{BB}$ and $\PBB$ require to validate true (line~\ref{line: validate true} Figure~\ref{algo: cr-loki algorithms}), the same holds for ballots in set $\mathbb{T}$. Denote $\mathbb{C}=\{(id^A_{i},o^{A}_{i},*)\}_{i=1}^{n_A}$. Now we prove that each $\beta_{id}=(id, usk_{id},(ct_{o_{id}},ct_{cr_{id}},\pi))\in\mathbb{T}$ must be a re-randomization of a ballot in one of the following sets:
\begin{itemize}
    \item $\mathbb{CK}$: the last votes of honest voters who checked their ballots.
    \item $\mathbb{HB}\setminus\mathbb{CK}$: the votes of honest voters who did not check.
    \item $\mathbb{C}$: the last votes of corrupted voters.
\end{itemize}
The knowledge soundness of the proof $\pi$ embedded in ballot $\beta_{id}$ ensures that $\beta_{id}$ is generated by the voter or the Voting Server.
During the registration, a default null ballot is added to the board. Hence $\beta_{id}$ can be a null vote if an honest voter does not take any actions at all (neither casting nor checking). Anyone can verify that initial ballot ciphertext is null with the embedded randomness (see line~\ref{line: embeded rand} Figure~\ref{algo: cr-loki algorithms}). \TZdone{What does ``then embedded randomness'' mean?}
Since the Voting Server is only allowed to re-randomize the ballot cast by the voter, which requires the secret signing key, $\beta_{id}$ must contain either the vote of an honest voter (in $\mathbb{CK}$ or $\mathbb{HB}\setminus\mathbb{CK}$) or a corrupted voter in $\mathbb{C}$. The Voting Server can re-randomize the voter's ballot based on the credential and the corresponding proof $\pi$ ensures that $\beta_{id}$ is either a re-encryption of the voter's new ballot or the second-last ballot on $\PBB^{id}$. Therefore, the adversary $\adv$ cannot generate a new ballot for honest voters, which means it cannot insert malicious votes more than the number of corrupted voter $|\mathbb{U}_{corr}|$.\TZdone{Do you mean that $\adv$ ``inserted a number of malicious votes that is larger than the number of corrupted voters''?}
In addition, the malicious Voting Server cannot manipulate the vote by skipping the voter's ballot as the voter with the real credential at hand can verify through $\pi$ which disjoint relation it falls into. \JQdone{This is where $\mathbb{CK}$ and $\mathbb{HB}\setminus\mathbb{CK}$ are different.} Furthermore, even if the voter cast multiple times with valid credential, $\pi$ ensures that the re-randomization is always of the last ballot. Overall, assuming a trusted registration authority, and the knowledge soundness of NIZKP, it proves\TZdone{where does ``it'' refer to?}\JQdone{The above reasoning.} that the probability that result $W$ outputted by $\mathsf{Tally}(\cdot)$ does not correspond to the result
of the tally function $\rho$ computed on all last votes by honest voters who checked their ballots, by at most $|\mathbb{U}_{corr}|$ votes cast by corrupted voters, and by a subset of votes cast by honest voters who did not check their ballots is smaller or equal \TZdone{who is smaller or equal? Some probability I presume.} than $\epsilon_{sound}(\lambda)$. Hence, $\Pr[\mathcal{G}ame^{\mathcal{A}}_{Ver}(1^{\lambda},\mathbf{param})=1] \leq \epsilon_{sound}(\lambda)$. \TZdone{What I do not understand in the proof is in which way we treat the votes in $\mathbb{CK}$ and $\mathbb{HB}\setminus\mathbb{CK}$ differently. It seems as checking or no checking has no impact on security.}\qed

\end{appendices}

\end{document}